\documentclass[fp]{jpsj3} 
\usepackage{txfonts} 

\title{Holonomy Analysis of Optical-polarization Temperature Trajectories in Stress-induced Ferroelectric SrTiO$_3$}

\author{Hirotaka Manaka$^1$\thanks{E-mail address: manaka@eee.kagoshima-u.ac.jp}, Kazuma Seike$^1$, and Yoko Miura$^2$} 

\inst{$^1$Graduate School of Science and Engineering, Kagoshima University, Korimoto, Kagoshima 890-0065, Japan\\ $^2$National Institute of Technology, Suzuka College, Shiroko-cho, Suzuka, Mie 510-0294, Japan } 

\abst{We develop a data-induced geometric framework for temperature trajectories in optical-polarization data and apply it to temperature-dependent birefringence imaging of stress-induced ferroelectric SrTiO$_3$. By treating the measured response as an observable projection of local material states, the optical-polarization response at each pixel is cast as a trajectory along the temperature axis. Singular value decomposition separates trajectory-induced geometric objects associated with optical-polarization-direction structure and temperature-evolution modes. Holonomies are then defined by transporting these objects around closed loops in real space. The resulting maps reveal spatially localized connection mismatches correlated with an enhanced ferroelectric transition temperature and stress-related optical anisotropy. Local order parameters and angular-gradient analyses confirm that these loop-level signals are distinct from orientational disorder and simple spatial variation. The signed holonomy of the temperature-evolution modes further resolves positive and negative connection structures under a fixed frame convention. These results demonstrate that the data-induced connection geometry of temperature trajectories provides an experimentally accessible diagnostic of electromechanical inhomogeneity in SrTiO$_3$ under stress, without explicitly reconstructing hidden strain or electric-polarization fields.}

\begin{document}

\maketitle

\section{Introduction}

In ferroic materials, the local material state is often spatially inhomogeneous because ferroelectric, ferroelastic, and related degrees of freedom couple to defects, dislocations, surfaces, interfaces, external stress, and thermal history \cite{I1,I2,I3}. Such inhomogeneity gives rise to domains, domain walls, local strain and stress concentrations, and spatially varying electromechanical responses. When elastic and electric degrees of freedom are coupled, local stress and strain can modify the electric polarization, dielectric response, and electromechanical properties through piezoelectric and flexoelectric responses \cite{Fl3-2,Fl3}. Real-space characterization of these spatially varying responses is therefore essential for understanding ferroic phase transitions and electromechanical functionality.

Optical birefringence imaging offers a useful approach to this problem because it visualizes spatial variations in optical anisotropy associated with stress, strain, and electric-polarization-related responses \cite{manaka-bf4,manaka-bf6,manaka-bf7,manaka-bf9,manaka-bf11}. Conventional analyses typically treat images recorded at different temperatures ($T$) as independent spatial maps. In our previous holonomy analysis, birefringence-derived director fields at fixed $T$ windows were analyzed through closed-loop connection mismatch, demonstrating that loop-level geometric information can capture features not reducible to local gradients \cite{manaka-bf17}. Although useful for identifying anomalous regions at a fixed $T$, this snapshot-based approach does not fully exploit the information contained in the temperature evolution of the optical response at each pixel. Recent advances in operando imaging make it possible to acquire spatially resolved sequential data over an entire field of view as functions of control parameters such as temperature, time, external field, or applied stress. These data can be regarded not merely as a sequence of images, but also as a set of local response trajectories associated with individual pixels. An important analytical task is therefore to retain the evolution of each local response while determining how these trajectories are organized and connected in real space.

In the present study, we take a complementary viewpoint: $T$-dependent imaging data are regarded as both spatial maps and a collection of local response trajectories. During the cooling process, the optical-polarization state at each pixel evolves under the influence of the local strain field, domain configuration, defects, boundary conditions, and electromechanical coupling. These internal degrees of freedom are generally not all observed directly, but their influence can appear in the measured optical response. Thus, the temperature series measured at each pixel is treated as an observable projection of a local material state evolving along the $T$ axis. The ensemble of such trajectories defines a data-induced geometric structure in the optical-response space \cite{Diffusion,Diffusion2}.

This viewpoint is conceptually related to state-space reconstruction in nonlinear time-series analysis \cite{Takens, LatentFlow}. The present $T$-dependent birefringence data do not constitute a direct application of Takens' embedding theorem because $T$ is a control parameter rather than time in an autonomous dynamical system. Nevertheless, the underlying idea is analogous: a measured response trajectory can retain geometric fingerprints of internal degrees of freedom that are only indirectly observed. Here, we use this idea to construct an effective geometry from optical-polarization temperature trajectories, without claiming a direct reconstruction of hidden strain or electric polarization fields.

We apply this framework to $T$-dependent birefringence imaging of SrTiO$_3$ exhibiting stress-induced ferroelectricity \cite{ESR, w30, Di, Mc, w28, Di2, w29, X1, Sidoruk,manaka-bf8,manaka-bf10}. This system serves as a suitable test case because the ferroelectric transition temperature, stress concentration, and electromechanical response are spatially inhomogeneous under stress. Even when the external force is applied macroscopically along one direction, the local internal stress and strain fields can vary owing to boundary constraints, contact conditions, defects, dislocations, slip structures, plastic deformation, and other microstructural factors \cite{manaka-bf8,manaka-bf10}. The measured temperature trajectories therefore provide an experimentally accessible representation of spatially varying local material responses \cite{PST}.

A natural first step in analyzing such trajectories is to define distances or dissimilarities between them and to classify pixels according to their response patterns. Distance-based analyses can identify groups of similar trajectories and relate them to real-space regions \cite{manaka-bf13}. However, distances alone do not reveal how local geometric objects attached to neighboring trajectories are connected or how these connections accumulate upon transport around a closed loop in real space. To access this loop-level information, a connection between trajectory-induced local frames must be constructed.

Holonomy provides a natural geometric tool for probing such connection structures \cite{h1,ho1}. It is defined as the residual rotation remaining after locally defined orientations or frames are transported along a closed path and returned to their starting point. Building on our previous single-$T$ director-field analysis \cite{manaka-bf17}, we extend holonomy to local geometric objects induced by temperature trajectories. Singular value decomposition (SVD) is used to extract right-singular-vector structures associated with director axes in optical-polarization space and left-singular-vector frames associated with temperature-evolution modes in temperature-series space \cite{SVD}. These trajectory-induced objects are then transported along closed loops in real space to evaluate loop-level connection mismatch.

Because the local frames obtained from temperature trajectories possess gauge freedom, the holonomy must be defined under a consistent convention. We fix the gauge by prescribing sign and orientation rules. The robustness of the resulting holonomy is examined with respect to gauge choice and Gaussian process regression (GPR) posterior resampling \cite{GPR1,GPR2,GPR3}. We compare the resulting trajectory-induced holonomies with local order parameters and local angular-variation measures. The holonomy obtained here does not directly identify topological defects, true singularities, or the hidden strain and electric-polarization fields themselves \cite{ho2,h2}. Rather, it serves as an effective connection-geometric descriptor constructed from measured $T$-dependent optical data, providing a data-induced route to diagnosing spatially inhomogeneous electromechanical responses \cite{Diffusion,Diffusion2}.

\section{Birefringence Dataset and Temperature-trajectory Representation}

This section describes the $T$-dependent birefringence dataset and introduces the temperature-trajectory representation of the measured optical-polarization response. We first summarize the physical background of stress-induced ferroelectricity in SrTiO$_3$ and the spatial inhomogeneity observed in birefringence images. We then explain how the $T$ evolution of the optical-polarization response at each pixel serves as the basic object for the SVD and holonomy analyses presented below.

\subsection{Birefringence in SrTiO$_3$ under stress}

This study analyzes $T$-dependent birefringence imaging data of SrTiO$_3$ measured under a uniaxially applied force along [001], corresponding to a nominal stress of 231~MPa, at $\lambda=575$~nm \cite{manaka-bf8,manaka-bf10}. Here, the nominal stress is defined as the applied load divided by the specimen cross-sectional area normal to the loading direction and does not imply a spatially uniform local internal stress field. SrTiO$_3$ is a prototypical quantum paraelectric that undergoes a cubic-to-tetragonal structural phase transition at $T_{\rm c}=105$ K in the absence of stress \cite{STO1,STO2,STO3,Courtens}. Below $T_{\rm c}$, quantum fluctuations suppress ferroelectric order, and the quantum paraelectric state persists under zero stress. However, ferroelectricity can be induced by applying external perturbations such as an electric field \cite{STOFE1, STOFE2, w32, Hemberger, STOFE3, manaka-bf3, manaka-bf5} or mechanical stress \cite{ESR, w30, Di, Mc, w28, Di2, w29, X1, Sidoruk, manaka-bf8,manaka-bf10}, which weaken this suppression. In particular, under uniaxial stress, a stress-induced ferroelectric state appears below approximately 30~K \cite{manaka-bf8,manaka-bf10,manaka-bf13,manaka-bf12}. Although the external force is applied macroscopically along one direction, the local internal stress and strain fields can be spatially inhomogeneous owing to boundary conditions, contact conditions, defects, dislocations, slip structures, and plastic deformation \cite{manaka-bf12,manaka-bf15}. Consequently, the local ferroelectric transition temperature, optical anisotropy, and electromechanical response can vary with position within the specimen.

In birefringence imaging, spatial variations in local optical anisotropy are evaluated from the optical-polarization state of light transmitted through the specimen \cite{azz,manaka-bf1}. The data used here were acquired during continuous cooling from 300.0~K to 14.1~K at a constant rate \cite{manaka-bf8,manaka-bf10}. Circularly polarized light was incident on the specimen, and the transmitted polarization state was recorded at each pixel as a Stokes vector $\mbox{\boldmath $S$}$. The original image size was $384\times288$ pixels, and a total of 3,362 $T$-dependent imaging frames were acquired. In conventional birefringence analysis, the optical retardance and fast-axis direction are obtained from the incident and transmitted Stokes vectors.

Figure~1(a) shows the folded optical retardance map $\delta_{\rm fold}$ at the lowest measured temperature of 14.1~K \cite{manaka-bf8}. This map is obtained directly from the measured Stokes vectors without assigning the retardance branch beyond $\lambda/2$. Because the optical-polarization state alone cannot uniquely distinguish retardance branches related by $\delta$ and ($\lambda-\delta$), retardance values exceeding $\lambda/2$ are folded into the range $0\le \delta_{\rm fold}\le \lambda/2$ \cite{azz,manaka-bf1,manaka-bf2}. The auxiliary scale ($\lambda-\delta_{\rm fold}$) in Fig.~1(a) indicates the corresponding high-retardance branch. Thus, Fig.~1(a) should be interpreted as a folded representation of the measured optical response rather than as a monotonic scalar map of retardance magnitude. In the present study, we do not use scalar retardance maps; instead, we regard the $T$ evolution of the measured Stokes vector at each pixel as an optical-polarization temperature trajectory.

Previous analyses of the same $T$-dependent birefringence data demonstrated that $T_{\rm F}$ can be estimated at the single-pixel level from the temperature series \cite{manaka-bf13}. Figure~1(b) shows the spatial distribution of $T_{\rm F}$ obtained from the sample region extracted from Fig.~1(a), corresponding to a $302\times140$ pixel area. The retardance branch assignment was determined from the $\lambda$ dependence in the previous analysis \cite{manaka-bf8}. This assignment revealed that the region of enhanced $T_{\rm F}$ spatially coincides with the high-retardance branch, particularly in the lower-right region of the field of view. The boundaries of the images shown here and in the following spatial maps correspond to the boundaries of the field of view, not to the external boundaries or geometrical corners of the specimen. Previous statistical causal inference further suggested that this local enhancement of $T_{\rm F}$ is associated with piezoelectric and flexoelectric contributions \cite{manaka-bf14}. These findings suggest that the folded birefringence contrast reflects differences in local electromechanical response rather than mere optical contrast variations.

The 231~MPa condition analyzed here is in the plastic regime, as established by the yielding behavior and the nonuniform deformation structure observed in our previous study \cite{manaka-bf8,manaka-bf10}. After loading to 231~MPa at room $T$, a nonuniform stripe-like deformation structure associated with $\langle111\rangle$ slip was observed prior to cooling. Therefore, the analyzed specimen cannot be considered a homogeneous linear-elastic body subjected to a spatially uniform uniaxial internal stress. Plastic deformation, residual strain, dislocations, and slip structures can redistribute the local stress field and generate nonuniform strain fields and strain gradients. Phenomenologically, the stress-induced electric polarization may be written to first order as 

\begin{equation}
P_i \simeq e_{ijk}\epsilon_{jk}+\mu_{ijkl}\partial_l\epsilon_{jk}, 
\end{equation}

\noindent where the first term represents piezoelectric coupling to strain and the second term represents flexoelectric coupling to strain gradients \cite{Fl3-2,Fl3,Fl2}. In regions where stress concentration or strain gradients are substantial, the local electric-polarization response may change, potentially leading to electromechanical inhomogeneity related to the bound charge density 

\begin{equation}
\rho_{\rm b}=-\nabla\cdot\mbox{\boldmath $P$}.
\end{equation}

\noindent From this perspective, our previous holonomy analysis of single-$T$ director fields established that loop-level geometric mismatch provides information distinct from local orientational gradients \cite{manaka-bf17}. In the present study, we extend this viewpoint to temperature trajectories traced by the optical-polarization state at each pixel.

\subsection{Temperature trajectories and data-space geometry}

For a pixel position $(x,y)$, the measured Stokes vector $\mbox{\boldmath $S$}^{(T)}(x,y)$ obtained by varying $T$ traces the $T$ evolution of the optical-polarization state in data space \cite{azz,manaka-bf1}. Beyond a single-$T$ response, each pixel is characterized by its temperature trajectory \cite{manaka-bf13}. This trajectory combines the optical polarization states measured over the full temperature range. Therefore, it represents the measured temperature dependence at that pixel, rather than the local state at a single temperature. Because these trajectories are influenced by phase transitions, stress distributions, strain fields, boundary conditions, and local electromechanical coupling, the trajectory ensemble is expected to encode information on spatially varying material environments \cite{manaka-bf4,manaka-bf6,manaka-bf7,manaka-bf9,manaka-bf11, manaka-bf8, manaka-bf10, manaka-bf3, manaka-bf5}. In this study, we treat this trajectory ensemble as a data-induced geometry in optical-response space \cite{Diffusion,Diffusion2}.

A temperature trajectory is a high-dimensional, multicomponent sequence; direct use of all measured $T$ points can be sensitive to noise, amplitude variations, and local fluctuations. We therefore represent the dominant structure of each trajectory using a low-dimensional subspace. Specifically, after the reference alignment and normalization described below, a response matrix is constructed from the optical-polarization trajectory at each pixel, and its dominant components are extracted using SVD \cite{SVD}. The resulting subspace is represented as a point on a Grassmann manifold \cite{Grassmann, Grassmann2, Grassmann3}. This representation provides a basis for describing the $T$ evolution as a geometry induced by the measured data rather than as a real-space geometry assumed a priori. The following section describes the smoothing of temperature trajectories via GPR, the construction of response matrices, and the extraction of subspaces via SVD. Detailed derivations are provided in Supplementary Materials \cite{Supply}.

The analysis was implemented in Python 3.11, with numerical calculations primarily performed using standard linear-algebra libraries. The analyzed data consist of $T$-dependent birefringence images under 231~MPa at a wavelength of $\lambda=575$~nm acquired during continuous cooling from 300.0~K to 14.1~K \cite{manaka-bf8}. From the original images, we extracted the sample region comprising $302\times140=42{,}280$ pixels as shown in Fig.~1 and used all 3,362 imaging frames \cite{manaka-bf12,manaka-bf13}. The Grassmann representation, holonomy calculations, and resampling analyses were performed on a workstation with an Intel Core i9-13900K CPU, 128 GB of memory, and an NVIDIA RTX A4000 GPU. The GPU accelerated the iterative calculations and resampling procedures for a large number of pixels.

\section{Results and Discussion}

Here, we analyze the connection geometry induced by the $T$-dependent optical-polarization trajectories in SrTiO$_3$ under stress. The measured Stokes vectors $\mbox{\boldmath $S$}^{(T)}(x,y)$ are first converted, after reference alignment and normalization, into optical-polarization trajectories on the Poincar\'e sphere $S^2$. We then apply SVD to each trajectory; the right singular vectors characterize local director-axis structures in optical-polarization-direction space, whereas the dominant left singular vectors define a representative local frame attached to the temperature-evolution-mode space. By connecting these trajectory-induced objects along closed loops in real space, we define the right-side holonomy $\omega_{\mathcal R}$ and the left-side holonomy $\omega_{\mathcal L}$. We first examine the spatial distributions and resampling stability of these holonomies. We then compare them with local geometric quantities and stress-related inhomogeneity. In this construction, $\omega_{\mathcal R}$ and $\omega_{\mathcal L}$ probe complementary aspects of the temperature trajectory: $\omega_{\mathcal R}$ reflects the closed-loop mismatch of the dominant optical-polarization axis, whereas $\omega_{\mathcal L}$ reflects the corresponding mismatch of the local temperature-evolution frame.

\subsection{Construction of temperature trajectories}

To describe the $T$ evolution of the transmitted optical-polarization state at each pixel $(x,y)$, we denote the measured $T$ points by $\{T_m\}_{m=1}^{M_T}$, where $M_T$ is the total number of measured $T$ points. At each $T_m$, the measured incident and transmitted Stokes vectors are denoted by $\boldsymbol S_{\rm In}^{(T_m)}(x,y)$ and $\boldsymbol S_{\rm Out}^{(T_m)}(x,y)$, respectively \cite{azz,manaka-bf1}. The observed incident optical polarization is close to the circular reference state $\boldsymbol e_{\rm Base}$ but contains small position-dependent deviations. To suppress extraneous geometric fluctuations arising from these deviations, we determine at each measured point the shortest rotation mapping the normalized incident Stokes vector onto $\boldsymbol e_{\rm Base}$ and apply the same rotation to the normalized transmitted Stokes vector. This auxiliary reference-alignment rotation is implemented by a unit quaternion and is distinct from the holonomy defined below. After renormalization, we obtain the transmitted optical-polarization direction $\widehat{\boldsymbol n}(x,y;T_m)\in S^2$, $(m=1,\ldots,M_T)$. Each pixel is represented by a point sequence on the Poincar\'e sphere, $T_m\mapsto \widehat{\boldsymbol n}(x,y;T_m)$, which we call the temperature trajectory on the measured $T$ points. At this stage, the temperature trajectory is a sequence of normalized optical-polarization directions on $S^2$; projective or director-type identifications are introduced only later, when sign-indeterminate singular-vector axes are analyzed.

The measured $T$ points $\{T_m\}$ are not perfectly equally spaced and exhibit small fluctuations from temperature control. In addition, each component of $\boldsymbol S_{\rm In}^{(T_m)}$ and $\boldsymbol S_{\rm Out}^{(T_m)}$ contains measurement errors, residual errors from incident optical-polarization correction, and small temperature-control-related fluctuations. Using the measured series as input, we therefore apply GPR along the $T$ direction and evaluate the posterior mean and posterior samples on a common $T$ grid, $\mathcal T_{\rm grid}=\{\tau_i\}_{i=1}^{M_{\rm grid}}$, where $M_{\rm grid}$ is the number of grid points. Here, $\tau_i$ is not $T_m$ itself, but an evaluation point used to compare the GPR-smoothed series on a common grid for all pixels. In this study, $\tau_i$ was taken as an equally spaced grid with an interval of 0.5~K. GPR is performed independently for each component of the unit direction vector $\widehat{\boldsymbol n}(x,y;T_m)$. Hereafter, we omit the pixel position $(x,y)$ and write the $j$-th Cartesian component of $\widehat{\boldsymbol n}(x,y;T_m)\in S^2$ as $y_{j,m}$, with $j=\mathsf{x},\mathsf{y},\mathsf{z}$ \cite{Supply}. For each component, we introduce a latent function $f_j(T)$ that varies smoothly with $T$:

\begin{eqnarray}
y_{j,m}
&=&
f_j(T_m)+\varepsilon_{j,m},\\
\varepsilon_{j,m} &\sim& \mathcal N(0,\sigma_{\rm e}^2),
\end{eqnarray}

\noindent and the likelihood is

\begin{equation}
p(\boldsymbol y_j|\boldsymbol f_j)
=
\prod_{m=1}^{M_T}
\mathcal N
\left(
y_{j,m};
f_j(T_m),
\sigma_{\rm e}^2
\right),
\end{equation}

\noindent where $\boldsymbol y_j=(y_{j,1},\ldots,y_{j,M_T})^{\mathsf T}$ and $\boldsymbol f_j=(f_j(T_1),\ldots,f_j(T_{M_T}))^{\mathsf T}$. For the latent function, we assume a Gaussian process prior with a radial basis function (RBF) kernel,

\begin{eqnarray}
f_j(T) &\sim& {\rm GP}(0,k(T,T')),\\
k(T,T') &=&
\sigma_f^2
\exp\left[
-\frac{(T-T')^2}{2\ell_T^2}
\right].
\end{eqnarray}

\noindent We set $\ell_T=15.0~{\rm K}$, $\sigma_f^2=1.0$, and $\sigma_{\rm e}^2=1.0\times10^{-3}$. The length scale was chosen with reference to our previous temperature-trajectory analysis using long short-term memory (LSTM) \cite{manaka-bf15}, and the amplitude reflects the fact that each component $\widehat n_j(T_m)$ lies in the range $[-1,1]$. The effective fluctuation variance corresponds to $\pm3\sigma_{\rm e}\simeq \pm0.1$, which accommodates small experimental variations while avoiding excessive smoothing. The GPR step is used only to obtain a smooth common-grid representation of each measured temperature trajectory together with its posterior resampling uncertainty; the same hyperparameters are used for all pixels, and no pixel-dependent tuning is introduced.

The GPR model provides the posterior mean $\boldsymbol \mu(\tau_i) = (\mu_{\mathsf{x}}(\tau_i), \mu_{\mathsf{y}}(\tau_i), \mu_{\mathsf{z}}(\tau_i))^{\mathsf T}$ on the evaluation grid. Because the three components are regressed independently, $\boldsymbol \mu(\tau_i)$ does not generally preserve unit length \cite{GPR1,GPR2,GPR3}. We renormalize it at each grid point and define the representative smoothed trajectory as

\begin{equation}
\widehat{\boldsymbol n}^{\rm ref}(\tau_i)
=
\widehat{\boldsymbol n}^{(b=1)}(\tau_i)
:=
\frac{
\boldsymbol \mu(\tau_i)
}{
\|\,\boldsymbol \mu(\tau_i)\,\|
}
\in S^2,
\end{equation}

\noindent where the superscript ``ref'' denotes the posterior-mean series, assigned the index $b=1$ in the resampling analysis.

To propagate trajectory uncertainty into the holonomy analysis, we further generate posterior-resampled trajectories. Because the GPR posterior covariance contains correlations along the $T$ direction, the resampled fluctuations represent correlated deformations of the trajectory rather than independent noise at each $T$ point. These fluctuations are generated in the spherical tangent plane of the reference series at each $\tau_i$ and renormalized \cite{GPR1,GPR2,GPR3}, giving

\begin{equation}
\widehat{\boldsymbol n}^{(b)}(\tau_i)\in S^2,
\qquad
b=2,\ldots,1000.
\end{equation}

\noindent In what follows, results are mainly shown for the reference series $(b=1)$, whereas the posterior-resampled series are used to evaluate the resampling stability of the holonomy quantities. The detailed formulation of GPR and posterior resampling is given in Supplementary Materials \cite{Supply}.

\subsection{SVD representation of temperature trajectories}

We decompose the temperature trajectory $\{\widehat{\boldsymbol n}^{(b)}(x,y;\tau_i)\}$ $(b=1,\ldots,1000)$ at each pixel into $T$-direction and optical-polarization-direction components. Unless stated otherwise, the same operation is applied independently for each $b$, and the superscript $(b)$ is omitted below. To focus on the low-$T$ response, which encompasses both the structural phase transition and the stress-induced ferroelectric transition, we use the evaluation-grid points satisfying $14.0\,{\rm K}\le \tau_i \le 135.0\,{\rm K}$ and define

\begin{equation}
\mathcal{T}_{\le 135\,\mathrm{K}}
=
\left\{
\tau_i\in\mathcal{T}_{\rm grid}
\,\middle|\,
14.0~\mathrm{K}\le \tau_i \le 135.0~\mathrm{K}
\right\},
\end{equation}

\noindent where 14.0~K is the lower edge of the common 0.5~K evaluation grid, while the lowest measured $T$ in the original data is 14.1~K. For notational simplicity, the elements of this set are written as $\{\tau_1,\ldots,\tau_{|\mathcal{T}_{\le 135\,\mathrm{K}}|}\}$, and the pixel position $(x,y)$ is labeled by ${\mathsf P}$ and used as a subscript. The temperature trajectory at pixel ${\mathsf P}$ is then

\begin{equation}
X_{\mathsf P}
=
\begin{pmatrix}
\widehat{\boldsymbol n}_{\mathsf P}(\tau_1)^{\mathsf T}\\
\widehat{\boldsymbol n}_{\mathsf P}(\tau_2)^{\mathsf T}\\
\vdots\\
\widehat{\boldsymbol n}_{\mathsf P}
\left(
\tau_{|\mathcal{T}_{\le 135\,\mathrm{K}}|}
\right)^{\mathsf T}
\end{pmatrix}
\in
\mathbb{R}^{|\mathcal{T}_{\le 135\,\mathrm{K}}|\times 3},
\end{equation}

\noindent where each row corresponds to one $T$ point and the three columns correspond to the components in the direction space in which the Poincar\'e sphere is embedded. Thus, $X_{\mathsf P}$ is a matrix representation of the trajectory traced by the transmitted optical-polarization state as $T$ varies at pixel ${\mathsf P}$.

We apply reduced SVD without subtracting the column mean. This choice preserves the directional structure of the entire optical-polarization temperature trajectory on the Poincar\'e sphere, including its absolute direction in optical-response space, rather than highlighting only fluctuations around a local average direction. We decompose

\begin{equation}
X_{\mathsf P}
=
U_{\mathsf P}\Sigma_{\mathsf P}V_{\mathsf P}^{\mathsf T},
\end{equation}

\noindent where $U_{\mathsf P}\in\mathbb{R}^{|\mathcal{T}_{\le 135\,\mathrm{K}}|\times 3}$ contains the left singular vectors in the $T$ direction, $\Sigma_{\mathsf P}\in\mathbb{R}^{3\times 3}$ is the diagonal singular-value matrix, and $V_{\mathsf P}\in\mathbb{R}^{3\times 3}$ contains the right singular vectors in the optical-polarization-direction space. Singular values are non-negative and ordered in descending order. The $r\text{-th}$ singular value is denoted by $\sigma_{{\mathsf P},r}$, and the corresponding left and right singular vectors by $\boldsymbol{\mathsf u}_{{\mathsf P},r}$ and $\boldsymbol{\mathsf v}_{{\mathsf P},r}$ $(r=1,2,3)$, respectively, so that

\begin{eqnarray}
U_{\mathsf P}
&=&
(
\boldsymbol{\mathsf u}_{{\mathsf P},1},
\boldsymbol{\mathsf u}_{{\mathsf P},2},
\boldsymbol{\mathsf u}_{{\mathsf P},3}
),\\
V_{\mathsf P}
&=&
(
\boldsymbol{\mathsf v}_{{\mathsf P},1},
\boldsymbol{\mathsf v}_{{\mathsf P},2},
\boldsymbol{\mathsf v}_{{\mathsf P},3}
),
\end{eqnarray}

\noindent where $\boldsymbol{\mathsf v}_{{\mathsf P},r}\in\mathbb{R}^{3}$ belongs to the optical-polarization-direction space and $\boldsymbol{\mathsf u}_{{\mathsf P},r}\in \mathbb{R}^{|\mathcal{T}_{\le 135\,\mathrm{K}}|}$ is a vector representing a temperature-evolution mode. From the SVD,

\begin{equation}
X_{\mathsf P}\boldsymbol{\mathsf v}_{{\mathsf P},r}
=
\sigma_{{\mathsf P},r}
\boldsymbol{\mathsf u}_{{\mathsf P},r},
\end{equation}

\noindent where $X_{\mathsf P}\boldsymbol{\mathsf v}_{{\mathsf P},r}$ is the temperature series obtained by projecting the trajectory onto the right singular direction $\boldsymbol{\mathsf v}_{{\mathsf P},r}$, expressed as $\sigma_{{\mathsf P},r}\boldsymbol{\mathsf u}_{{\mathsf P},r}$. Thus, $V_{\mathsf P}$ encodes the dominant directions or subspaces occupied by the trajectory in optical-polarization-direction space \cite{Supply}. In this sense, the right singular vectors represent the dominant geometric direction of the optical-polarization trajectory. In contrast, $U_{\mathsf P}$ describes how the corresponding components are distributed and evolve along the temperature axis. Therefore, the left singular vectors represent the temperature evolution of the trajectory. Here, the term ``temperature-evolution mode'' refers to a data-derived temperature profile extracted from the measured trajectory and not to a microscopic excitation of the material.

The global contribution ratios, evaluated by summing squared singular values over all analyzed pixels, are approximately $84.7\%$ for the first component and exceed $99\%$ for the first two components combined \cite{Supply}. This global low-rank structure motivates representing the temperature trajectories by a dominant principal axis and an associated two-dimensional trajectory plane in the three-dimensional optical-polarization-direction space. As the object of the right-side holonomy, we use the first right singular vector $\boldsymbol{\mathsf v}_{{\mathsf P},1}$ rather than the trajectory plane. Because singular vectors have a sign ambiguity, $\boldsymbol{\mathsf v}_{{\mathsf P},1}$ and $-\boldsymbol{\mathsf v}_{{\mathsf P},1}$ represent the same one-dimensional subspace. The right-side object is a director-axis field,

\begin{equation}
{\mathsf P}
\longmapsto
[
\boldsymbol{\mathsf v}_{{\mathsf P},1}
]
\in
\mathrm{Gr}(1,3)
\cong
\mathbb{R}P^2,
\end{equation}

\noindent where $[\cdot]$ denotes the identification $\boldsymbol{\mathsf v}_{{\mathsf P},1} \sim -\boldsymbol{\mathsf v}_{{\mathsf P},1}$. The right-side holonomy $\omega_{\mathcal R}$, defined below, is evaluated by connecting this director-axis field along closed loops in real space.

The first two right singular vectors additionally define the local plane in which the temperature trajectory is primarily distributed,

\begin{equation}
\mathcal W_{\mathsf P}
=
{\rm span}
\left(
\boldsymbol{\mathsf v}_{{\mathsf P},1},
\boldsymbol{\mathsf v}_{{\mathsf P},2}
\right)
\in
\mathrm{Gr}(2,3).
\label{W_mathsfP}
\end{equation}

\noindent In three-dimensional space, a two-dimensional plane and its one-dimensional orthogonal complement are in one-to-one correspondence. Because $\mathcal W_{\mathsf P}$ can equivalently be represented by its normal director rather than by the in-plane principal axis $[\boldsymbol{\mathsf v}_{{\mathsf P},1}]$, we define

\begin{equation}
\left[
\widehat{\boldsymbol{\mathsf n}}_{\rm plane}({\mathsf P})
\right]
=
\left[
\frac{
\boldsymbol{\mathsf v}_{{\mathsf P},1}
\times
\boldsymbol{\mathsf v}_{{\mathsf P},2}
}{
\left\|
\boldsymbol{\mathsf v}_{{\mathsf P},1}
\times
\boldsymbol{\mathsf v}_{{\mathsf P},2}
\right\|
}
\right],
\label{n_plane_P}
\end{equation}

\noindent where $[\widehat{\boldsymbol{\mathsf n}}_{\rm plane}({\mathsf P})]$ is the normal director of the trajectory plane. In what follows, $[\boldsymbol{\mathsf v}_{{\mathsf P},1}]$ is used for the right-side holonomy, while $\mathcal W_{\mathsf P}$ or its normal director is used to evaluate the local order parameter $S_{\rm plane}$.

We next turn to the left-singular-vector side. The left singular vectors belong to the temperature-series space. Because the first two singular components account for the dominant global contribution, the corresponding left singular vectors $\boldsymbol{\mathsf u}_{{\mathsf P},1}$ and $\boldsymbol{\mathsf u}_{{\mathsf P},2}$ are used to represent the two dominant modes of $T$ evolution. We focus on the two-dimensional subspace

\begin{equation}
\mathcal U_{\mathsf P}
=
{\rm span}
\left(
\boldsymbol{\mathsf u}_{{\mathsf P},1},
\boldsymbol{\mathsf u}_{{\mathsf P},2}
\right)
\in
\mathrm{Gr}
\left(
2,
|\mathcal{T}_{\le 135\,\mathrm{K}}|
\right).
\end{equation}

\noindent Unlike the right-side plane in three-dimensional space, this two-dimensional subspace cannot be naturally replaced by a one-dimensional normal axis. The meaningful object is therefore the temperature-evolution-mode space $\mathcal U_{\mathsf P}$ itself, not the sign or orientation of each individual left singular vector. Indeed, the transformation

\begin{equation}
(
\boldsymbol{\mathsf u}_{{\mathsf P},1},
\boldsymbol{\mathsf u}_{{\mathsf P},2}
)
\rightarrow
(
\boldsymbol{\mathsf u}_{{\mathsf P},1},
\boldsymbol{\mathsf u}_{{\mathsf P},2}
)H,
\qquad
H\in \mathrm{O}(2),
\end{equation}

\noindent leaves the spanned subspace unchanged. To compare temperature-evolution-mode spaces between neighboring pixels and to define a connection along closed loops in real space, we choose a representative orthonormal frame

\begin{equation}
\widetilde{U}_{\mathsf P}
=
(
\boldsymbol{\mathsf u}_{{\mathsf P},1},
\boldsymbol{\mathsf u}_{{\mathsf P},2}
)
\in
\mathbb{R}^{|\mathcal{T}_{\le 135\,\mathrm{K}}|\times 2}.
\end{equation}

\noindent The subspace $\mathcal U_{\mathsf P}$ is invariant under an $\mathrm{O}(2)$ change of basis, and $\widetilde{U}_{\mathsf P}$ is one representative frame selected under the SVD convention. The full $\mathrm{O}(2)$ freedom encompasses both orientation-preserving and orientation-reversing transformations. In two dimensions, the orientation-preserving sector is $\mathrm{SO}(2)=\{H\in\mathrm{O}(2)\mid \det H=+1\}$, while transformations with $\det H=-1$ include a reflection that reverses the orientation of the local frame. Because the left-side holonomy is defined as a signed in-plane rotation angle, such orientation reversal would alter the sign convention of this angle. To define the signed left-side holonomy, we fix an orientation convention and compare neighboring representative frames through the closest orientation-preserving $\mathrm{SO}(2)$ rotation \cite{Supply}. These rotations accumulate along closed loops in real space. Thus, $\omega_{\mathcal L}$ does not measure the sign of individual left singular vectors but rather the signed residual rotation of local $\mathrm{SO}(2)$ frames attached to the temperature-evolution-mode spaces under the fixed frame convention. The scalar magnitude $|\omega_{\mathcal L}|$ serves as a robust measure of left-frame mismatch strength. The signed field $\omega_{\mathcal L}(x,y)$ is interpreted only after fixing the real-space coordinate system, loop orientation, and orientation-preserving $\mathrm{SO}(2)$ frame convention. Under this convention, its positive and negative regions represent a signed connection structure rather than an absolute physical sign at an isolated point.

Overall, SVD is used here for dimensional reduction and to separate two complementary aspects of each temperature trajectory. The right singular vectors define the dominant direction or subspace in optical-polarization-direction space, whereas the left singular vectors describe the corresponding temperature evolution. These trajectory-derived objects are subsequently compared between neighboring pixels and transported around closed loops in real space to evaluate their residual connection mismatch.

\subsection{Closed-loop connections and holonomy}

Holonomy is defined as the residual rotation obtained by connecting the trajectory-induced objects along closed loops in real space. At each step, the direction, subspace, or frame at one pixel is compared with that at a neighboring pixel through the minimum rotation required to align them. These local rotations are then accumulated sequentially around a closed real-space loop. If the rotations cancel after one traversal, the transported object returns to its initial orientation. Otherwise, the remaining rotation defines the holonomy. We first consider the right-side director-axis field. Here, returning from the pixel index ${\mathsf P}$ to the real-space coordinate $(x,y)$, we denote the local director axis by $\widehat{\boldsymbol n}_{\mathcal R}(x,y)\in S^2$. $\widehat{\boldsymbol n}(x,y;T)$ denotes a point on the optical-polarization trajectory at a given $T$, whereas $\widehat{\boldsymbol n}_{\mathcal R}(x,y)$ is the representative axis extracted from the entire temperature trajectory. The connection of the director field is defined as the minimum rotation between nearest-neighbor pixels under the identification $\widehat{\boldsymbol n}_{\mathcal R}\sim-\widehat{\boldsymbol n}_{\mathcal R}$. For each edge, the signs of the two unit vectors are chosen so that their inner product is non-negative, and the resulting minimum rotation is represented by a unit quaternion. We denote the quaternions for nearest-neighbor edges in the $+x$ and $+y$ directions by $q_x(x,y)$ and $q_y(x,y)$, respectively. For edges in the opposite directions, the inverse quaternion is used. The detailed formulation is given in Supplementary Materials \cite{Supply}.

As in our previous studies, the origin of the real-space coordinate system is set at the lower-right corner of the field of view \cite{manaka-bf17}. Under this convention, the $+x$ direction points to the left in the image and the $+y$ direction points upward. For each pixel $(x,y)$, we define an $L\times L$ square loop $\partial\Box_L(x,y)$ with $(x,y)$ as its lower-right base point. The loop proceeds in the order $+x$, $+y$, $-x$, and $-y$, which fixes the loop orientation. The loop product of the right-side director field is then defined by multiplying the edge quaternions in the traversal order along the loop:

\begin{equation}
Q_L(x,y)
=
\prod_{e\in\overrightarrow{\partial\Box_L(x,y)}} q_e
=
\left(
w_L(x,y),\boldsymbol v_L(x,y)
\right),
\label{Q_L_xy}
\end{equation}

\noindent where $q_e$ denotes the quaternion assigned to each edge, or its inverse for an oppositely oriented edge \cite{Supply}. The order of multiplication is essential because quaternion multiplication is generally noncommutative. The loop product $Q_L(x,y)$ is a unit quaternion, that is, an element of ${\rm SU}(2)$ and represents the three-dimensional residual rotation accumulated by completing the closed loop. Since $Q_L$ and $-Q_L$ correspond to the same ${\rm SO}(3)$ rotation, we choose the representative with $w_L(x,y)\ge 0$. The right-side holonomy angle is then defined as

\begin{equation}
\omega_{\mathcal R}(x,y)
=
2\arctan2
\left(
\|\boldsymbol v_L(x,y)\|,
w_L(x,y)
\right),
\end{equation}

\noindent where $\omega_{\mathcal R}(x,y)$ is a non-negative rotation angle. The condition $\omega_{\mathcal R}=0$ indicates that the connection along the closed loop is trivial, whereas $\omega_{\mathcal R}>0$ indicates a geometric mismatch in the local director-axis field within the loop. In the numerical implementation and figures, this angle is expressed in degrees.

We next define the left-side holonomy $\omega_{\mathcal L}$. The temperature-evolution-mode space has an ${\rm O}(2)$ basis freedom, and $\widetilde{U}(x,y)$ is a representative local frame chosen under the fixed SVD convention. The meaningful object is therefore not the sign of each individual left singular vector, but the connection of the representative frames attached to the temperature-evolution-mode spaces of neighboring pixels. Between neighboring pixels, the relative arrangement of two such frames is represented by

\begin{eqnarray}
G_x(x,y) &=&
\widetilde{U}(x,y)^{\mathsf T}
\widetilde{U}(x+1,y), \\
G_y(x,y) &=&
\widetilde{U}(x,y)^{\mathsf T}
\widetilde{U}(x,y+1).
\end{eqnarray}

\noindent In general, $G_x(x,y)$ and $G_y(x,y)$ are not exactly orthogonal. We define $R_x(x,y)$ and $R_y(x,y)$ as the orthogonal parts of their polar decompositions \cite{Supply}. For the present dataset, the polar factors were found to satisfy $\det R_x(x,y)=\det R_y(x,y)=+1$ for all valid nearest-neighbor edges; these matrices are therefore treated as elements of ${\rm SO}(2)$. For oppositely oriented edges, the inverse matrices are used. Multiplying these local ${\rm SO}(2)$ connections in order along the same square loop $\partial\Box_L(x,y)$ yields

\begin{equation}
R_L(x,y) = 
\prod_{e\in\overrightarrow{\partial\Box_L(x,y)}} R_e
\in {\rm SO}(2),
\end{equation}

\noindent where $R_e$ denotes $R_x$, $R_y$, or the appropriate inverse. Since $R_L(x,y)\in{\rm SO}(2)$, its $(1,1)$ and $(2,1)$ components are the cosine and sine of the residual in-plane rotation angle, respectively \cite{Supply}. The left-side holonomy angle is defined as 

\begin{equation}
\omega_{\mathcal L}(x,y)
=
\arctan2
\left(
(R_L(x,y))_{21},
(R_L(x,y))_{11}
\right),
\end{equation}

\noindent where $\omega_{\mathcal L}$ is a signed angle (expressed in degrees) representing the residual in-plane rotation obtained by connecting the local temperature-evolution frames along the closed loop. Because the left-side holonomy is evaluated in the orientation-preserving $\mathrm{SO}(2)$ sector with $\det R_e=+1$, the sign of $\omega_{\mathcal L}$ should be interpreted as the orientation of the accumulated frame mismatch under the fixed loop-orientation and frame convention, not as an absolute local electric polarity.

In summary, $\omega_{\mathcal R}$ measures the magnitude of the three-dimensional residual rotation of the director-axis field, whereas $\omega_{\mathcal L}$ measures the signed in-plane residual rotation of representative local ${\rm SO}(2)$ frames attached to two-dimensional temperature-evolution-mode spaces. The latter does not represent the sign of individual left singular vectors. Rather, under the fixed frame convention, it represents the connection mismatch of the temperature-evolution frames, reflecting a trajectory-induced geometric structure that is absent from single-$T$ birefringence images.

\subsection{Local order parameters associated with holonomy}

We next define two local orientational order parameters associated with the trajectory-induced geometric quantities. The first, $S_{\mathcal R}$, quantifies the local order of the residual rotation axis obtained from the right-side holonomy. The second, $S_{\rm plane}$, quantifies the local orientational order of the temperature-trajectory plane $\mathcal W_{\mathsf P}$ on the right-singular-vector side.

The residual rotation axis associated with $\omega_{\mathcal R}$ is extracted from the right-side loop product $Q_L(x,y)$ in Eq.~(\ref{Q_L_xy}). Here, $\boldsymbol v_L(x,y)$ is the vector part of the loop-product quaternion, distinct from the right singular vector $\boldsymbol{\mathsf v}_{{\mathsf P},r}$. A unit quaternion can be written as

\begin{equation}
Q_L =
\left(
\cos\frac{\omega_{\mathcal R}}{2},
\widehat{\boldsymbol a}_{\mathcal R}
\sin\frac{\omega_{\mathcal R}}{2}
\right).
\end{equation}

\noindent For pixels where the residual rotation axis is retained, we define

\begin{equation}
\widehat{\boldsymbol a}_{\mathcal R}(x,y)
=
\frac{
\boldsymbol v_L(x,y)
}{
\|\boldsymbol v_L(x,y)\|
},
\end{equation}

\noindent where $\widehat{\boldsymbol a}_{\mathcal R}(x,y)$ is the axis of the three-dimensional residual rotation given by the loop product $Q_L(x,y)$. When $\omega_{\mathcal R}$ is very small, the loop product is close to the identity rotation and the residual rotation axis becomes numerically ill-defined. The axis-based local-order analysis is therefore restricted to the high-holonomy pixel set corresponding to the top $p\%$ of the $\omega_{\mathcal R}$ distribution within the valid holonomy region.

To quantify the local alignment of $\widehat{\boldsymbol a}_{\mathcal R}(x,y)$, we consider an $L_{\rm loc}\times L_{\rm loc}$ local square region around a base point $(x,y)$. The set of pixels within this region for which $\widehat{\boldsymbol a}_{\mathcal R}$ is retained is denoted $\Omega^{\rm valid}_{\mathcal R,L_{\rm loc}}$. As detailed in Supplementary Materials \cite{Supply}, the high-holonomy selection serves two purposes: only the residual rotation axes on the top $p\%$ high-$\omega_{\mathcal R}$ pixel set are retained as input samples for the local second-moment tensor, and the displayed and summarized $S_{\mathcal R}$ values are restricted to base pixels in the same high-holonomy set. The local window size $L_{\rm loc}$ can be chosen independently of the holonomy loop size $L$, while $\Omega^{\rm valid}_{\mathcal R,L_{\rm loc}}$ is further constrained by the region in which the $L\times L$ holonomy loop can be constructed \cite{Supply}. We define the second-moment matrix

\begin{equation}
\boldsymbol A_{\mathcal R}(x,y)
=
\frac{
\sum_{(x',y')\in
\Omega^{\rm valid}_{\mathcal R,L_{\rm loc}}}
\widehat{\boldsymbol a}_{\mathcal R}(x',y')\,
\widehat{\boldsymbol a}_{\mathcal R}(x',y')^{\mathsf T}
}{
\left|
\Omega^{\rm valid}_{\mathcal R,L_{\rm loc}}
\right|
}.
\end{equation}

\noindent Letting $\lambda_{\max}^{(\mathcal R)}(x,y)$ denote the largest eigenvalue of $\boldsymbol A_{\mathcal R}(x,y)$, the right-side local order parameter is defined as

\begin{equation}
S_{\mathcal R}(x,y)
=
\frac{
3\lambda_{\max}^{(\mathcal R)}(x,y)-1
}{2}.
\end{equation}

\noindent This nematic-type order parameter quantifies the local orientational order of the residual rotation axes under the identification $\widehat{\boldsymbol a}_{\mathcal R}\sim -\widehat{\boldsymbol a}_{\mathcal R}$ \cite{S1,S2,S3}. $S_{\mathcal R}\simeq 1$ thus indicates well-aligned residual rotation axes, while small $S_{\mathcal R}$ indicates weak local orientational order.

We next define $S_{\rm plane}$ as a complementary local order parameter for the trajectory plane on the right-singular-vector side. Because $\omega_{\mathcal L}$ is an in-plane rotation angle in the temperature-series space, it does not directly yield a three-dimensional residual rotation axis analogous to that of the right-side holonomy. We therefore use the normal director $[\widehat{\boldsymbol{\mathsf n}}_{\rm plane}(x,y)]$ defined in Eq.~(\ref{n_plane_P}) to characterize the local orientational order of the trajectory planes. Around a base point $(x,y)$, we take an $L_{\rm loc}\times L_{\rm loc}$ local region and denote by $\Omega^{\rm valid}_{{\rm plane},L_{\rm loc}}$ the set of pixels where $\widehat{\boldsymbol{\mathsf n}}_{\rm plane}$ is validly defined. The second-moment matrix is then

\begin{equation}
\boldsymbol A_{\rm plane}(x,y)
=
\frac{
\sum_{(x',y')\in
\Omega^{\rm valid}_{{\rm plane},L_{\rm loc}}}
\widehat{\boldsymbol{\mathsf n}}_{\rm plane}(x',y')\,
\widehat{\boldsymbol{\mathsf n}}_{\rm plane}(x',y')^{\mathsf T}
}{
\left|
\Omega^{\rm valid}_{{\rm plane},L_{\rm loc}}
\right|
},
\end{equation}

\noindent where $\boldsymbol A_{\rm plane}(x,y)$ depends on $L_{\rm loc}$ but not on the holonomy loop size $L$. Letting $\lambda_{\max}^{({\rm plane})}(x,y)$ denote the largest eigenvalue of $\boldsymbol A_{\rm plane}(x,y)$, we define

\begin{equation}
S_{\rm plane}(x,y)
=
\frac{
3\lambda_{\max}^{({\rm plane})}(x,y)-1
}{2}.
\end{equation}

\noindent This quantity has the same nematic-type form as $S_{\mathcal R}$ and measures the degree of alignment of the local temperature-trajectory-plane normals within the local region.

\subsection{Holonomy maps and local order-parameter maps}

Using the quantities defined above, we visualize the closed-loop connection mismatch contained in the temperature trajectories of SrTiO$_3$ under stress in real space. Representative results are shown for $L=10$ and $L_{\rm loc}=5$. For $S_{\mathcal R}(x,y)$, the analysis is restricted to the high-holonomy region where $\widehat{\boldsymbol a}_{\mathcal R}$ is numerically well defined. Specifically, the top $p=20\%$ pixel set of $\omega_{\mathcal R}(x,y)$ is used both to retain the residual-axis samples entering the local second-moment tensor and to select the displayed base pixels. The validity of these representative values, together with results for other values of $L$, $L_{\rm loc}$, and $p$, is discussed later and further detailed in Supplementary Materials \cite{Supply}.

Figure~2 presents the holonomy maps for $L=10$. Figure~2(a) shows $\omega_{\mathcal R}(x,y)$ obtained from the right-side director field. The right singular direction represents the dominant optical-polarization axis extracted from the temperature trajectory at each pixel. Therefore, $\omega_{\mathcal R}$ measures the magnitude of the residual mismatch of this dominant axis after its ordered transport around a closed loop in real space. Large $\omega_{\mathcal R}$ indicates that the dominant optical-polarization-axis structure cannot be connected compatibly around the loop. The large values are concentrated in the stress-concentrated region in the lower-right part of the field of view, whereas they remain small in most other regions. Figure~2(b) shows $\omega_{\mathcal L}(x,y)$ obtained from the left-side frame field. The left-side frame represents the principal form of the temperature evolution of the optical-polarization response. Thus, $\omega_{\mathcal L}$ measures the residual in-plane rotation of the temperature-evolution frame after transport around the same real-space loop. Unlike $\omega_{\mathcal R}$, which is represented by a non-negative mismatch magnitude, $\omega_{\mathcal L}$ is signed. These signs do not merely reflect differences in holonomy magnitude. Rather, under fixed real-space coordinates, loop orientation, and frame convention, the signs identify regions in which the temperature-evolution frames attached to the temperature trajectories in data space accumulate in-plane connection residuals of opposite orientations after transport around the real-space loop. Therefore, the signs distinguish the orientation of the loop-induced residual rotation, rather than the sign of an absolute local physical quantity.

Quantitatively, we compared two binary masks: the top $p=20\%$ mask of $\omega_{\mathcal R}$ and the stress-concentrated mask previously identified by the $K$-shape method \cite{manaka-bf13}. Among the pixels in the top-$20\%$ $\omega_{\mathcal R}$ mask, 64.5\% are located inside the stress-concentrated mask. The overlap between these two masks, evaluated using the intersection over union (IoU) and Dice coefficient, gives IoU = 0.396 and Dice = 0.567 \cite{Supply}. This partial overlap indicates that $\omega_{\mathcal R}$ is sensitive to stress-related inhomogeneity but does not simply reproduce the previous pixel-wise clustering result. As shown in Supplementary Materials \cite{Supply}, changing $L$ modifies the spatial smoothing of the holonomy maps, but the localization of large holonomy in the lower-right region of the field of view and the positive/negative structures in $\omega_{\mathcal L}$ are preserved.

For the representative condition $L=10$, the resampling stability of the holonomy strength was evaluated using GPR posterior samples. For each posterior sample ($b=2,\ldots,1000$), we calculated the mean over the top $p=20\%$ pixels of $\omega_{\mathcal R}(x,y)$ and over the top $p=20\%$ pixels of $|\omega_{\mathcal L}(x,y)|$. The top-$20\%$ pixel set was determined separately from the corresponding holonomy map for each $b$. We denote these sample-wise mean values by $\overline{\omega_{\mathcal R}}_{20}^{(b)}$ and $\overline{|\omega_{\mathcal L}|}_{20}^{(b)}$. The absolute value is used on the left side because positive and negative contributions would otherwise cancel. Figure~3 shows histograms of these statistics. For the right side, the mean of $\overline{\omega_{\mathcal R}}_{20}^{(b)}$ was $5.091^\circ$, with a 95\% resampling interval of $[5.077^\circ,5.105^\circ]$. For the left side, the mean of $\overline{|\omega_{\mathcal L}|}_{20}^{(b)}$ was $1.196^\circ$, with a 95\% resampling interval of $[1.180^\circ,1.212^\circ]$. Both distributions are narrow and unimodal, and the values obtained from the reference series $(b=1)$ lie within the typical ranges of the corresponding resampling distributions. Together with the spatial maps in Fig.~2, these results indicate that the characteristic holonomy strength and its lower-right localization are not accidental features of a particular posterior sample but are stable against GPR posterior resampling. The stability of the signed spatial pattern itself is further examined in Supplementary Materials \cite{Supply}.

Figure~4(a) shows the local order parameter $S_{\mathcal R}(x,y)$ evaluated within an $L_{\rm loc}=5$ local square region using $\widehat{\boldsymbol a}_{\mathcal R}$ obtained from square loops with $L=10$. The same top-$20\%$ high-$\omega_{\mathcal R}$ selection described above was used. The map reveals that $S_{\mathcal R}$ is not uniform even within the high-$\omega_{\mathcal R}$ region, with locally reduced areas present. This finding suggests that spatial inhomogeneity manifests not only in the magnitude of $\omega_{\mathcal R}$ but also in the orientational order of $\widehat{\boldsymbol a}_{\mathcal R}$. As shown in Supplementary Materials \cite{Supply}, a two-dimensional histogram of $S_{\mathcal R}$ and $\omega_{\mathcal R}$ further supports that these quantities do not exhibit a simple one-to-one correspondence.

Figure~4(b) shows the orientational order $S_{\rm plane}(x,y)$ of the temperature-trajectory plane evaluated within an $L_{\rm loc}=5$ local square region. In most regions, $S_{\rm plane}\simeq 1$, indicating that the trajectory planes in data space are spatially coherent across neighboring real-space pixels. In contrast, $S_{\rm plane}$ decreases in the stress-concentrated region, implying that the orientation of the data-space temperature-trajectory plane becomes locally inhomogeneous as a function of real-space position. A narrow stripe-like structure extending from the lower-right edge toward the upper-left is also observed. This structure more closely resembles the gradient map from our previous single-$T$ holonomy analysis than either the $\delta_{\rm fold}$ map in Fig.~1(a) or the $T_{\rm F}$ map in Fig.~1(b) \cite{manaka-bf17}. As shown in Supplementary Materials \cite{Supply}, increasing $L_{\rm loc}$ tends to smooth $S_{\mathcal R}(x,y)$ and $S_{\rm plane}(x,y)$, but the overall spatial trends are preserved.

The spatial patterns in Figs.~2 and 4 should not be interpreted as direct images of conventional ferroelectric or ferroelastic domain morphology. The holonomy maps in Fig.~2 represent closed-loop connection mismatches derived from the full optical-polarization temperature trajectories. Figure~4(a) is directly associated with the right-side holonomy: it shows the local orientational order of the residual rotation axes obtained from the right-side loop products. In contrast, Fig.~4(b) is not a holonomy map and is not constructed from a holonomy residual. It shows the local orientational order of the normals to the temperature-trajectory planes defined directly from the right-singular-vector subspaces at individual pixels. The sharp stripe-like structure in Fig.~4(b) marks a localized reduction in the orientational order of the temperature-trajectory planes. These quantities characterize distinct geometric aspects of the full temperature trajectories rather than direct domain morphology at a single temperature.

\subsection{Comparison with local angular variation}

To examine whether holonomy can be distinguished from a local axis-variation measure, we introduce a local variation measure $g(x,y)$ based on nearest-neighbor angular differences. This quantity can be computed for $\widehat{\boldsymbol a}_{\mathcal R}$ and $\widehat{\boldsymbol{\mathsf n}}_{\rm plane}$ defined above. The detailed procedure is given in Supplementary Materials \cite{Supply}; here, $g(x,y)$ is defined in order to clarify its difference from holonomy. For an arbitrary director field $\widehat{\boldsymbol d}(x,y)$, $|\widehat{\boldsymbol d}|=1$, with the identification $\widehat{\boldsymbol d}\sim-\widehat{\boldsymbol d}$, the angular distance between two directors is defined as

\begin{equation}
\theta
\left(
\widehat{\boldsymbol d}_i,
\widehat{\boldsymbol d}_j
\right)
=
\arccos
\left(
\left|\,
\widehat{\boldsymbol d}_i\cdot
\widehat{\boldsymbol d}_j
\right|
\right).
\end{equation}

\noindent This angular distance is non-negative and lies in the range $0$ to $\pi/2$. Denoting the set of nearest neighbors of $(x,y)$ by $\mathcal G_{\rm valid}(x,y)$, we define the local variation measure as

\begin{equation}
g(x,y)
=
\frac{1}{|\mathcal G_{\rm valid}(x,y)|}
\sum_{(x',y')\in\mathcal G_{\rm valid}(x,y)}
\theta
\left(\,
\widehat{\boldsymbol d}(x,y),
\widehat{\boldsymbol d}(x',y')
\right),
\end{equation}

\noindent where $g(x,y)$ is a pixel-centered average of local angular differences and reflects how rapidly the director field changes around that pixel. Because the nearest-neighbor distance is one pixel, we display the resulting quantities as angular gradients.

Figure~5 shows the spatial distributions of the local angular-variation measures $g_{\mathcal R}(x,y)$ and $g_{\rm plane}(x,y)$, calculated from $\widehat{\boldsymbol a}_{\mathcal R}$ and $\widehat{\boldsymbol{\mathsf n}}_{\rm plane}$, respectively. Both maps are enhanced near the stress-concentrated region and emphasize a stripe-like structure extending from the lower-right edge toward the upper left. This behavior is close to the gradient-like response observed in our previous single-$T$ holonomy analysis \cite{manaka-bf17}, indicating that $g_{\mathcal R}$ and $g_{\rm plane}$ mainly capture local spatial variation of the corresponding director fields. However, these local angular-variation maps do not replicate the holonomy maps. In particular, the positive and negative structures in $\omega_{\mathcal L}(x,y)$ cannot be accounted for by the non-negative local quantities $g_{\mathcal R}(x,y)$ and $g_{\rm plane}(x,y)$. This difference arises because holonomy is a loop-level connection-geometric quantity: $\omega_{\mathcal R}$ and $\omega_{\mathcal L}$ measure the residual rotation accumulated along closed loops, whereas $g_{\mathcal R}$ and $g_{\rm plane}$ measure only nearest-neighbor angular variation. Quantitative comparisons using correlation coefficients, IoU/Dice coefficients, and threshold-$p$ dependence are given in Supplementary Materials \cite{Supply}.

\subsection{Validity of the representative analysis conditions}

In this study, $L=10$, $L_{\rm loc}=5$, and $p=20\%$ were used as representative analysis conditions to visualize characteristic holonomy structures and compare them with local quantities and the stress-concentrated region. These values were selected on the basis of the loop-size dependence, the threshold dependence of the high-$|\omega|$ pixel set, and the window-size dependence of the local order parameters, rather than to highlight any particular figure. Here, we briefly summarize the rationale for these choices, with detailed parameter-dependence analyses given in Supplementary Materials \cite{Supply}.

We first consider the loop size $L$. Because holonomy is obtained by accumulating local connections along a closed loop, its magnitude depends on $L$. If $L$ is too small, the result is strongly influenced by pixel-scale variations; if $L$ is too large, the spatial structure is excessively coarse-grained. To characterize this dependence, we examine the empirical scaling.

\begin{equation}
\overline{|\omega_\chi|}(L;p) \propto L^{\alpha_\chi(p)}, \qquad \chi\in\{\mathcal R,\mathcal L\},
\label{power-law}
\end{equation}

\noindent where $\overline{|\omega_\chi|}(L;p)$ denotes the mean absolute holonomy strength over the top $p\%$ pixel set within the nonzero holonomy-strength set, using $|\omega_{\mathcal R}|$ for the right side and $|\omega_{\mathcal L}|$ for the left side. Figure~6 shows the $L$ dependence of $\overline{|\omega_{\mathcal R}|}(L;p)$ and $\overline{|\omega_{\mathcal L}|}(L;p)$ for $p=10,20,40,60,$ and $100\%$. For both the right-side and left-side holonomies, the mean absolute strength increases with $L$ and approximately follows the power-law-like trend in Eq.~(\ref{power-law}), although weak curvature and gradual fluctuations are present. In Fig.~6, the empirical exponent $\alpha_\chi(p)$ was estimated by a least-squares fit of $\log \overline{|\omega_\chi|}(L;p)$ against $\log L$ for each fixed $p$; this estimate is denoted by $\alpha_\chi^{\rm ls}(p)$. As shown in Supplementary Materials \cite{Supply}, different estimation methods for $\alpha_\chi(p)$ give slightly different numerical values, but they consistently indicate that no clear critical loop size is observed. Based on the comparison of holonomy maps for different $L$, we adopted $L=10$ as an intermediate representative scale between pixel-scale fluctuations and excessive coarse graining.

We next consider the threshold fraction $p$. The insets of Fig.~6 show the $p$ dependence of $\alpha_{\mathcal R}^{\rm ls}(p)$ and $\alpha_{\mathcal L}^{\rm ls}(p)$. A bend is observed near $p=10$--$20\%$ for both the right- and left-side holonomies. The low-$p$ regime is dominated by a small number of high-$|\omega|$ pixels, whereas the high-$p$ regime encompasses a larger contribution from background regions with $|\omega|\simeq0$. This bend can therefore be interpreted as a crossover from high-$|\omega|$-dominated behavior to behavior inclusive of the background. On this basis, $p=20\%$ was adopted as a representative threshold: it captures the high-$|\omega|$ regions while avoiding excessive sensitivity to a small number of extreme pixels.

Finally, we consider the local window size $L_{\rm loc}$ used for $S_{\mathcal R}$ and $S_{\rm plane}$. If $L_{\rm loc}$ is too small, the local order parameters become sensitive to fluctuations from only a few pixels; if it is too large, fine structures such as the stress-concentrated region and stripe-like features are smoothed out. Based on the window-size dependence examined in Supplementary Materials \cite{Supply}, $L_{\rm loc}=5$ was adopted as a representative value that suppresses pixel-scale fluctuations without excessively smoothing the spatial structures.

\subsection{Physical meaning of temperature-trajectory holonomy}

The geometric quantities extracted from the temperature trajectories reflect spatial inhomogeneity within the specimen, but they are not assumed to correspond one-to-one to a specific microscopic structure or unobserved physical quantity. Therefore, the shape of a trajectory-derived map is not expected to directly reproduce the shape of a domain, domain wall, slip structure, or other local feature observed in an image taken at a single temperature. Here, we summarize the physical meaning of $\omega_{\mathcal R}$, $\omega_{\mathcal L}$, $S_{\mathcal R}$, and $S_{\rm plane}$. Because the positive and negative structures in $\omega_{\mathcal L}$ carry information distinct from that in an ordinary intensity map, their interpretation requires particular care.

$\omega_{\mathcal R}$ identifies regions where the local axis field extracted from the temperature trajectories in data space cannot be connected consistently around a closed loop in real space. These regions are located in the same lower-right region of the field of view, where $T_{\rm F}$ increases and stress-related optical anisotropy is observed. Thus, $\omega_{\mathcal R}$ reflects spatial inhomogeneity of the temperature trajectories in the specimen. In contrast, $S_{\mathcal R}$ measures the local orientational order of the residual rotation axes $\widehat{\boldsymbol a}_{\mathcal R}$, not the magnitude of holonomy itself. In this sense, $S_{\mathcal R}$ provides information complementary to $\omega_{\mathcal R}$: it distinguishes regions where the residual rotations have a coherent local axis from regions where the residual axes are locally disordered, even within high-$\omega_{\mathcal R}$ regions. The folded optical retardance $\delta_{\rm fold}$ map in Fig.~1(a) provides the optical anisotropy at a single $T$ and is related to lattice strain and electric polarization. In contrast, $S_{\rm plane}$ is obtained from the entire $T$ sweep and represents the local orientational order of the temperature-trajectory plane. The spatial correspondence between reduced $S_{\rm plane}$ and the high-retardance branch $(\lambda-\delta_{\rm fold})$ regions suggests that regions with strong single-$T$ optical anisotropy also possess inhomogeneous geometric structures in their temperature trajectories. In other words, spatial inhomogeneity associated with strain and electric polarization is reflected not only in $\delta_{\rm fold}$, but also in disorder of the temperature-trajectory plane.

$\omega_{\mathcal L}$ is the signed ${\rm SO}(2)$ rotation angle that remains when the local frame $\widetilde{U}$ attached to the two-dimensional temperature-evolution-mode space $\mathcal U$ is connected along a closed loop in real space. Regions with $\omega_{\mathcal L}>0$ and $\omega_{\mathcal L}<0$ are not merely regions with different holonomy magnitudes; rather, they indicate that the temperature-evolution frames accumulate in-plane connection residuals of opposite orientations under the fixed coordinate and loop-orientation convention. Thus, the sign of $\omega_{\mathcal L}$ represents a convention-dependent connection structure, not an absolute physical sign at an isolated point. This signed connection structure indicates that the temperature trajectories do not follow a uniform response path throughout the specimen. In the presence of stress- and strain-related real-space inhomogeneity, the trajectory at each pixel can evolve differently in data space, and the associated temperature-evolution frames need not be connected uniformly in real space. Accordingly, $\omega_{\mathcal L}$ reflects not the local brightness or gradient in a single-$T$ birefringence image but rather how the optical-polarization response evolves as a temperature trajectory and how the associated frame is connected in real space.

The positive and negative structures in $\omega_{\mathcal L}$ may be related to stress-induced electric polarization, strain gradients, and associated electromechanical inhomogeneity. However, this study does not directly observe or reconstruct the electric-polarization field. The sign of $\omega_{\mathcal L}$ should therefore not be equated with the sign of electric polarization, the bound-charge density $\rho_{\rm b}=-\nabla\cdot\boldsymbol P$, or tensile or compressive strain. As shown in Supplementary Materials \cite{Supply}, the signed structure remains stable under local ${\rm SO}(2)$ gauge transformations, GPR posterior resampling, and shuffle-based null tests. Within the fixed real-space coordinate and loop-orientation convention, it is therefore unlikely to be an artifact of the SVD basis choice or posterior fluctuations, but rather a stable geometric feature of the temperature-evolution-frame connection.

These results also clarify the distinction between the present analysis and conventional approaches. Single-temperature imaging shows the local optical response at a selected temperature, and local angular-variation measures quantify neighboring differences only. In contrast, the present analysis retains the full temperature-dependent response at each pixel and evaluates the closed-loop connection of the trajectory-derived directions or frames in real space. Therefore, it provides information on the spatial organization of the local response trajectories without explicitly reconstructing the underlying strain or electric polarization fields.

In summary, the holonomy obtained in this study does not itself represent a topological invariant or an electric charge \cite{ho2,h2}. Rather, it indicates that stress- and strain-related real-space inhomogeneity leaves geometric fingerprints both in the temperature-response paths in data space and in the corresponding connection structure. The right-side holonomy captures the closed-loop mismatch of the local director-axis field in optical-polarization-direction space, whereas the left-side holonomy captures the signed connection mismatch of the two-dimensional temperature-evolution frames in temperature-series space. Together, the two holonomies extract geometric information embedded in the full $T$ evolution that is inaccessible from single-$T$ birefringence images or simple local gradients.

\section{Conclusion}

We analyzed $T$-dependent birefringence imaging data of stress-induced ferroelectric SrTiO$_3$ under 231~MPa by treating the $T$ evolution of the optical-polarization state at each pixel as a temperature trajectory in data space. By treating these trajectories as data-space objects, we extracted two complementary trajectory-induced objects: a right-side director-axis field in optical-polarization-direction space and a left-side representative local frame attached to the temperature-evolution-mode space. We then defined $\omega_{\mathcal R}$ and $\omega_{\mathcal L}$ as residual rotations obtained by connecting these objects along closed loops in real space. The resulting maps revealed localized connection mismatches in the stress-concentrated region in the lower-right part of the field of view.

$\omega_{\mathcal R}$ captures the magnitude of the closed-loop mismatch of the director-axis field, whereas $\omega_{\mathcal L}$ captures the signed ${\rm SO}(2)$ connection mismatch of the temperature-evolution frames under a fixed frame convention. Comparison with local order parameters and angular-gradient measures confirms that these holonomy structures are not merely local-gradient features but represent loop-level connection-geometric information embedded in the temperature trajectories. The data-induced connection geometry of temperature trajectories thus provides a diagnostic descriptor of stress- and strain-related real-space inhomogeneity in SrTiO$_3$ under stress, going beyond what single-$T$ birefringence maps can capture. Although the present study focused on optical-polarization trajectories, the same framework is applicable to other temperature- or parameter-dependent multicomponent responses, provided that a trajectory or local subspace can be constructed in an appropriate data space. This extension is particularly relevant to recent operando imaging techniques, which acquire spatially resolved sequential responses over an entire field of view as functions of temperature, time, external field, or applied stress.

\section*{Acknowledgments}

This work was partially supported by Grants-in-Aid for Scientific Research (KAKENHI; Grant Nos. JP23K03283 and JP25K08487) from the Japan Society for the Promotion of Science.

\clearpage

\begin{figure}[b]
\begin{center}
\includegraphics[width=12cm]{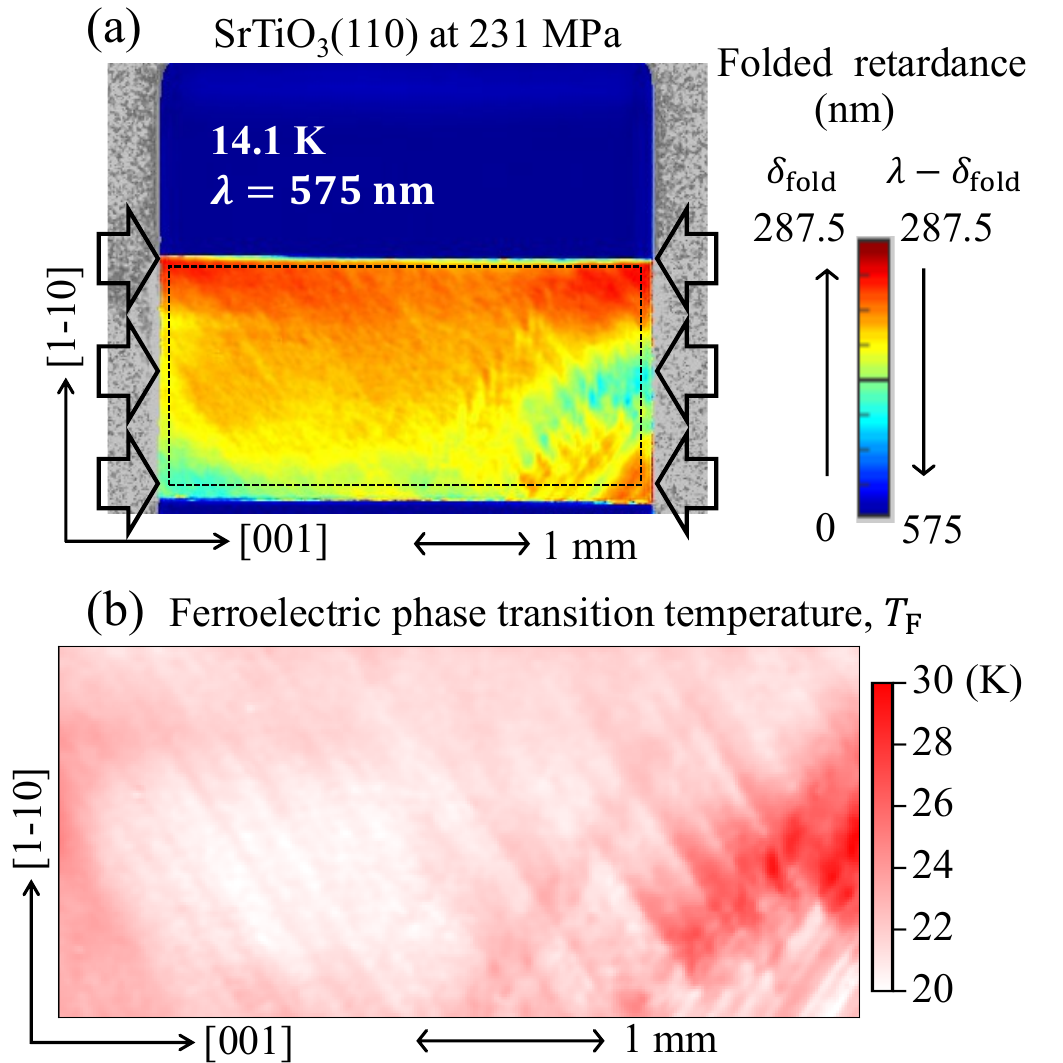}
\end{center}

\caption{(Color online) (a) Folded optical retardance image ($384\times288$ pixels) of SrTiO$_3$(110) under 231~MPa along [001] at 14.1~K for $\lambda=575$~nm, plotted from the data reported in Ref.~(25). The color scale shows the folded retardance $\delta_{\rm fold}$ in the range $0\le \delta_{\rm fold}\le \lambda/2$; the auxiliary scale indicates the corresponding high-retardance branch, $\lambda-\delta_{\rm fold}$. The large open arrows schematically indicate the direction of the external force. (b) Spatial distribution of the ferroelectric transition temperature $T_{\rm F}$ ($302\times140$ pixels) obtained from the sample region indicated in panel (a), plotted from the results reported in Ref.~(28).}

\end{figure}

\begin{figure}[b]
\begin{center}
\includegraphics[width=12cm]{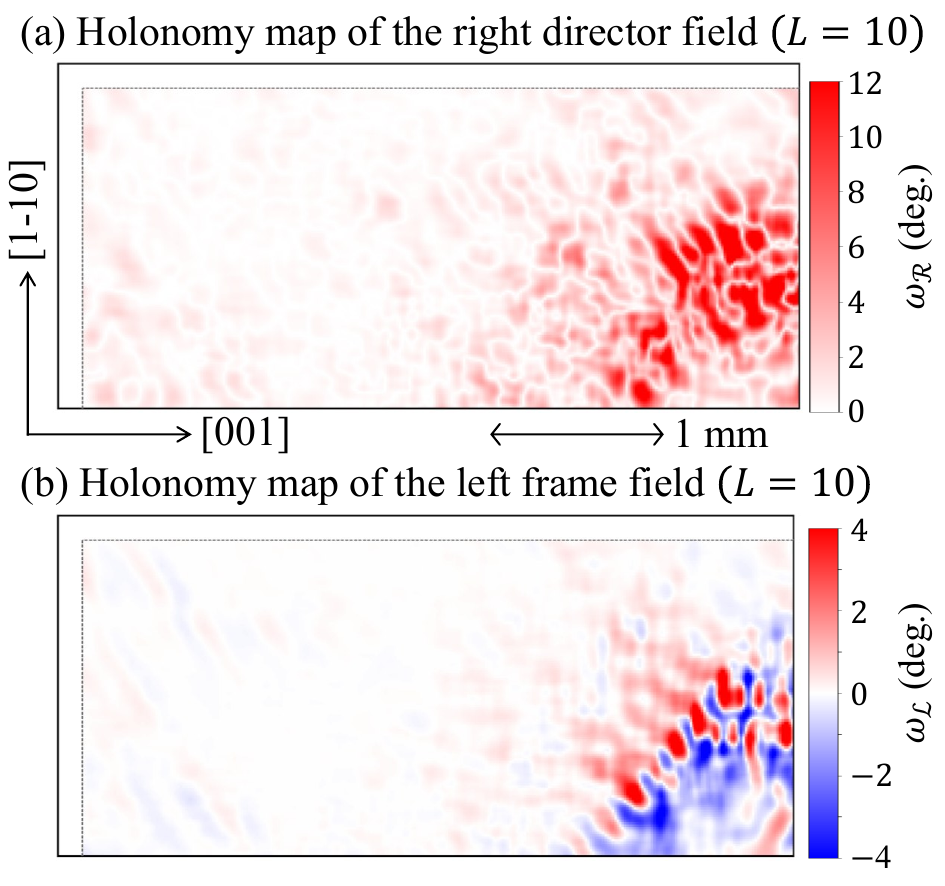}
\end{center}
\caption{(Color online) Holonomy maps obtained from the temperature trajectories at the representative loop size $L=10$. (a) Right-side holonomy angle $\omega_{\mathcal R}$ calculated from the director field in the optical-polarization-direction space. (b) Signed left-side holonomy angle $\omega_{\mathcal L}$ calculated from the local $\mathrm{SO}(2)$ frame field in temperature-series space. The color scales are shown in degrees. Regions above and to the left of the dashed lines are uncomputable because the $L\times L$ square loop extends beyond the field of view.}
\end{figure}

\begin{figure}[b]
\begin{center}
\includegraphics[width=12cm]{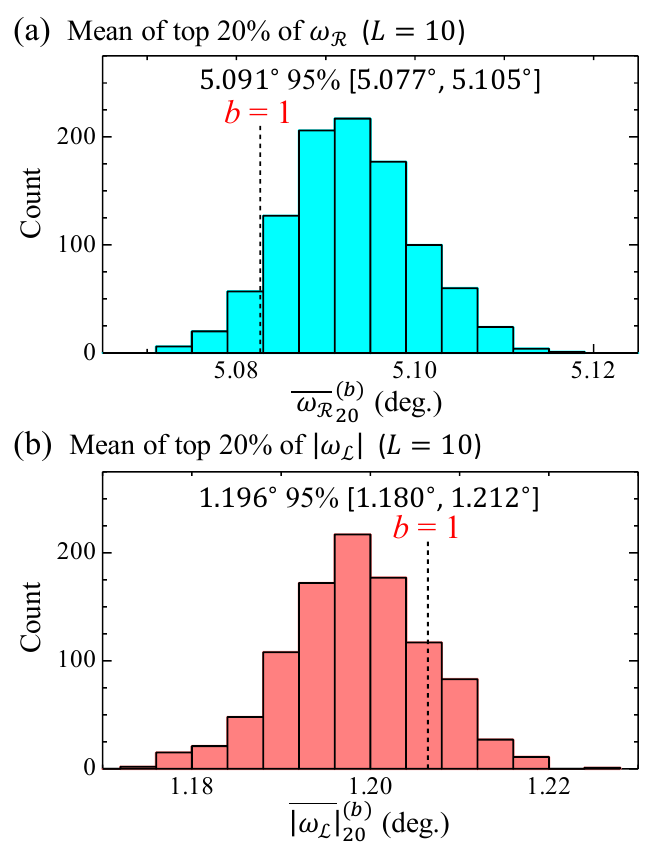}
\end{center}
\caption{(Color online) Resampling distributions of the mean top-20\% holonomy strength for $L=10$ obtained from the GPR posterior samples. (a) $\overline{\omega_{\mathcal R}}_{20}^{(b)}$, defined as the mean of the top 20\% values of the non-negative right-side holonomy angle $\omega_{\mathcal R}(x,y)$. (b) $\overline{|\omega_{\mathcal L}|}_{20}^{(b)}$, defined as the mean of the top 20\% values of $|\omega_{\mathcal L}(x,y)|$. The values shown in the panels indicate the mean and the central 95\% empirical interval of each resampling distribution, obtained from the 2.5 and 97.5 percentiles. Dashed vertical lines mark the values obtained from the reference series $(b=1)$.}
\end{figure}

\begin{figure}[b]
\begin{center}
\includegraphics[width=12cm]{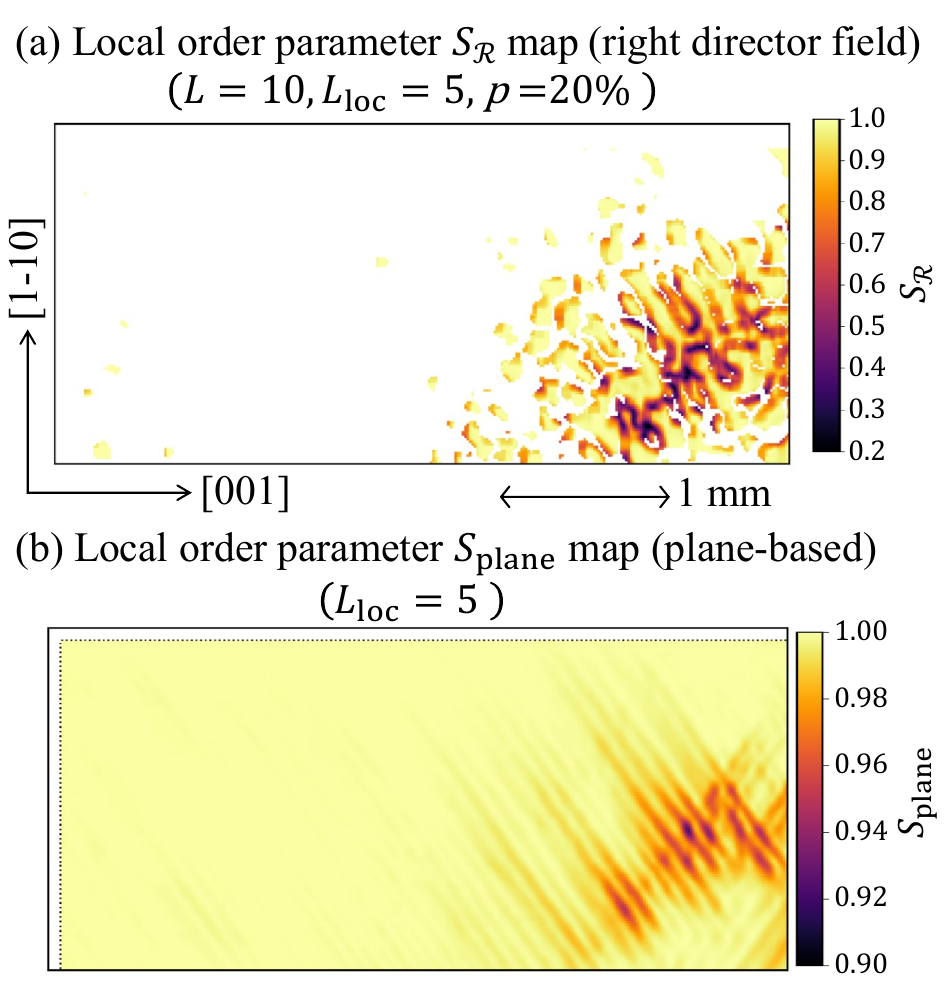}
\end{center}
\caption{(Color online) Local order parameter maps derived from trajectory-induced geometric objects. (a) Local order parameter $S_{\mathcal R}$ of the residual rotation axes obtained from the right-side holonomy. The map was calculated for $L=10$ and $L_{\rm loc}=5$. The top 20\% high-$\omega_{\mathcal R}$ pixel set was used both to retain the residual-axis samples entering the local second-moment tensor and to select the displayed base pixels. (b) Plane-based local order parameter $S_{\rm plane}$ calculated from the normal director of the local optical-polarization-trajectory plane with $L_{\rm loc}=5$. Regions above and to the left of the dashed lines are not shown because the corresponding $L_{\rm loc}\times L_{\rm loc}$ local windows are not fully included in the analyzed field of view.}
\end{figure}

\begin{figure}[b]
\begin{center}
\includegraphics[width=12cm]{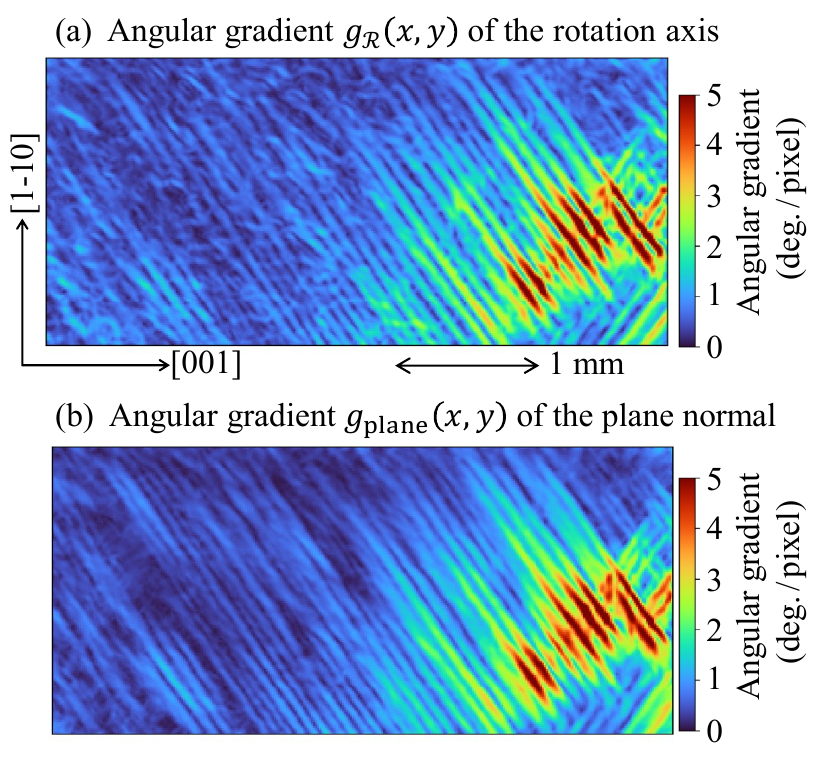}
\end{center}
\caption{(Color online) Spatial distributions of the local angular-variation measure $g(x,y)$ calculated from the trajectory-derived director fields. (a) Angular gradient $g_{\mathcal R}(x,y)$ of $\widehat{\boldsymbol a}_{\mathcal R}(x,y)$ obtained from the right-side holonomy. (b) Angular gradient $g_{\rm plane}(x,y)$ of $\widehat{\boldsymbol{\mathsf n}}_{\rm plane}(x,y)$ associated with the local temperature-trajectory plane. }
\end{figure}

\begin{figure}[b]
\begin{center}
\includegraphics[width=12cm]{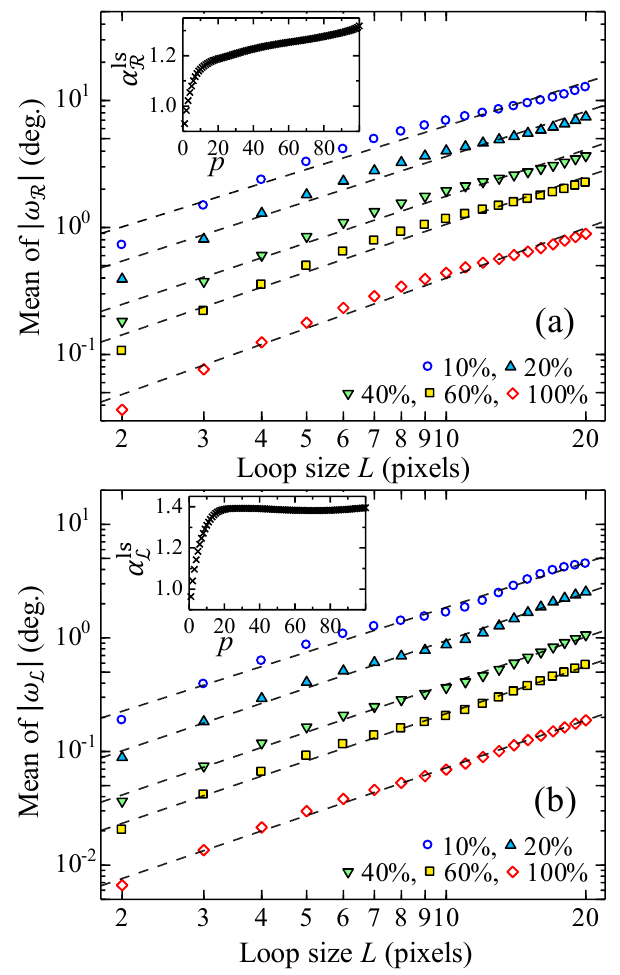}
\end{center}
\caption{(Color online) Loop-size dependence of the mean holonomy strength. (a) Mean absolute right-side holonomy strength $\overline{|\omega_{\mathcal R}|}(L;p)$ and (b) mean absolute left-side holonomy strength $\overline{|\omega_{\mathcal L}|}(L;p)$ as functions of the loop size $L$. For each $L$, the average was taken over the top $p=10,20,40,60,$ and $100\%$ of pixels within the nonzero holonomy-strength set, selected according to $|\omega_{\mathcal R}|$ in (a) and $|\omega_{\mathcal L}|$ in (b). Dashed lines indicate least-squares fits on a log--log scale using Eq.~(\ref{power-law}). Insets show the empirical exponents $\alpha_{\mathcal R}^{\rm ls}(p)$ and $\alpha_{\mathcal L}^{\rm ls}(p)$ estimated from these least-squares fits for each selected $p$.}
\end{figure}

\clearpage

\linespread{0.8}

\setcounter{equation}{0}
\setcounter{figure}{0}

\renewcommand{\thefigure}{S\arabic{figure}} 
\renewcommand{\theequation}{S\arabic{equation}} 
\renewcommand{\thetable}{S\arabic{table}}

\maketitle

\section*{Supplementary Materials}

\section*{S1.~Reference Alignment of Optical-polarization Directions}

\vspace{0.3cm}

Here, we describe how direction vectors that can be compared under a common reference are constructed from the incident and transmitted optical-polarization states, represented by Stokes vectors $\mbox{\boldmath $S$}=(S_1,S_2,S_3)$, obtained at each pixel and temperature ($T$). The observed incident optical-polarization state satisfies $(S_{1,{\rm In}},S_{2,{\rm In}},S_{3,{\rm In}})\simeq(0,0,-1)$, but it exhibits small pixel-to-pixel variations. In our previous analyses, because this incident optical-polarization deviated only slightly from the reference circular optical-polarization $\mbox{\boldmath $e$}_{\rm Base}=(0,0,-1)$, this deviation was treated by a simple linearized reference-alignment procedure based on component-wise differences.  Although this approximation works well in practice, the present study treats small differences in optical-polarization state between neighboring pixels as finite three-dimensional rotations. A more accurate alignment procedure is therefore required. We introduce a rotation operation that aligns the observed incident optical-polarization with $\mbox{\boldmath $e$}_{\rm Base}$ and apply the same rotation to the observed transmitted optical-polarization $(S_{1,{\rm Out}},S_{2,{\rm Out}},S_{3,{\rm Out}})$. This operation is an auxiliary optical-polarization reference alignment used before constructing the temperature trajectories; it is not itself a holonomy and is not accumulated along real-space loops. Below, we describe the procedure for applying this rotation to the transmitted optical-polarization using unit quaternions.

At each $T$, the observed incident and transmitted Stokes vectors are defined as

\begin{eqnarray}
\boldsymbol{S}^{(T)}_{\mathrm{In}}(x,y)
&:=&
\bigl(
S^{(T)}_{1,\mathrm{In}}(x,y),\,
S^{(T)}_{2,\mathrm{In}}(x,y),\,
S^{(T)}_{3,\mathrm{In}}(x,y)
\bigr)^{\mathsf T}
\in \mathbb{R}^3, \\
\boldsymbol{S}^{(T)}_{\mathrm{Out}}(x,y)
&:=&
\bigl(
S^{(T)}_{1,\mathrm{Out}}(x,y),\,
S^{(T)}_{2,\mathrm{Out}}(x,y),\,
S^{(T)}_{3,\mathrm{Out}}(x,y)
\bigr)^{\mathsf T}
\in \mathbb{R}^3.
\end{eqnarray}

\noindent The normalized Stokes vectors are therefore given by

\begin{eqnarray}
\widehat{\boldsymbol{S}}^{(T)}_{\mathrm{In}}(x,y) &=&
\frac{\boldsymbol{S}^{(T)}_{\mathrm{In}}(x,y)}
{\left\|\boldsymbol{S}^{(T)}_{\mathrm{In}}(x,y)\right\|} \in S^2, \\
\widehat{\boldsymbol{S}}^{(T)}_{\mathrm{Out}}(x,y) &=&
\frac{\boldsymbol{S}^{(T)}_{\mathrm{Out}}(x,y)}
{\left\|\boldsymbol{S}^{(T)}_{\mathrm{Out}}(x,y)\right\|} \in S^2.
\end{eqnarray}

\noindent These vectors represent points on the Poincar\'e sphere $S^2$.

Three-dimensional rotations are described using unit quaternions. A quaternion is written as

\begin{equation} 
q=(w,\boldsymbol{u}), \qquad \boldsymbol{u}=({\mathsf x}, {\mathsf y}, {\mathsf z})^{\mathsf T}, 
\label{eq:q_scalar_vector} 
\end{equation} 

\noindent where $w\in\mathbb{R}$ is the scalar part and $\boldsymbol{u}\in\mathbb{R}^3$ is the vector part. It should be noted that the orthogonal coordinate system $[{\mathsf x},{\mathsf y},{\mathsf z}]$ is defined on the Poincar\'e sphere $S^2$ and is unrelated to the pixel position $(x,y)$ in the birefringence image. The product of two quaternions is defined as

\begin{equation} 
(w_1,\boldsymbol{u}_1)(w_2,\boldsymbol{u}_2) = 
\Bigl( w_1w_2-\boldsymbol{u}_1\cdot\boldsymbol{u}_2,\,
w_1\boldsymbol{u}_2+w_2\boldsymbol{u}_1+\boldsymbol{u}_1\times\boldsymbol{u}_2 \Bigr).
\label{eq:quaternion_product} 
\end{equation} 

\noindent In component form, for $q_1=(w_1,{\mathsf x}_1,{\mathsf y}_1,{\mathsf z}_1)$ and $q_2=(w_2,{\mathsf x}_2,{\mathsf y}_2,{\mathsf z}_2)$, we have

\begin{eqnarray}
q_1q_2 &=& (w,{\mathsf x},{\mathsf y}, {\mathsf z}), \\
w &=& w_1w_2-{\mathsf x}_1{\mathsf x}_2-{\mathsf y}_1{\mathsf y}_2-{\mathsf z}_1{\mathsf z}_2, \\
{\mathsf x} &=& w_1{\mathsf x}_2+{\mathsf x}_1w_2+{\mathsf y}_1{\mathsf z}_2-{\mathsf z}_1{\mathsf y}_2, \\
{\mathsf y} &=& w_1{\mathsf y}_2-{\mathsf x}_1{\mathsf z}_2+{\mathsf y}_1w_2+{\mathsf z}_1{\mathsf x}_2, \\
{\mathsf z} &=& w_1{\mathsf z}_2+{\mathsf x}_1{\mathsf y}_2-{\mathsf y}_1{\mathsf x}_2+{\mathsf z}_1w_2.
\end{eqnarray}

\noindent The conjugate of a quaternion is given by $q^\ast=(w,-\boldsymbol{u})$, and its norm is

\begin{equation} 
|q| = \sqrt{w^2+\|\boldsymbol{u}\|^2}.
\label{eq:q_norm} 
\end{equation} 

\noindent A quaternion satisfying $|q|=1$ is called a unit quaternion $\widehat{q}$.

Unit quaternions provide a compact representation of three-dimensional rotations. Let $\boldsymbol{v}=(v_1,v_2,v_3)^{\mathsf T}\in\mathbb{R}^3$ be an arbitrary three-dimensional vector. Embedding it in a pure imaginary quaternion with zero scalar part gives

\begin{equation} 
\widetilde{\boldsymbol{v}} := (0,\boldsymbol{v}) = (0,v_1,v_2,v_3).
\label{eq:pure_quaternion_embedding} 
\end{equation} 

\noindent The rotation of this pure imaginary quaternion by a unit quaternion $\widehat{q}$ is then

\begin{equation} 
\widetilde{\boldsymbol{v}}\,' = \widehat{q}~\widetilde{\boldsymbol{v}}~\left(\widehat{q}\right)^\ast = (0, v_1',v_2',v_3'),
\label{eq:q_rotate_formula} 
\end{equation}

\noindent where $\widetilde{\boldsymbol{v}}\,'$ is again a pure imaginary quaternion with vector part

\begin{equation}
\boldsymbol{v}' = \operatorname{Rot}_{\widehat{q}}\,(\boldsymbol{v}) \in \mathbb{R}^3.
\label{eq:Rot_q_def}
\end{equation} 

\noindent For a rotation axis $\widehat{\boldsymbol{a}}\in S^2$ and rotation angle $\theta$, the corresponding unit quaternion is

\begin{equation} 
\widehat{q}= \left( \cos\frac{\theta}{2}, \,\widehat{\boldsymbol{a}}\sin\frac{\theta}{2} \right).
\label{eq:axis_angle_quaternion} 
\end{equation} 

\noindent A unit quaternion thus compactly encodes the axis and angle of a three-dimensional rotation.

The purpose of this procedure is to rotate the observed incident optical-polarization $\widehat{\boldsymbol{S}}^{(T)}_{\mathrm{In}}$ to align with the reference $\mbox{\boldmath $e$}_{\rm Base}$. For two unit vectors $\boldsymbol{\eta},\,\boldsymbol{\xi}\in S^2$, the shortest rotation mapping $\boldsymbol{\eta}$ to $\boldsymbol{\xi}$ is obtained by setting

\begin{eqnarray} 
\boldsymbol{u} &=& \boldsymbol{\eta}\times\boldsymbol{\xi}, 
\label{eq:rotation_axis_cross} \\
w&=&1+\boldsymbol{\eta}\cdot\boldsymbol{\xi}, 
\label{eq:rotation_scalar_1plusdot} 
\end{eqnarray} 

\noindent and the corresponding unit quaternion is given by

\begin{eqnarray} 
\widehat{q}\,(\boldsymbol{\eta}\to\boldsymbol{\xi}) &=& 
\frac{(w,\boldsymbol{u})}{\sqrt{w^2+\|\boldsymbol{u}\|^2}} \nonumber \\
&=&
\frac{
\bigl(
1+\boldsymbol{\eta}\cdot\boldsymbol{\xi},\,
\boldsymbol{\eta}\times\boldsymbol{\xi}
\bigr)
}{
\sqrt{
\bigl(1+\boldsymbol{\eta}\cdot\boldsymbol{\xi}\bigr)^2
+
\|\boldsymbol{\eta}\times\boldsymbol{\xi}\|^2
}
}.
\label{eq:q_from_two_vectors_basic} 
\end{eqnarray} 

\noindent In numerical implementation, when $\boldsymbol{\eta}$ and $\boldsymbol{\xi}$ are nearly antiparallel ($\boldsymbol{\eta}\cdot\boldsymbol{\xi}\simeq -1$), the factor $(1+\boldsymbol{\eta}\cdot\boldsymbol{\xi})$ in Eq.~\eqref{eq:q_from_two_vectors_basic} approaches zero, causing numerical instability. To prevent this, we introduce

\begin{equation}
\boldsymbol{\xi}' =
\operatorname{sign}(\boldsymbol{\eta}\cdot\boldsymbol{\xi})\,\boldsymbol{\xi},
\label{eq:xi_prime_def}
\end{equation}

\noindent where $\operatorname{sign}(0)=+1$, so that

\begin{equation}
\boldsymbol{\eta}\cdot\boldsymbol{\xi}' \ge 0,
\end{equation}

\noindent always holds, preventing the denominator in Eq.~\eqref{eq:q_from_two_vectors_basic} from becoming too small. This reflects the fact that, in the present birefringence measurement and data representation, antipodal representatives of the optical-polarization direction are not distinguished. The reference optical-polarization is therefore treated as an unoriented direction for alignment purposes rather than as a strictly oriented vector. The replacement $\boldsymbol{\xi}\rightarrow\boldsymbol{\xi}'$ is a representative choice used to define the shortest alignment rotation stably; it does not introduce an additional physical degree of freedom. The rotation actually applied is thus the shortest rotation mapping $\boldsymbol{\eta}$ to $\boldsymbol{\xi}'$. In the present dataset, the incident optical-polarization is close to $\boldsymbol e_{\rm Base}$, so this sign adjustment acts solely as a numerical safeguard. We apply this general construction to the observed incident Stokes vector by setting

\begin{eqnarray}
\boldsymbol{\eta} &=& \widehat{\boldsymbol{S}}^{(T)}_{\mathrm{In}}(x,y),
\label{eq:eta_is_Sin} \\
\boldsymbol{\xi} &=& \boldsymbol{e}_{\mathrm{Base}} = (0,0,-1)^{\mathsf T}.
\label{eq:eref_def} 
\end{eqnarray} 

\noindent Therefore, the alignment unit quaternion actually used is

\begin{eqnarray} 
\widehat{q}_{\mathrm{align}}^{(T)}(x,y) &=&
\frac{
\bigl(
1+\widehat{\boldsymbol{S}}^{(T)}_{\mathrm{In}}(x,y)\cdot\boldsymbol{\xi}',\,
\widehat{\boldsymbol{S}}^{(T)}_{\mathrm{In}}(x,y)\times\boldsymbol{\xi}'
\bigr)
}{
\sqrt{
\bigl(
1+\widehat{\boldsymbol{S}}^{(T)}_{\mathrm{In}}(x,y)\cdot\boldsymbol{\xi}'
\bigr)^2
+
\bigl\|
\widehat{\boldsymbol{S}}^{(T)}_{\mathrm{In}}(x,y)\times\boldsymbol{\xi}'
\bigr\|^2
}
},
\label{eq:q_align_sign_adjusted}
\\
\boldsymbol{\xi}'(x,y) &=&
\operatorname{sign}
\left(
\widehat{\boldsymbol{S}}^{(T)}_{\mathrm{In}}(x,y)
\cdot
\boldsymbol{e}_{\mathrm{Base}}
\right)
\boldsymbol{e}_{\mathrm{Base}}.
\label{eq:xi_prime_ref} 
\end{eqnarray} 

\noindent This reference alignment suppresses the influence of pixel-dependent variations in the incident optical-polarization when comparing transmitted optical-polarization states under a common reference.

We next apply the rotation by $\widehat{q}_{\mathrm{align}}^{(T)}(x,y)$ to $\widehat{\boldsymbol{S}}^{(T)}_{\mathrm{Out}}(x,y)$ so that the transmitted optical-polarization direction can be evaluated consistently under the common reference. First, the observed transmitted optical-polarization vector is written as a pure imaginary quaternion,

\begin{equation}
\widetilde{\boldsymbol{S}}^{(T)}_{\mathrm{Out}}(x,y)
=
(0,\,\widehat{\boldsymbol{S}}^{(T)}_{\mathrm{Out}}(x,y)).
\label{eq:Sout_pure_quaternion}
\end{equation}

\noindent Applying this alignment rotation yields

\begin{equation}
\widetilde{\boldsymbol{S}}^{\prime(T)}_{\mathrm{Out}}(x,y)
=
\widehat{q}_{\mathrm{align}}^{(T)}(x,y)\,
\widetilde{\boldsymbol{S}}^{(T)}_{\mathrm{Out}}(x,y)\,
\bigl(\widehat{q}_{\mathrm{align}}^{(T)}(x,y)\bigr)^\ast.
\label{eq:Sout_rotated_quaternion}
\end{equation}

\noindent Since $\widetilde{\boldsymbol{S}}^{\prime(T)}_{\mathrm{Out}}(x,y)$
is again a pure imaginary quaternion, its vector part is

\begin{equation}
\boldsymbol{S}^{\prime(T)}_{\mathrm{Out}}(x,y)
=
\operatorname{Rot}_{\widehat{q}_{\mathrm{align}}^{(T)}(x,y)}
\left(
\widehat{\boldsymbol{S}}^{(T)}_{\mathrm{Out}}(x,y)
\right)
\in \mathbb{R}^3,
\label{eq:Sout_rotated_vector}
\end{equation}

\noindent where $\boldsymbol{S}^{\prime(T)}_{\mathrm{Out}}(x,y)$ represents the transmitted optical-polarization state in the coordinate system aligned with the common reference. The quaternion $\widehat{q}_{\mathrm{align}}^{(T)}(x,y)$ is an auxiliary alignment rotation, not the holonomy defined later. Although this rotation preserves the vector norm in principle, we renormalize to guard against finite numerical precision and define

\begin{equation}
\widehat{\boldsymbol{n}}(x,y;T)
:=
\frac{
\boldsymbol{S}^{\prime(T)}_{\mathrm{Out}}(x,y)
}{
\left\|
\boldsymbol{S}^{\prime(T)}_{\mathrm{Out}}(x,y)
\right\|
}.
\label{eq:n_final_def}
\end{equation}

\noindent Thus, $\widehat{\boldsymbol{n}}(x,y;T)\in S^2\subset\mathbb{R}^3$ and $\|\widehat{\boldsymbol{n}}(x,y;T)\|=1$. This aligned output direction, rather than the auxiliary alignment quaternion itself, is used as the input direction vector in the subsequent analysis. The procedure is applied independently at each pixel $(x,y)$ and at all temperatures.

\vspace{0.5cm}

\section*{S2.~GPR Common-grid Representation and Tangent-plane Posterior Resampling}

\vspace{0.3cm}

Birefringence measurements were performed at a discrete set of temperatures $\{T_m\}_{m=1}^{M_T}$, where $m$ denotes the index of the measured $T$ point and $M_T$ is the total number of measured $T$ points. We apply Gaussian process regression (GPR) to the sequence of direction vectors $\{\,\widehat{\boldsymbol{n}}(x,y;T_m)\}_{m=1}^{M_T}$ obtained along this measured temperature series in order to evaluate the trajectory on a common temperature grid and to propagate the uncertainty associated with this representation. The posterior mean is used as a representative temperature trajectory, whereas posterior-resampled trajectories are used to assess the resampling stability of the holonomy quantities and signed spatial patterns. The direction vector cannot be treated as a simple unconstrained three-component real-valued sequence because it is constrained to lie on the unit sphere $S^2$. Even when common-grid evaluation and resampling are performed along the $T$ direction, the resulting vector sequence must remain on $S^2$. The GPR common-grid representation and posterior resampling procedures satisfying this condition are described below.

In the GPR analysis of temperature trajectories, the measurement at each pixel $(x,y)$ is treated independently; the pixel position is therefore omitted from the notation below. The direction vector at $T_m$ is written as

\begin{equation}
\widehat{\boldsymbol{n}}(T_m) =
\bigl(
n_{\mathsf x}(T_m),\,
n_{\mathsf y}(T_m),\,
n_{\mathsf z}(T_m)
\bigr)^{\mathsf T}
\in S^2
\qquad (m=1,\ldots,M_T).
\end{equation}

\noindent In the implementation, we use 3,362 measured $T$ points obtained during cooling from 300.0~K to 14.1~K. Although there are no duplicated measured $T$ points, small local back-and-forth variations in $T$ occur owing to the $T$-control process. We therefore use the series sorted in descending order of measured $T$. The data observed on $\{T_m\}_{m=1}^{M_T}$ are used as the input, and we separately introduce an evaluation temperature-grid set

\begin{equation}
\mathcal{T}_{\mathrm{grid}}
=
\{\tau_i\}_{i=1}^{M_{\mathrm{grid}}}
\subset \mathbb{R},
\label{eq:Tgrid_def}
\end{equation}

\noindent on which the posterior is evaluated. In the present implementation, $\tau_i$ is equally spaced with an interval of 0.5~K. Thus, the role of GPR in this analysis is to provide a common-grid representation of all pixel-wise temperature trajectories, together with a posterior covariance used for resampling, rather than to introduce additional spatial structure.

The GPR hyperparameters are fixed throughout the analysis, with the same kernel, length scale, and noise variance applied to all GPR calculations. To describe the component-wise GPR formula, we temporarily omit the Cartesian index $j$ and consider a single scalar component. The observed value at temperature $T_m$ is denoted $y_m$, which is treated as a noisy observation of a latent smooth function $f(T)$:

\begin{eqnarray}
y_m
&=&
f(T_m)+\varepsilon_m,\\
\varepsilon_m
&\sim&
\mathcal{N}(0,\sigma_{\rm e}^2).
\end{eqnarray}

\noindent The observed scalar series on the measured temperature points is

\begin{equation}
\boldsymbol{y}
=
\bigl(
y_1,\ldots,y_{M_T}
\bigr)^{\mathsf T}
\in \mathbb{R}^{M_T}.
\end{equation}

\noindent However, the values to be inferred by GPR on the evaluation temperature grid are

\begin{equation}
\boldsymbol{f}_{\mathrm{grid}}
=
\bigl(
f(\tau_1),\ldots,f(\tau_{M_{\mathrm{grid}}})
\bigr)^{\mathsf T}
\in \mathbb{R}^{M_{\mathrm{grid}}}.
\end{equation}

\noindent Collecting the observed data as

\begin{equation}
\mathcal{D}
=
\left\{
(T_m, y_m)
\right\}_{m=1}^{M_T},
\end{equation}

\noindent the GPR posterior on the evaluation grid takes the form

\begin{eqnarray}
p\,\left(
\boldsymbol{f}_{\mathrm{grid}}
\,\middle|\,
\mathcal{D}
\right)
&=&
p\,\left(
\bigl(f(\tau_1),\ldots,f(\tau_{M_{\mathrm{grid}}})\bigr)^{\mathsf T}
\,\middle|\,
\mathcal{D}
\right) \nonumber \\
&=&
\mathcal{N}(\boldsymbol{\mu},\Sigma),
\label{eq:gpr_posterior}
\end{eqnarray}

\noindent where $\boldsymbol{\mu}\in\mathbb{R}^{M_{\mathrm{grid}}}$ is the posterior mean vector and $\Sigma\in\mathbb{R}^{M_{\mathrm{grid}}\times M_{\mathrm{grid}}}$ is the posterior covariance matrix on the evaluation grid. The posterior mean is used as the representative common-grid series, referred to as the reference series with posterior-resampling index $b=1$:

\begin{equation}
\boldsymbol{f}_{\mathrm{grid}}^{\rm ref}
=
\boldsymbol{f}_{\mathrm{grid}}^{(b=1)}
:=
\boldsymbol{\mu},
\end{equation}

\noindent where ``ref'' explicitly denotes the reference series ($b=1$). Below, ``ref'' is used when referring to the reference series itself, whereas $b=1$ is used when comparing across all posterior samples.

Restoring the Cartesian component index, the above scalar GPR formulation is applied independently to the three components of the direction vector. The observed values at $T_m$ are

\begin{eqnarray}
y_{1,m} &:=& n_{\mathsf x}(T_m),\nonumber\\
y_{2,m} &:=& n_{\mathsf y}(T_m),\\
y_{3,m} &:=& n_{\mathsf z}(T_m), \nonumber
\end{eqnarray}

\noindent for $m=1,\ldots,M_T$. An independent one-dimensional Gaussian process prior is assumed for each latent function $f_j(T)$,

\begin{equation}
f_j(T)\sim \mathrm{GP}\left(0,k(T,T')\right),
\end{equation}

\noindent with zero mean function and radial basis function (RBF) kernel,

\begin{equation}
k(T,T')
=
\sigma_f^2
\exp\left(
-\frac{(T-T')^2}{2\ell_T^2}
\right).
\end{equation}

\noindent The observation model for each component takes the form

\begin{eqnarray}
y_{j,m} &=& f_j(T_m)+\varepsilon_{j,m}, \\
\varepsilon_{j,m}&\sim&\mathcal{N}(0,\sigma_{\rm e}^2),
\end{eqnarray}

\noindent where $y_{j,m}$ is the observed component value at $T_m$ and $f_j(T)$ is the latent smooth function representing the $T$ dependence of that component. In our previous analysis using long short-term memory, one low-$T$ point was predicted from high-$T$ data within a window of approximately 30~K. To make the present implementation comparable to that condition, we set the RBF-kernel length scale to $\ell_T=15.0~\mathrm{K}$ and the kernel amplitude to $\sigma_f^2=1.0$. Since $\widehat{\boldsymbol{n}}(T_m)$ is a unit vector, the effective fluctuation variance is fixed at $\sigma_{\rm e}^2=1\times 10^{-3}$, giving $\sigma_{\rm e}=\sqrt{10^{-3}}\simeq0.032$ and a $\pm 3\sigma_{\rm e}$ width of approximately $\pm0.1$. These hyperparameters are applied uniformly to all pixels and for both the reference series ($b=1$) and the posterior-resampled series $b=2,\ldots,1000$; no pixel-dependent tuning is introduced.

For each component $j=1,2,3$, the observed series on $T_m$ is written as

\begin{equation}
\boldsymbol{y}_j
=
\bigl(
y_{j,1},\ldots,y_{j,M_T}
\bigr)^{\mathsf T}
\in\mathbb{R}^{M_T}.
\end{equation}

\noindent The kernel matrix $K_{TT}\in\mathbb{R}^{M_T\times M_T}$ on $\{T_m\}_{m=1}^{M_T}$ is

\begin{equation}
(K_{TT})_{mn}
=
k(T_m,T_n)
\qquad
(m,n=1,\ldots,M_T),
\end{equation}

\noindent or in explicit form,

\begin{equation}
K_{TT}
=
\begin{pmatrix}
k(T_1,T_1) & k(T_1,T_2) & \cdots & k(T_1,T_{M_T}) \\
k(T_2,T_1) & k(T_2,T_2) & \cdots & k(T_2,T_{M_T}) \\
\vdots & \vdots & \ddots & \vdots \\
k(T_{M_T},T_1) & k(T_{M_T},T_2) & \cdots & k(T_{M_T},T_{M_T})
\end{pmatrix}.
\end{equation}

\noindent Including the measurement noise, the covariance matrix for the observed series $\boldsymbol{y}_j$ is

\begin{equation}
K_{\mathrm{obs}}
=
K_{TT}+\sigma_{\rm e}^2 I_{M_T},
\end{equation}

\noindent where $I_{M_T}$ is the $M_T\times M_T$ identity matrix. $K_{\mathrm{obs}}$ is the covariance matrix of the observed data on $T_m$ and is distinct from the posterior covariance on the evaluation grid.

The cross-kernel matrix $K_{T\tau}\in\mathbb{R}^{M_T\times M_{\mathrm{grid}}}$ between $\{T_m\}_{m=1}^{M_T}$ and the evaluation grid $\{\tau_i\}_{i=1}^{M_{\mathrm{grid}}}$ is

\begin{equation}
(K_{T\tau})_{mi}
=
k(T_m,\tau_i)
\qquad
(m=1,\ldots,M_T,\; i=1,\ldots,M_{\mathrm{grid}}).
\end{equation}

\noindent The kernel matrix $K_{\tau\tau}\in\mathbb{R}^{M_{\mathrm{grid}}\times M_{\mathrm{grid}}}$ on the evaluation grid is 

\begin{equation}
(K_{\tau\tau})_{ij}
=
k(\tau_i,\tau_j)
\qquad
(i,j=1,\ldots,M_{\mathrm{grid}}).
\end{equation}

\noindent From the standard GPR conditional distribution, the posterior mean vector and covariance matrix for component $f_j$ on the evaluation grid are

\begin{eqnarray}
\boldsymbol{\mu}_j
&=&
K_{T\tau}^{\mathsf T}K_{\mathrm{obs}}^{-1}\boldsymbol{y}_j,
\\
\Sigma_j
&=&
K_{\tau\tau}
-
K_{T\tau}^{\mathsf T}
K_{\mathrm{obs}}^{-1}
K_{T\tau}.
\end{eqnarray}

\noindent In the present implementation, the same $T_m$, evaluation grid, and GPR hyperparameters are used for all three Cartesian components. Although the posterior means $\boldsymbol{\mu}_j$ differ among components, the posterior covariance matrix is common to all three because it is independent of the observed values $\boldsymbol{y}_j$ under fixed hyperparameters. This common covariance is denoted 

\begin{equation} 
\Sigma_{\rm com} := \Sigma_1=\Sigma_2=\Sigma_3 \in \mathbb{R}^{M_{\mathrm{grid}}\times M_{\mathrm{grid}}}.
\end{equation}

\noindent At each evaluation grid point $\tau_i$, the component-wise posterior mean is

\begin{equation} 
\boldsymbol{\mu}(\tau_i)
=
\bigl(
\mu_{\mathsf{x}}(\tau_i),\,
\mu_{\mathsf{y}}(\tau_i),\,
\mu_{\mathsf{z}}(\tau_i)
\bigr)^{\mathsf T}
\in \mathbb{R}^3
\qquad
(i=1,\ldots,M_{\mathrm{grid}}).
\end{equation}

\noindent Because GPR is applied independently to the three components, $\boldsymbol{\mu}(\tau_i)$ does not generally have unit norm. Projecting back onto the unit sphere gives

\begin{equation}
\widehat{\boldsymbol{n}}(\tau_i)
:=
\frac{
\boldsymbol{\mu}(\tau_i)
}{
\left\|
\boldsymbol{\mu}(\tau_i)
\right\|
}
\in S^2
\qquad
(i=1,\ldots,M_{\mathrm{grid}}).
\end{equation}

\noindent The sequence $\{\widehat{\boldsymbol{n}}(\tau_i)\}_{i=1}^{M_{\mathrm{grid}}}$ constitutes the reference series, corresponding to posterior-resampling index $b=1$:

\begin{equation}
\widehat{\boldsymbol n}^{\rm ref}(\tau_i)
=
\widehat{\boldsymbol n}^{(b=1)}(\tau_i)
:=
\widehat{\boldsymbol n}(\tau_i)
\qquad
(i=1,\ldots,M_{\mathrm{grid}}).
\end{equation}

\noindent The reference series ($b=1$) is thus the representative common-grid direction-vector series obtained by evaluating the component-wise GPR posterior mean on the evaluation grid and renormalizing to $S^2$.

The posterior-resampled series $(b=2,\ldots,1000)$ represent plausible trajectory deformations around the reference series $\widehat{\boldsymbol n}^{\rm ref}$ within the tangent plane of $S^2$. Adding fluctuations directly to the Cartesian components in $\mathbb{R}^3$ would generally move the resulting vectors off the sphere, making their interpretation ambiguous. We therefore work in the tangent plane at each evaluation grid point $\tau_i$,

\begin{equation}
T_{\widehat{\boldsymbol{n}}^{\rm ref}(\tau_i)} S^2
=
\left\{
\boldsymbol{u}\in\mathbb{R}^3
\,\middle|\,
\boldsymbol{u}\cdot \widehat{\boldsymbol{n}}^{\rm ref}(\tau_i)=0
\right\}
\qquad
(i=1,\ldots,M_{\mathrm{grid}}),
\end{equation}

\noindent where $T$ in $T_{\widehat{\boldsymbol{n}}^{\rm ref}(\tau_i)}S^2$ denotes the tangent space of the sphere, not temperature. On this tangent plane, an orthonormal basis $\{\boldsymbol e_1(\tau_i),\boldsymbol e_2(\tau_i)\}$ is chosen satisfying

\begin{eqnarray}
\boldsymbol e_a(\tau_i)
&\in&
T_{\widehat{\boldsymbol{n}}^{\rm ref}(\tau_i)} S^2
\qquad
(a=1,2),\\
\boldsymbol e_1(\tau_i)\cdot\boldsymbol e_2(\tau_i)
&=&
0,\\
\|\boldsymbol e_1(\tau_i)\|
&=&
\|\boldsymbol e_2(\tau_i)\|=1.
\end{eqnarray}

\noindent In practice, this tangent-plane basis is chosen by a fixed deterministic convention common to all posterior samples. A trial Cartesian basis vector not parallel to $\widehat{\boldsymbol n}^{\rm ref}(\tau_i)$ is projected onto the tangent plane and normalized to define $\boldsymbol e_1(\tau_i)$; the second basis vector follows as $\boldsymbol e_2(\tau_i)=\widehat{\boldsymbol n}^{\rm ref}(\tau_i)\times\boldsymbol e_1(\tau_i)$. Basis-vector signs are chosen consistently along the temperature grid to prevent artificial sign jumps between adjacent $\tau_i$ points.

Fluctuations in the tangent plane are parameterized by scalar coefficients along the two basis directions. The displacement from $\widehat{\boldsymbol n}^{\rm ref}(\tau_i)$ along the $a$-th direction is denoted $c_a(\tau_i)$, $a=1,2$, collected as

\begin{equation}
\boldsymbol{c}_a
=
\bigl(
c_a(\tau_1),\ldots,c_a(\tau_{M_{\mathrm{grid}}})
\bigr)^{\mathsf T}
\in\mathbb{R}^{M_{\mathrm{grid}}}
\qquad
(a=1,2).
\end{equation}

\noindent Since $\boldsymbol c_a$ represents fluctuations around $\widehat{\boldsymbol n}^{\rm ref}$, its mean is zero. The $T$-direction covariance of the tangent-plane coefficients equals the common GPR posterior covariance $\Sigma_{\rm com}$ defined above, so that

\begin{equation}
\boldsymbol{c}_a
\sim
\mathcal{N}
\left(
\boldsymbol{0},
\Sigma_{\rm com}
\right)
\qquad
(a=1,2),
\label{eq:tangent_coeff_cov}
\end{equation}

\noindent where the processes for $a=1$ and $a=2$ are treated as independent with a common covariance $\Sigma_{\rm com}$. Because $\Sigma_{\rm com}$ contains correlations along the temperature grid, these posterior samples represent correlated trajectory deformations rather than independent noise at individual $\tau_i$ points. Specifically, we use the eigenvalue decomposition

\begin{equation}
\Sigma_{\rm com}
=
E_{\rm com}\Lambda_{\rm com} E_{\rm com}^{\mathsf T},
\end{equation}

\noindent where $E_{\rm com}$ is the orthogonal matrix whose columns are the eigenvectors of $\Sigma_{\rm com}$, and $\Lambda_{\rm com}$ is the diagonal matrix of the corresponding eigenvalues. Using random vectors drawn independently from the standard normal distribution,

\begin{equation}
\boldsymbol{z}_a^{(b)}
\sim
\mathcal{N}
\left(
\boldsymbol{0},
I_{M_{\mathrm{grid}}}
\right),
\end{equation}

\noindent we generate coefficient samples as 

\begin{equation}
\boldsymbol{c}_a^{(b)}
=
E_{\rm com}\Lambda_{\rm com}^{1/2}\boldsymbol{z}_a^{(b)}
\qquad
(a=1,2,\; b=2,\ldots,1000),
\end{equation}

\noindent where $I_{M_{\mathrm{grid}}}$ is the $M_{\mathrm{grid}}\times M_{\mathrm{grid}}$ identity matrix, and $\boldsymbol{z}_a^{(b)}$ are drawn independently for $a=1,2$ and $b=2,\ldots,1000$. Using these coefficient samples, the perturbed direction vectors at each evaluation grid point $\tau_i$ are

\begin{equation} 
\widetilde{\boldsymbol{n}}^{(b)}(\tau_i)
=
\widehat{\boldsymbol{n}}^{\rm ref}(\tau_i)
+
c_1^{(b)}(\tau_i)\,\boldsymbol{e}_1(\tau_i)
+
c_2^{(b)}(\tau_i)\,\boldsymbol{e}_2(\tau_i)
\qquad
(i=1,\ldots,M_{\mathrm{grid}},\; b=2,\ldots,1000).
\end{equation}

\noindent At this stage, $\widetilde{\boldsymbol{n}}^{(b)}(\tau_i)$ is a vector in $\mathbb{R}^3$ obtained by adding tangent-plane fluctuations to $\widehat{\boldsymbol{n}}^{\rm ref}(\tau_i)$, and in general it is not exactly a unit vector. We finally renormalize it at each temperature-grid point as

\begin{equation}
\widehat{\boldsymbol{n}}^{(b)}(\tau_i)
=
\frac{
\widetilde{\boldsymbol{n}}^{(b)}(\tau_i)
}{
\left\|
\widetilde{\boldsymbol{n}}^{(b)}(\tau_i)
\right\|
}
\in S^2
\qquad
(i=1,\ldots,M_{\mathrm{grid}},\; b=2,\ldots,1000).
\end{equation}

\noindent This renormalization ensures that each posterior sample is a sequence of direction vectors on $S^2$, so that all series satisfy

\begin{equation}
\left\{
\widehat{\boldsymbol{n}}^{(b)}(\tau_i)
\right\}_{i=1}^{M_{\mathrm{grid}}}
\subset S^2
\qquad
(b=1,\ldots,1000).
\end{equation}

\noindent In summary, $b=1$ is the representative common-grid series obtained by normalizing the component-wise GPR posterior mean on the evaluation grid. The series $b=2,\ldots,1000$ are posterior-resampled series obtained by drawing zero-mean tangent-plane deformation coefficients, using the GPR posterior covariance $\Sigma_{\rm com}$, to generate deformations around $\widehat{\boldsymbol n}^{\rm ref}$ in a linearized tangent-plane representation, followed by renormalization onto $S^2$. These resampled series propagate trajectory uncertainty into the subsequent holonomy analysis and enable assessment of the resampling stability of holonomy quantities and signed spatial patterns.

\vspace{0.5cm}

\section*{S3.~SVD-Based Grassmann Representation of Temperature Trajectories}

\vspace{0.3cm}

Here, we describe how the $T$-dependent vector series obtained at each pixel is summarized as a local geometric structure at each pixel $(x,y)$. The temperature series at each pixel is collected into a single matrix, and the dominant data-space directions are extracted via SVD. In the implementation, the analyzed spatial grid size is $302\times140$ pixels. To focus on the low-temperature response, we use only the data points in $\mathcal{T}_{\mathrm{grid}}$ that satisfy $14.0~\mathrm{K}\le \tau_i \le 135.0~\mathrm{K}$. We define this set of $T$ points as

\begin{equation}
\mathcal{T}_{\le 135\,\mathrm{K}}
=
\left\{
\tau_i\in\mathcal{T}_{\mathrm{grid}}
\,\middle|\,
14.0~\mathrm{K}\le \tau_i \le 135.0\,\mathrm{K}
\right\}
\subset \mathcal{T}_{\mathrm{grid}}.
\end{equation}

\noindent For notational simplicity, the elements of $\mathcal{T}_{\le 135\,\mathrm{K}}$ are again denoted as $\{\tau_1,\ldots,\tau_{|\mathcal{T}_{\le 135\,\mathrm{K}}|}\}$. Hereafter, each pixel $(x,y)$ is represented by a single index ${\mathsf P}$. For each pixel ${\mathsf P}$, the direction vector at $\tau_i\in\mathcal{T}_{\le 135\,\mathrm{K}}$ is expressed as

\begin{equation}
\widehat{\boldsymbol{n}}_{\mathsf P}^{(b)}(\tau_i) :=
\bigl(
\widehat{n}_{\mathsf x}^{(b)}(\mathsf P;\tau_i),\,
\widehat{n}_{\mathsf y}^{(b)}(\mathsf P;\tau_i),\,
\widehat{n}_{\mathsf z}^{(b)}(\mathsf P;\tau_i)
\bigr)^{\mathsf T}
\in S^2
\qquad
(i=1,\dots,\left|\mathcal{T}_{\le 135\,\mathrm{K}}\right|).
\end{equation}

\noindent For notational simplicity, the superscript $(b)$ indicating the posterior sample number ($b=1$--1,000) is omitted below. All quantities are defined for a fixed $b$, with the same procedure applied independently to all $b$. The temperature-series matrix for pixel ${\mathsf P}$ is

\begin{equation}
X_{\mathsf P}
=
\begin{pmatrix}
\left(\widehat{\boldsymbol{n}}_{\mathsf P}(\tau_1)\right)^{\mathsf T}\\
\left(\widehat{\boldsymbol{n}}_{\mathsf P}(\tau_2)\right)^{\mathsf T}\\
\vdots\\
\left(
\widehat{\boldsymbol{n}}_{\mathsf P}
\left(
\tau_{\left|\mathcal{T}_{\le 135\,\mathrm{K}}\right|}
\right)
\right)^{\mathsf T}
\end{pmatrix}
\in
\mathbb{R}^{\left|\mathcal{T}_{\le 135\,\mathrm{K}}\right|\times 3}.
\end{equation}

\noindent Each row represents the three-component direction vector at one $T$ point, and the columns represent the $T$ dependence of the $\mathsf{x}$, $\mathsf{y}$, and $\mathsf{z}$ components.

For the temperature-series matrix $X_{\mathsf P}$, we compute the reduced (thin) SVD,

\begin{eqnarray}
X_{\mathsf P}
&=&
U_{\mathsf P}\Sigma_{\mathsf P}V_{\mathsf P}^{\mathsf T} \in
\mathbb{R}^{|\mathcal{T}_{\le 135\,\mathrm{K}}|\times 3},
\\
U_{\mathsf P}
&=&
(\boldsymbol{\mathsf u}_{{\mathsf P},1},
 \boldsymbol{\mathsf u}_{{\mathsf P},2},
 \boldsymbol{\mathsf u}_{{\mathsf P},3}),
\\
V_{\mathsf P}
&=&
(\boldsymbol{\mathsf v}_{{\mathsf P},1},
 \boldsymbol{\mathsf v}_{{\mathsf P},2},
 \boldsymbol{\mathsf v}_{{\mathsf P},3}),
\label{V_mathsf_P}
\end{eqnarray}

\noindent where $U_{\mathsf P}\in\mathbb{R}^{|\mathcal{T}_{\le 135\,\mathrm{K}}|\times 3}$ has left singular vectors as columns, $\Sigma_{\mathsf P}\in\mathbb{R}^{3\times 3}$ is the diagonal singular-value matrix with non-negative entries in descending order, and $V_{\mathsf P}\in\mathbb{R}^{3\times 3}$ has right singular vectors as columns. The $r$-th left and right singular vectors are $\boldsymbol{\mathsf u}_{{\mathsf P},r}$ and $\boldsymbol{\mathsf v}_{{\mathsf P},r}$, respectively.

The matrices $V_{\mathsf P}$ and $U_{\mathsf P}$ obtained by SVD play complementary roles. The matrix $V_{\mathsf P}$ describes the directions and subspaces in which the sequence of direction vectors is mainly distributed in the three-dimensional direction space over the entire temperature series. In contrast, as shown by the relation $X_{\mathsf P}\boldsymbol{\mathsf v}_{{\mathsf P},r} = \sigma_{{\mathsf P},r}\boldsymbol{\mathsf u}_{{\mathsf P},r}$, the vector $\sigma_{{\mathsf P},r}\boldsymbol{\mathsf u}_{{\mathsf P},r}$ describes how the coefficient along the corresponding right-singular-vector direction varies along the $T$ direction. In other words, $V_{\mathsf P}$ gives the geometric structure in the optical-polarization-direction space, whereas $U_{\mathsf P}$ describes the corresponding normalized temperature-series profiles. Because

\begin{equation}
U_{\mathsf P}\Sigma_{\mathsf P}
=
X_{\mathsf P}V_{\mathsf P},
\end{equation}

\noindent the matrix $U_{\mathsf P}\Sigma_{\mathsf P}$ is the coefficient matrix that describes how the projection coefficients onto the corresponding right singular directions vary with $T$. If one only wants to extract a local axis or local plane, it is sufficient to focus on the right singular vectors, which belong to the three-dimensional direction space. The first right singular vector of $V_{\mathsf P}$ denotes the direction of maximum contribution, while the subspace spanned by the first two right singular vectors defines the local two-dimensional plane wherein the temperature series predominantly lies. However, our objective extends beyond merely extracting a principal axis or plane per pixel. Crucially, we aim to preserve the variation structure of the temperature series for inter-pixel comparison, and to trace the response approaching the ferroelectric transition within a local frame. Therefore, we retain both the local direction-space structure from $V_{\mathsf P}$ and the corresponding information from $U_{\mathsf P}$.

We begin by examining the right singular vectors. In the SVD representation, the first right singular vector $\boldsymbol{\mathsf v}_{{\mathsf P},1}\in\mathbb{R}^3$ indicates the directional component with the largest contribution over the temperature series at pixel ${\mathsf P}$. This local axis functions as a director, where sign reversal is identified, rather than an ordinary oriented vector. This aligns with the folded birefringence measurement employed here. Because optical retardances exceeding $\lambda/2$ render the distinction between phase advance and phase delay experimentally ambiguous, the measured optical-polarization state alone cannot uniquely specify the retardance branch. In the present folded experimental representation, the antipodal vectors $\widehat{\boldsymbol n}$ and $-\widehat{\boldsymbol n}$ represent the identical director-like optical response, rather than distinct measured branches. Consequently, the local axes derived from the temperature trajectories represent an axis, not an arrow. Specifically, the sign of the SVD vector is arbitrary; $\boldsymbol{\mathsf v}_{{\mathsf P},1}$ and $-\boldsymbol{\mathsf v}_{{\mathsf P},1}$ denote the same singular direction. Thus, the geometric object derived from the first right singular vector constitutes a one-dimensional subspace in $\mathbb{R}^3$,

\begin{equation}
{\mathsf P}
\longmapsto
[\boldsymbol{\mathsf v}_{{\mathsf P},1}]
\in
\mathrm{Gr}(1,3)\cong\mathbb{R}P^2.
\end{equation}

\noindent The first two right singular vectors also define the local plane in the three-dimensional direction space, 

\begin{equation}
\mathcal W_{\mathsf P}
=
\mathrm{span}
\left(
\boldsymbol{\mathsf v}_{{\mathsf P},1},
\boldsymbol{\mathsf v}_{{\mathsf P},2}
\right)
\in
\mathrm{Gr}(2,3).
\end{equation}

\noindent In three dimensions, a two-dimensional plane and its one-dimensional orthogonal complement are in one-to-one correspondence, so $\mathcal W_{\mathsf P}$ is equivalently represented by the normal director

\begin{equation}
\left[
\widehat{\boldsymbol{\mathsf n}}_{{\rm plane},{\mathsf P}}
\right]
=
\left[
\frac{
\boldsymbol{\mathsf v}_{{\mathsf P},1}
\times
\boldsymbol{\mathsf v}_{{\mathsf P},2}
}{
\left\|
\boldsymbol{\mathsf v}_{{\mathsf P},1}
\times
\boldsymbol{\mathsf v}_{{\mathsf P},2}
\right\|
}
\right].
\label{widehat_boldsymbol_n}
\end{equation}

\noindent This plane-based director is used below to define the local order parameter $S_{\rm plane}$; the right-side holonomy is defined using the one-dimensional director $[\boldsymbol{\mathsf v}_{{\mathsf P},1}]$.

We next describe the left singular vectors. $U_{\mathsf P}\in\mathbb{R}^{|\mathcal{T}_{\le 135\,\mathrm{K}}|\times 3}$ has column vectors $\boldsymbol{\mathsf{u}}_{{\mathsf P},r}\in\mathbb{R}^{|\mathcal{T}_{\le 135\,\mathrm{K}}|}$, $(r=1,2,3)$, representing series along the $T$ direction. Whereas the right singular vectors belong to the three-dimensional direction space and characterize which optical-polarization-direction components are present in the temperature series, the left singular vectors describe how each component varies as a temperature-dependent series on $\mathcal{T}_{\le 135\,\mathrm{K}}$. Although the left and right singular vectors correspond through the same singular values, they belong to different spaces and carry distinct information: each left singular vector paired with its singular value encodes the mode structure of the corresponding directional component along $T$. Here, the term ``temperature-evolution mode'' refers to a temperature profile derived from a measured trajectory, rather than to a microscopic excitation of the material.

We focus on the two-dimensional subspace spanned by the first two left singular vectors, retaining the two dominant temperature modes at each pixel. Reducing this to a one-dimensional quantity (the first mode only) would discard the dominant secondary variation in the temperature series that captures differences in onset behavior near the phase transition and in the response across $T$ regions. Although SVD yields a particular ordered orthonormal frame $(\boldsymbol{\mathsf u}_{{\mathsf P},1}, \boldsymbol{\mathsf u}_{{\mathsf P},2})$ up to the usual sign ambiguity for nondegenerate singular values, the two-dimensional subspace is represented by any orthonormal frame spanning the same space. Treating the subspace as the geometric object means the replacement

\begin{equation}
(\boldsymbol{\mathsf u}_{{\mathsf P},1},
 \boldsymbol{\mathsf u}_{{\mathsf P},2})
\mapsto
(\boldsymbol{\mathsf u}_{{\mathsf P},1},
 \boldsymbol{\mathsf u}_{{\mathsf P},2})H,
\qquad
H\in \mathrm{O}(2),
\end{equation}

\noindent does not change the point in the Grassmann manifold. The geometric object on the left side is therefore the two-dimensional subspace in temperature-series space spanned by the first two left singular vectors, not the individual basis vectors. The choice of local frame within this subspace is treated as a gauge freedom to be fixed below.

The SVD used here corresponds to the eigendecomposition of $X_{\mathsf P}^{\mathsf T}X_{\mathsf P}$, with the right singular vectors giving the directions that best represent the squared norm of the temperature series in the three-dimensional direction space. This is equivalent to non-centered principal component analysis (PCA), without explicit subtraction of the column mean. The squared singular value for each component therefore equals the sum of squared projections of the temperature series in the corresponding right singular direction. To quantify the global contribution of each singular component, the squared singular values are summed over all pixels to define the global PCA-type contribution ratio:

\begin{equation}
r_k
=
\frac{
\sum_{\mathsf P}\sigma_{{\mathsf P},k}^2
}{
\sum_{\mathsf P}\sum_{r=1}^{3}\sigma_{{\mathsf P},r}^2
} \qquad (k=1,2,3).
\end{equation}

\noindent The first component accounts for 84.7\% of the total contribution, and the cumulative contribution of the first two components exceeds 99\%. This global contribution ratio shows that the analyzed temperature-series matrices have a strong low-rank tendency at the ensemble level. This motivates the use of a dominant right-side axis and a two-dimensional mode space as data-induced geometric descriptors, without implying that the local singular-vector structure is equally stable at every pixel.

The geometric objects extracted from the left and right sides are different. On the right side, the target is a local optical-polarization axis representing the entire temperature series, extracted as the first right singular vector (maximum-contribution direction), $[\boldsymbol{\mathsf v}_{{\mathsf P},1}]\in \mathrm{Gr}(1,3)\cong\mathbb{R}P^2$. The plane $\mathcal W_{\mathsf P}$ spanned by the first two right singular vectors is used separately to characterize local plane order through $S_{\rm plane}$. In the present work, the right-side holonomy is defined from the one-dimensional director $[\boldsymbol{\mathsf v}_{{\mathsf P},1}]$, while $\mathcal W_{\mathsf P}$ is used only for the plane-order analysis.

On the left side, the aim is not to retain a single maximum mode but to capture the form of the $T$ evolution associated with the phase transition. The first left singular vector represents the dominant temperature mode, while differences in onset behavior and response across $T$ regions contribute to secondary modes. Reducing to the first mode alone would discard this secondary variation. We therefore retain the subspace

\begin{equation}
\mathcal U_{\mathsf P}
=
\mathrm{span}
(\boldsymbol{\mathsf u}_{{\mathsf P},1},
\boldsymbol{\mathsf u}_{{\mathsf P},2}),
\end{equation}

\noindent as a compact two-dimensional mode space containing the main response and its dominant secondary component.

Since the first two components account for over 99\% of the total contribution, this two-dimensional mode space compactly represents the dominant ensemble-level variation of the temperature series. Whereas the right side uses a one-dimensional director to represent the dominant optical-polarization axis, the left side uses the two-dimensional subspace spanned by the first two left singular vectors to represent the principal temperature-evolution structure. The left-side holonomy defined below evaluates the signed residual rotation of representative local frames attached to this temperature-evolution-mode space after transport around a closed loop in real space.

\vspace{0.5cm}

\section*{S4.~Construction of holonomy and local order parameters}

\vspace{0.3cm}

This section describes how holonomy is constructed as a residual rotation along closed loops from the right-side director field and the left-side local frame at each pixel. Local order parameters are also defined for the orientational order of the residual rotation axes associated with $\omega_{\mathcal R}$, as well as for the orientational order of the optical-polarization temperature-trajectory planes. Because the geometric objects on the right and left sides differ, the corresponding holonomies have different definitions and interpretations. For $\omega_{\mathcal R}$, the representative unit vector $\widehat{\boldsymbol{n}}_{\mathcal R}(x,y)\in S^2$ spanning the first right singular direction is used; pixel-position notation $(x,y)$ is restored here to describe the loop product in real space. For $\omega_{\mathcal L}$, the local orthonormal frame used is

\begin{equation}
\widetilde{U}(x,y)
=
\left(
\boldsymbol{\mathsf u}_{1}(x,y),
\boldsymbol{\mathsf u}_{2}(x,y)
\right)
\in
\mathbb{R}^{|\mathcal{T}_{\le 135\,\mathrm{K}}|\times 2}.
\end{equation}

\noindent This is a representative orthonormal frame attached to the two-dimensional temperature-evolution-mode space at pixel $(x,y)$, where $\boldsymbol{\mathsf u}_{1}(x,y)$ and $\boldsymbol{\mathsf u}_{2}(x,y)$ are mutually orthogonal unit vectors in $\mathbb{R}^{|\mathcal{T}_{\le 135\,\mathrm{K}}|}$. The mode space has an $\mathrm{O}(2)$ basis freedom, and $\widetilde U(x,y)$ is one frame selected under the SVD convention. We describe $\omega_{\mathcal R}$, then $\omega_{\mathcal L}$, and finally the local order parameters. As in our previous analysis, the origin of the real-space coordinate system is placed at the lower-right corner of the field of view. This makes it easier to compare with previous birefringence-based holonomy maps, as well as with the region of stress concentration in the lower-right part of the field of view. The pixel coordinate $(x,y)$ is thus defined relative to this origin, with $+x$ pointing leftward in the displayed image and $+y$ pointing upward. Near the left and upper image edges, the $L\times L$ loop extends outside the field of view; such regions are excluded as uncomputable. Their boundaries are indicated by dashed lines in all distribution maps.

The derivation of $\omega_{\mathcal R}$ uses the right-side director field. Although the intrinsic geometric object on the right side is an $\mathbb{R}P^2$-valued director field, a representative unit vector $\widehat{\boldsymbol{n}}_{\mathcal R}(x,y)\in S^2$ is selected at each pixel for the calculation. To define the minimum rotation between neighboring pixels, it is first necessary to determine, for each nearest-neighbor edge, which sign representative minimizes the angular distance. Following the quaternion definition described above, for the nearest-neighbor edge in the $+x$ direction, $(x,y)\to(x+1,y)$, we define

\begin{eqnarray}
s_x(x,y)
&=&
\mathrm{sign}\left[
\widehat{\boldsymbol{n}}_{\mathcal R}(x,y)\cdot
\widehat{\boldsymbol{n}}_{\mathcal R}(x+1,y)
\right], \\
\widehat{\boldsymbol{n}}_{\mathcal R}'(x+1,y)
&=&
s_x(x,y)\,
\widehat{\boldsymbol{n}}_{\mathcal R}(x+1,y).
\end{eqnarray}

\noindent
With $\mathrm{sign}(0)=+1$, this gives

\begin{equation}
\widehat{\boldsymbol{n}}_{\mathcal R}(x,y)\cdot
\widehat{\boldsymbol{n}}_{\mathcal R}'(x+1,y)
\ge 0,
\end{equation}

\noindent so the angular distance between $\widehat{\boldsymbol{n}}_{\mathcal R}(x,y)$ and $\widehat{\boldsymbol{n}}_{\mathcal R}'(x+1,y)$ is always evaluated using the minimum-angle representative. Similarly, for the nearest-neighbor edge in the $+y$ direction, $(x,y)\to(x,y+1)$,

\begin{eqnarray}
s_y(x,y)
&=&
\mathrm{sign}\left[
\widehat{\boldsymbol{n}}_{\mathcal R}(x,y)\cdot
\widehat{\boldsymbol{n}}_{\mathcal R}(x,y+1)
\right], \\
\widehat{\boldsymbol{n}}_{\mathcal R}'(x,y+1)
&=&
s_y(x,y)\,
\widehat{\boldsymbol{n}}_{\mathcal R}(x,y+1).
\end{eqnarray}

\noindent Then

\begin{equation}
\widehat{\boldsymbol{n}}_{\mathcal R}(x,y)\cdot
\widehat{\boldsymbol{n}}_{\mathcal R}'(x,y+1)
\ge 0.
\end{equation}

\noindent By this construction, a minimum-angle representative rotation is defined for each nearest-neighbor edge of the $\mathbb{R}P^2$ line field after removing the sign ambiguity. The sign-adjusted pair is denoted $(\widehat{\boldsymbol n}_{\mathcal R},\widehat{\boldsymbol n}'_{\mathcal R})$, and the minimum rotation between these two representatives is encoded as a unit quaternion. Specifically, following the definitions in Eqs.~(\ref{eq:rotation_axis_cross})--(\ref{eq:q_from_two_vectors_basic}), we define

\begin{eqnarray}
\boldsymbol{u}
&=&
\widehat{\boldsymbol{n}}_{\mathcal R}
\times
\widehat{\boldsymbol{n}}'_{\mathcal R}, \\
w
&=&
1+
\widehat{\boldsymbol{n}}_{\mathcal R}
\cdot
\widehat{\boldsymbol{n}}'_{\mathcal R}, \\
q(\widehat{\boldsymbol{n}}_{\mathcal R},
\widehat{\boldsymbol{n}}'_{\mathcal R})
&=&
\frac{(w,\boldsymbol{u})}
{\sqrt{w^2+\|\boldsymbol{u}\|^2}},
\end{eqnarray}

\noindent as the unit quaternion for the nearest-neighbor minimum rotation from $\widehat{\boldsymbol{n}}_{\mathcal R}$ to $\widehat{\boldsymbol{n}}'_{\mathcal R}$. This minimum-rotation quaternion is assigned to each nearest-neighbor edge: for the $+x$ edge $(x,y)\to(x+1,y)$ and the $+y$ edge $(x,y)\to(x,y+1)$, 

\begin{eqnarray}
q_x(x,y)
&:=&
q\left(
\widehat{\boldsymbol{n}}_{\mathcal R}(x,y),
\widehat{\boldsymbol{n}}_{\mathcal R}'(x+1,y)
\right), \\
q_y(x,y)
&:=&
q\left(
\widehat{\boldsymbol{n}}_{\mathcal R}(x,y),
\widehat{\boldsymbol{n}}_{\mathcal R}'(x,y+1)
\right).
\end{eqnarray}

\noindent For oppositely oriented edges, the inverse is used: $q_{-x}(x,y)=q_x(x-1,y)^{-1}$ and $q_{-y}(x,y)=q_y(x,y-1)^{-1}$. The holonomy along a closed loop is obtained by multiplying, in loop-orientation order, the quaternions for forward edges and their inverses for backward edges. For an $L\times L$ square loop $\partial \Box_L(x,y)$ with lower-right base point $(x,y)$, the order of multiplication is essential because quaternion multiplication is noncommutative. The residual rotation over the entire loop is therefore defined as

\begin{equation}
Q_L(x,y)
:=
\prod_{e\in \overrightarrow{\partial \Box_L(x,y)}} q_e
\in {\rm SU}(2).
\end{equation}

\noindent The product is taken in order around the loop starting from the lower-right base point:

\[
+L\ \text{steps in } +x
\;\rightarrow\;
+L\ \text{steps in } +y
\;\rightarrow\;
+L\ \text{steps in } -x
\;\rightarrow\;
+L\ \text{steps in } -y.
\]

\noindent Note that $\boldsymbol u$ above denoted the vector part of individual edge quaternions; after multiplication, the loop product $Q_L(x,y)$ has its own scalar and vector parts. The canonical representative is

\begin{equation}
Q_L(x,y)
=
\bigl(
w_L(x,y),
\boldsymbol v_L(x,y)
\bigr),
\qquad
w_L(x,y)\ge 0.
\label{eq:right_canonical_Q}
\end{equation}

\noindent The condition $w_L(x,y)\ge0$ selects the canonical representative in the double cover, since $Q_L$ and $-Q_L$ correspond to the same ${\rm SO}(3)$ rotation. The right-side holonomy angle is defined as

\begin{equation}
\omega_{\mathcal R}(x,y)=
2\arctan2\left(\|\boldsymbol{v}_L(x,y)\|,\,w_L(x,y)\right)
\in [0,\pi].
\label{omega_mathcal_R}
\end{equation}

\noindent In all figures and tables, this angle is expressed in degrees. $\omega_{\mathcal R}$ is a non-negative scalar representing residual rotation magnitude; $\omega_{\mathcal R}=0$ means the loop connection is trivial, while $\omega_{\mathcal R}>0$ signals a closed-loop connection mismatch in the director-axis field. When $\|\boldsymbol{v}_L(x,y)\|>0$, the rotation axis is

\begin{equation}
\widehat{\boldsymbol a}_{\mathcal R}(x,y)=
\frac{\boldsymbol{v}_L(x,y)}
{\|\boldsymbol{v}_L(x,y)\|}.
\label{a_rotation}
\end{equation}

\noindent This is the residual rotation axis, treated as a director under the identification $\widehat{\boldsymbol a}_{\mathcal R}\sim -\widehat{\boldsymbol a}_{\mathcal R}$ when used to define the local order parameter. Thus, $\omega_{\mathcal R}$ is the non-negative residual rotation angle accumulated by the nearest-neighbor minimum rotations of the $\mathbb{R}P^2$ director field around a closed loop.

We next construct $\omega_{\mathcal L}$ from the left-side local two-dimensional frame. Each pixel on the left side is assigned not a single unit vector but a frame $\widetilde{U}(x,y)$ spanning the local two-dimensional temperature-evolution-mode space. Unlike on the right side, nearest-neighbor rotations are not compared directly on $\mathbb{R}P^2$; instead, the relative transformation between neighboring local frames in their respective two-dimensional temperature-evolution-mode space bases is evaluated. Because $\omega_{\mathcal L}$ is a signed angle, the orientation of the local frame matters. Although the two-dimensional subspace is invariant under $\mathrm{O}(2)$ basis changes, reflections within $\mathrm{O}(2)$ reverse the sign of the signed rotation angle. The left-side holonomy is therefore restricted to the orientation-preserving $\mathrm{SO}(2)$ sector; as shown below, the holonomy angle along a closed loop is invariant under local basis rotations within this sector.

For neighboring pixels $(x,y)$ and $(x+1,y)$, we introduce the $2\times2$ matrix

\begin{equation}
G_x(x,y)
=
\widetilde{U}(x,y)^{\mathsf T}\widetilde{U}(x+1,y),
\label{eq:left_Gx}
\end{equation}

\noindent representing the relative arrangement of the two local frames. Similarly, for neighboring pixels in the $+y$ direction,

\begin{equation}
G_y(x,y)
=
\widetilde{U}(x,y)^{\mathsf T}\widetilde{U}(x,y+1).
\label{eq:left_Gy}
\end{equation}

\noindent These matrices capture the relative arrangement of two local frames in their two-dimensional bases. In general, $G_x(x,y)$ and $G_y(x,y)$ are not exactly orthogonal. Only their orthogonal parts are extracted and used as the nearest in-plane orthogonal transformations in the Frobenius-norm sense:

\begin{eqnarray}
G_x(x,y)
&=&
\widetilde{U}_x^{(G)}(x,y)\,\Sigma_x^{(G)}(x,y)\,
\left(V_x^{(G)}(x,y)\right)^{\mathsf T}, \\
G_y(x,y)
&=&
\widetilde{U}_y^{(G)}(x,y)\,\Sigma_y^{(G)}(x,y)\,
\left(V_y^{(G)}(x,y)\right)^{\mathsf T},
\end{eqnarray}

\noindent with corresponding orthogonal parts

\begin{eqnarray}
R_x(x,y)
&=&
\widetilde{U}_x^{(G)}(x,y)
\left(V_x^{(G)}(x,y)\right)^{\mathsf T},
\label{eq:left_Rxy1}\\
R_y(x,y)
&=&
\widetilde{U}_y^{(G)}(x,y)
\left(V_y^{(G)}(x,y)\right)^{\mathsf T}.
\label{eq:left_Rxy2}
\end{eqnarray}

\noindent These matrices are adopted as the nearest-neighbor in-plane transformations. Here $\widetilde{U}_x^{(G)}$, $V_x^{(G)}$, $\widetilde{U}_y^{(G)}$, and $V_y^{(G)}$ are $2\times2$ orthogonal matrices from the respective SVDs, and $R_x$, $R_y$ are the Frobenius-norm-closest orthogonal matrices to $G_x$, $G_y$. We numerically confirmed that $\det R_x(x,y)=\det R_y(x,y)=+1$ for all nearest-neighbor edges in the present data; we therefore treat $R_x(x,y), R_y(x,y)\in \mathrm{SO}(2)$.

This construction extracts the orthogonal part of the polar decomposition that best aligns two local frames. On the left side, the nearest-neighbor transformation is an orthogonal transformation within the local two-dimensional temperature-evolution-mode space rather than a quaternion. For oppositely oriented edges, inverse matrices are used: $R_{-x}(x,y)=R_x(x-1,y)^{-1}$ and $R_{-y}(x,y)=R_y(x,y-1)^{-1}$. Multiplying the edge transformations in order along the same $L\times L$ loop $\partial \Box_L(x,y)$ defines the left-side holonomy matrix as

\begin{equation}
R_L(x,y)
=
\prod_{e\in\overrightarrow{\partial\Box_L(x,y)}} R_e,
\label{eq:left_loop_product}
\end{equation}

\noindent where $R_e$ is the local orthogonal transformation assigned to each edge, ordered along the loop orientation. Since $\det R_e=+1$ for every valid edge, the loop product $R_L(x,y)\in\mathrm{SO}(2)$.

We now verify the invariance of the left-side holonomy under local $\mathrm{SO}(2)$ gauge transformations. The local frame at each pixel $(x,y)$ has the basis freedom

\begin{equation}
\widetilde{U}(x,y)
\rightarrow
\widetilde{U}(x,y)H(x,y),
\qquad
H(x,y)\in \mathrm{SO}(2).
\label{eq:left_SO2_gauge}
\end{equation}

\noindent Under this transformation, the nearest-neighbor transformations become

\begin{eqnarray}
R_x(x,y)
&\rightarrow&
H(x,y)^{\mathsf T}
R_x(x,y)
H(x+1,y),
\\
R_y(x,y)
&\rightarrow&
H(x,y)^{\mathsf T}
R_y(x,y)
H(x,y+1).
\label{eq:left_R_gauge}
\end{eqnarray}

\noindent When these edge transformations are multiplied in order along $\partial\Box_L(x,y)$, the gauge matrices at intermediate pixels cancel. The loop product transforms only by conjugation at the base point:

\begin{equation}
R_L(x,y)
\rightarrow
H(x,y)^{\mathsf T}
R_L(x,y)
H(x,y).
\label{eq:left_loop_gauge}
\end{equation}

\noindent Since $H(x,y),R_L(x,y)\in \mathrm{SO}(2)$ and $\mathrm{SO}(2)$ is Abelian,

\begin{equation}
H(x,y)^{\mathsf T}
R_L(x,y)
H(x,y)
=
R_L(x,y)
\label{eq:left_loop_gauge2}
\end{equation}

\noindent holds. Consequently, the loop product $R_L(x,y)$, and therefore the angle $\omega_{\mathcal L}(x,y)$ defined below, are independent of local basis rotations within the orientation-fixed $\mathrm{SO}(2)$ sector. The left-side holonomy angle is defined from the loop product as 

\begin{equation}
\omega_{\mathcal L}(x,y)
=
\arctan2\bigl((R_L(x,y))_{21},(R_L(x,y))_{11}\bigr)
\in [-\pi,\pi).
\label{eq:left_omega}
\end{equation}

\noindent The interval is taken as a principal branch; the endpoints represent the same $\mathrm{SO}(2)$ rotation, so only one of them is included. A possible discontinuity at the branch cut is a convention of the angle representation, not a singularity of $R_L(x,y)$. Since $R_L(x,y)\in \mathrm{SO}(2)$,

\begin{equation}
R_L(x,y)=
\begin{pmatrix}
\cos\omega_{\mathcal L}(x,y) & -\sin\omega_{\mathcal L}(x,y)\\
\sin\omega_{\mathcal L}(x,y) & \cos\omega_{\mathcal L}(x,y)
\end{pmatrix},
\end{equation}

\noindent from which the in-plane rotation angle is read from the $(1,1)$ and $(2,1)$ components. This angle is expressed in degrees throughout. The sign of $\omega_{\mathcal L}$ should be interpreted under the fixed real-space coordinate system, loop orientation, and $\mathrm{SO}(2)$ frame convention. It represents the orientation of the residual in-plane rotation and should not be identified with the sign of electric polarization, bound-charge density, or tensile or compressive strain. Reversing the loop orientation also reverses the sign of $\omega_{\mathcal L}$. When only the magnitude of the rotation residual is of interest, $|\omega_{\mathcal L}(x,y)|$ is used.

\begin{figure}[b]
\begin{center}
\includegraphics[width=15cm]{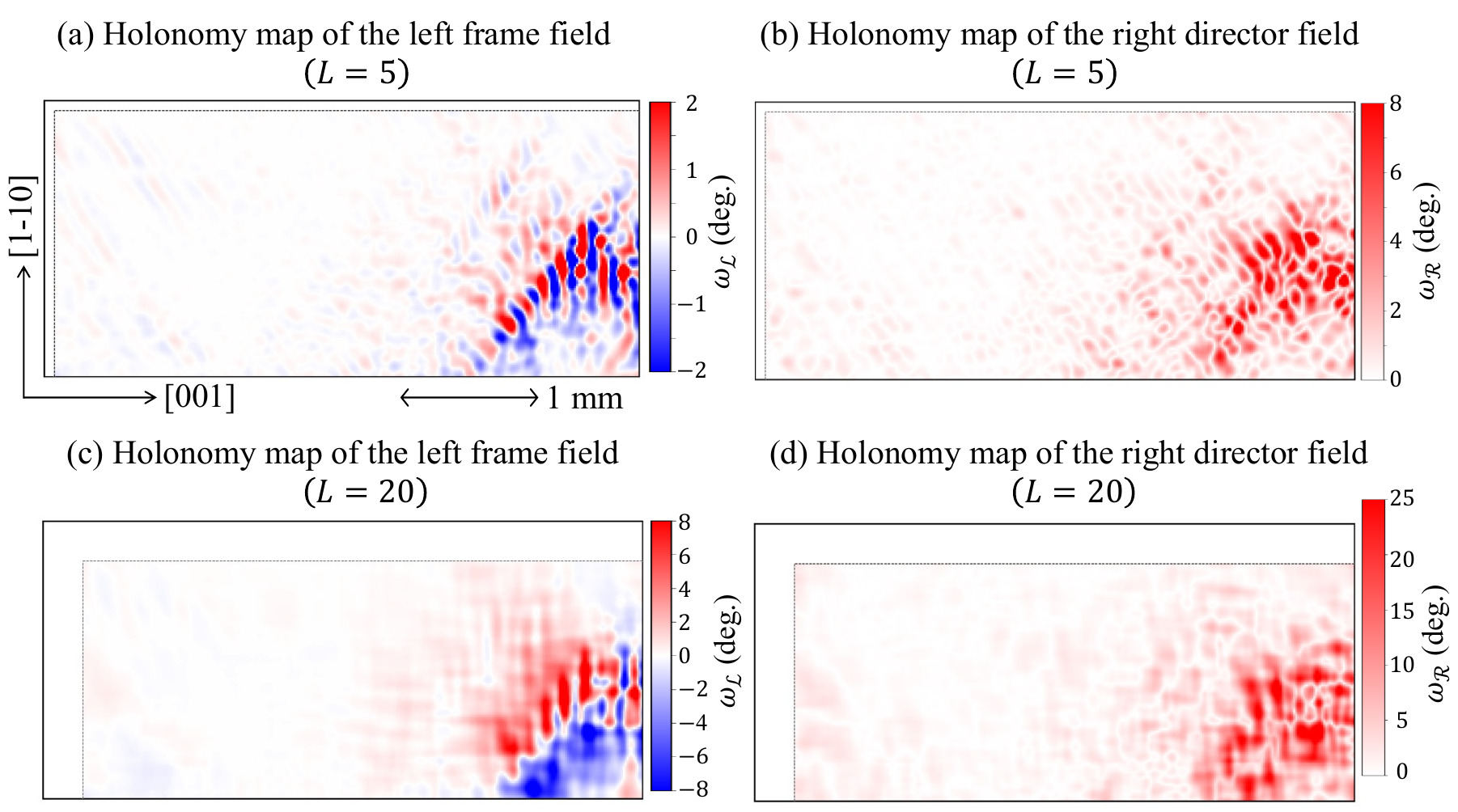}
\end{center}
\caption{Holonomy maps obtained from the left-side frame field and the right-side director field for various loop sizes. (a, c) Signed left-side holonomy angle $\omega_{\mathcal L}$ obtained from the local frame field in the temperature-series space for (a) $L=5$ and (c) $L=20$. (b, d) Right-side holonomy angle $\omega_{\mathcal R}$ obtained from the director field in the polarization-direction space for (b) $L=5$ and (d) $L=20$. Color scales are in degrees. Regions above and to the left of the dashed lines are uncomputable because the square loop extends beyond the field of view.}
\end{figure}

\begin{figure}[b]
\begin{center}
\includegraphics[width=11cm]{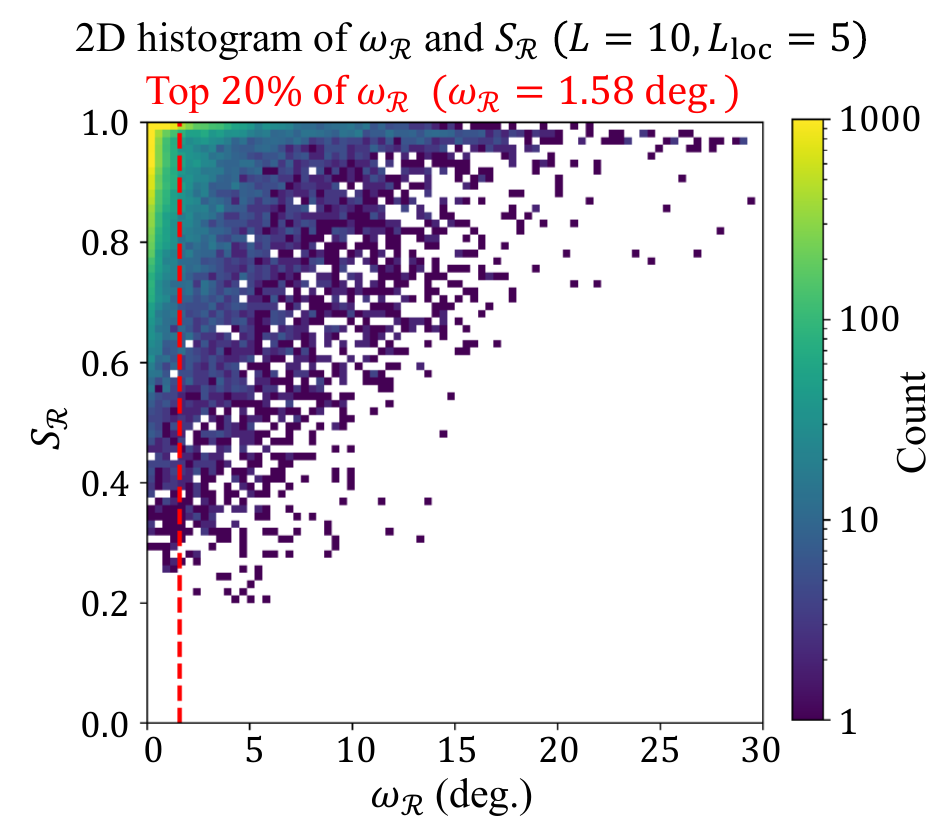}
\end{center}
\caption{Two-dimensional histogram of the right-side holonomy angle $\omega_{\mathcal R}$ and local order parameter $S_{\mathcal R}$ for $L=10$ and $L_{\rm loc}=5$. The color scale indicates the number of pixels per bin. The vertical dashed line marks the threshold for the top 20\% of $\omega_{\mathcal R}$.}
\end{figure}

\begin{figure}[b]
\begin{center}
\includegraphics[width=15cm]{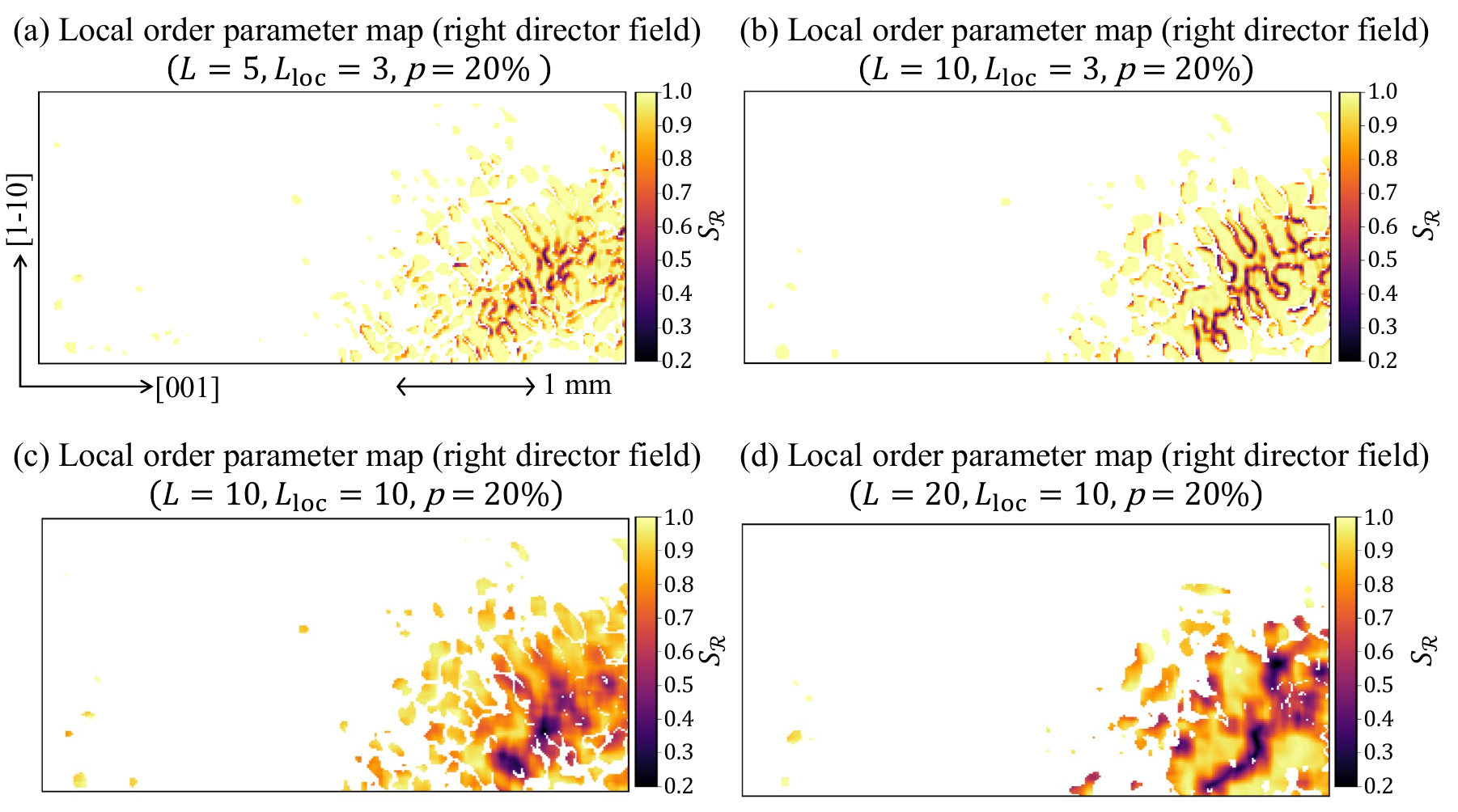}
\end{center}
\caption{Local order parameter maps $S_{\mathcal R}$ obtained from the right-side director field within the top 20\% high-$\omega_{\mathcal R}$ pixel set. Maps are shown for various combinations of loop size $L$ and local window size $L_{\rm loc}$: (a) $L=5$, $L_{\rm loc}=3$; (b) $L=10$, $L_{\rm loc}=3$; (c) $L=10$, $L_{\rm loc}=10$; and (d) $L=20$, $L_{\rm loc}=10$. For each base pixel, $S_{\mathcal R}$ was evaluated from the residual rotation axes in $\Omega^{\rm valid}_{\mathcal R,L_{\rm loc}} =\Omega_{\rm loc}\cap\Omega_{\omega,S}^{(20)}$, with the displayed base pixels restricted to $\Omega_{\omega,S}^{(20)}$.}
\end{figure}

The right and left sides share the common structure of accumulating nearest-neighbor minimum transformations along a loop, but differ in target space and transformation group. On the right side, the object is a line field in $\mathbb{R}P^2$, the nearest-neighbor rotation is a quaternion in three-dimensional space, and the loop product is an element of ${\rm SU}(2)$. After resolving the double-cover ambiguity, $\omega_{\mathcal R}$ is a non-negative angle in $[0,\pi]$. On the left side, the object is a frame in the local two-dimensional temperature-evolution-mode space. The nearest-neighbor transformation is a $2\times2$ orthogonal matrix, and the loop product is an element of $\mathrm{O}(2)$. Since $\det R_e=+1$ for every nearest-neighbor edge in the present data, the loop product lies in $\mathrm{SO}(2)$. The left-side holonomy angle $\omega_{\mathcal L}$ is accordingly a signed angle in the principal interval $[-\pi,\pi)$ and is interpreted as the loop residual of the in-plane orthogonal transformations associated with the local two-dimensional temperature-evolution-mode spaces obtained from the temperature series.

On the right side, the residual rotation axis $\widehat{\boldsymbol a}_{\mathcal R}(x,y)$ is obtained from the vector part of the loop-product quaternion as in Eq.~(\ref{a_rotation}). To evaluate local alignment of this axis, a local window size $L_{\mathrm{loc}}$ is introduced independently of the holonomy loop size $L$. With both $L$ and $L_{\mathrm{loc}}$ fixed, we write $S_{\mathcal R}^{(L,L_{\mathrm{loc}})}(x,y)$ simply as $S_{\mathcal R}(x,y)$ for brevity. Here, $\Omega_L$ is the set of base pixels for which the $L\times L$ holonomy loop lies entirely within the valid field of view. We first consider an $L_{\mathrm{loc}}\times L_{\mathrm{loc}}$ local square region extending from the base point $(x,y)\in\Omega_L$ in the $+x$ and $+y$ directions. The set of valid base points contained in this region is defined as

\begin{equation}
\Omega_{\mathrm{loc}}
=
\left\{
(x',y')\in\Omega_L
\,\middle|\,
x\leq x' < x+L_{\mathrm{loc}},\;
y\leq y' < y+L_{\mathrm{loc}}
\right\}.
\end{equation}

\noindent The sets $\Omega_{\rm loc}$, $\Omega^{\rm valid}_{\mathcal R,L_{\rm loc}}$, and $\Omega^{\rm valid}_{{\rm plane},L_{\rm loc}}$ depend on $(x,y)$, $L$, and $L_{\rm loc}$; this dependence is suppressed for readability. When $\omega_{\mathcal R}$ is very small, the residual rotation axis is numerically ill-defined. We therefore define the high-holonomy pixel set $\Omega_{\omega,S}^{(p)}$ as the set of pixels belonging to the top $p\%$ of the $\omega_{\mathcal R}$ distribution within $\Omega_L$. This set is used in two ways: first, only the residual rotation axes on $\Omega_{\omega,S}^{(p)}$ are retained as input samples for the local second-moment tensor; second, the displayed and summarized $S_{\mathcal R}$ values are restricted to base pixels belonging to the same set. Within $\Omega_{\mathrm{loc}}$, the local average is taken over the high-$\omega_{\mathcal R}$ pixels only. We therefore define

\begin{equation}
\Omega^{\mathrm{valid}}_{\mathcal R,L_{\mathrm{loc}}}
=
\Omega_{\mathrm{loc}}
\cap
\Omega_{\omega,S}^{(p)} ,
\end{equation}

\noindent with pixels satisfying $\|\boldsymbol v_L(x',y')\|=0$ excluded for numerical safety. If $\Omega^{\mathrm{valid}}_{\mathcal R,L_{\mathrm{loc}}}$ is empty, $S_{\mathcal R}(x,y)$ is not computed at that base point. The second-moment matrix of the axis vectors in $\Omega^{\mathrm{valid}}_{\mathcal R,L_{\mathrm{loc}}}$ is

\begin{equation}
\boldsymbol A_{\mathcal R}(x,y)
=
\frac{
\sum_{(x',y')\in
\Omega^{\rm valid}_{\mathcal R,L_{\rm loc}}}
\widehat{\boldsymbol a}_{\mathcal R}(x',y')
\widehat{\boldsymbol a}_{\mathcal R}(x',y')^{\mathsf T}
}{
\left|
\Omega^{\rm valid}_{\mathcal R,L_{\rm loc}}
\right|
}.\label{eq:AR_tensor}
\end{equation}

\noindent Letting $\lambda_{\max}^{(\mathcal R)}(x,y)$ be the largest eigenvalue of $\boldsymbol{A}_{\mathcal R}(x,y)$, the right-side local order parameter is

\begin{equation}
S_{\mathcal R}(x,y)
=
\frac{
3\lambda_{\max}^{(\mathcal R)}(x,y)-1
}{2}.
\label{eq:SR_definition}
\end{equation}

\noindent This nematic-type quantity quantifies the local orientational order of axis vectors under the identification $\widehat{\boldsymbol a}_{\mathcal R}\sim-\widehat{\boldsymbol a}_{\mathcal R}$: $S_{\mathcal R}\simeq 1$ indicates well-aligned residual rotation axes, while $S_{\mathcal R}\simeq 0$ indicates weak orientational order. Crucially, $S_{\mathcal R}$ measures the alignment of the rotation axes, not the magnitude of $\omega_{\mathcal R}$ itself.

We next define $S_{\rm plane}$ as a complementary local order parameter for the trajectory plane on the right-singular-vector side. Unlike $S_{\mathcal R}$, $S_{\rm plane}$ is not constructed from a holonomy residual and is not an order parameter of the left-side holonomy. The left-side holonomy $\omega_{\mathcal L}$ is a loop rotation angle associated with the representative local frame $\widetilde U(x,y)$ in the temperature-series space, and it does not directly provide a three-dimensional residual rotation axis analogous to $\widehat{\boldsymbol a}_{\mathcal R}(x,y)$. Instead, $S_{\rm plane}$ characterizes the local orientational order of the temperature-trajectory plane in the three-dimensional optical-polarization-direction space, obtained from the right singular vectors of the same trajectory. The local plane in which the trajectory is primarily distributed is

\begin{equation}
\mathcal W_{\mathsf P}
=
\mathrm{span}
\left(
\boldsymbol{\mathsf v}_{{\mathsf P},1},
\boldsymbol{\mathsf v}_{{\mathsf P},2}
\right)
\in
\mathrm{Gr}(2,3),
\end{equation}

\noindent where ${\mathsf P}$ corresponds to the pixel $(x,y)$ as in Eq.~(\ref{V_mathsf_P}). A representative orthonormal frame for this local plane is

\begin{equation}
W(x,y)
:=
\left(
\boldsymbol{\mathsf v}_{{\mathsf P},1},
\boldsymbol{\mathsf v}_{{\mathsf P},2}
\right)
\in
\mathbb{R}^{3\times2},
\label{Wxy_R3}
\end{equation}

\noindent so that $\mathcal W_{\mathsf P}=\mathrm{span}\,W(x,y)$. The representative normal vector $\widehat{\boldsymbol{\mathsf n}}_{\rm plane}(x,y)$ is defined in Eq.~(\ref{widehat_boldsymbol_n}). Because its sign depends on the sign choices of $\boldsymbol{\mathsf v}_{{\mathsf P},1}$ and $\boldsymbol{\mathsf v}_{{\mathsf P},2}$, we identify $\widehat{\boldsymbol{\mathsf n}}_{\rm plane}\sim -\widehat{\boldsymbol{\mathsf n}}_{\rm plane}$ and treat it as the normal director $[\widehat{\boldsymbol{\mathsf n}}_{\rm plane}(x,y)]$. For the $S_{\rm plane}$ analysis, we use the same local window as above but do not impose the high-$\omega_{\mathcal R}$ restriction. Thus, the valid set for the plane-normal directors is defined by

\begin{equation}
\Omega^{\mathrm{valid}}_{{\rm plane},L_{\mathrm{loc}}}
=
\Omega_{\mathrm{loc}}.
\end{equation}

\noindent Since the second-moment matrix is invariant under this sign flip, no global alignment of normal vectors is required. Using the normal directors in $\Omega^{\mathrm{valid}}_{{\rm plane},L_{\mathrm{loc}}}$, the second-moment matrix is

\begin{equation}
\boldsymbol A_{\rm plane}(x,y)
=
\frac{
\sum_{(x',y')\in
\Omega^{\rm valid}_{{\rm plane},L_{\rm loc}}}
\widehat{\boldsymbol{\mathsf n}}_{\rm plane}(x',y')
~\widehat{\boldsymbol{\mathsf n}}_{\rm plane}(x',y')^{\mathsf T}
}{
\left|
\Omega^{\rm valid}_{{\rm plane},L_{\rm loc}}
\right|
}.
\label{eq:AL_tensor}
\end{equation}

\noindent Letting $\lambda_{\max}^{(\rm plane)}(x,y)$ be the largest eigenvalue of $\boldsymbol{A}_{\rm plane}(x,y)$, the local order parameter of the temperature-trajectory plane is

\begin{equation}
S_{\rm plane}(x,y)
=
\frac{
3\lambda_{\max}^{(\rm plane)}(x,y)-1
}{2}.
\label{eq:SL_definition}
\end{equation}

\noindent This nematic-type orientational order parameter quantifies the alignment of the normal directions of the temperature-trajectory planes within the local region.

\begin{figure}[b]
\begin{center}
\includegraphics[width=15cm]{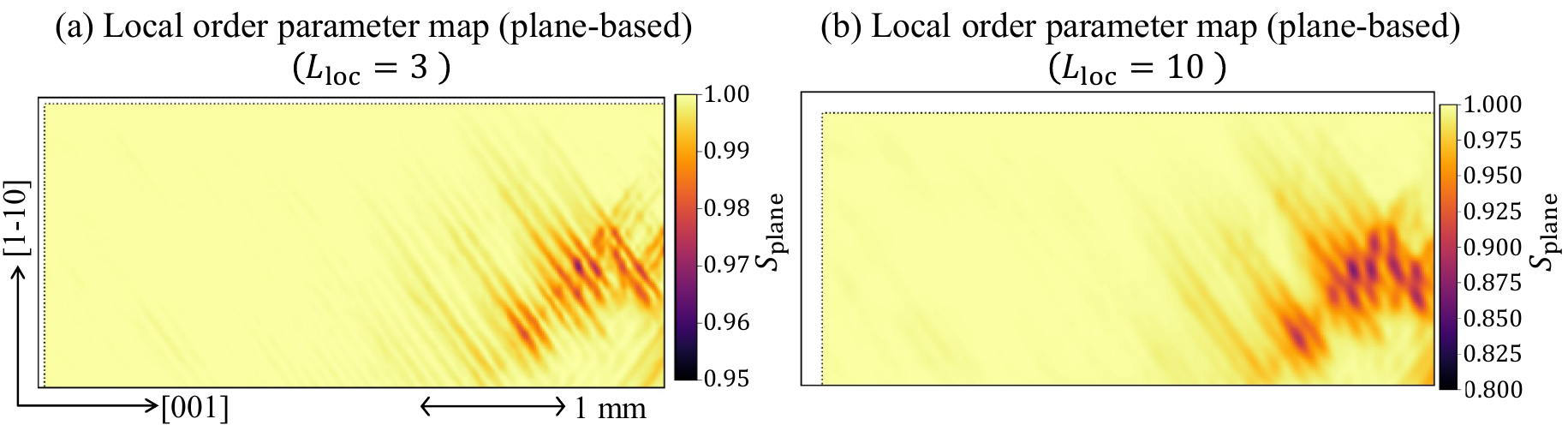}
\end{center}
\caption{Plane-based local order parameter maps $S_{\rm plane}$ calculated from the normal direction of the local optical-polarization temperature-trajectory plane on the right-singular-vector side. Maps were evaluated using local window sizes (a) $L_{\rm loc}=3$ and (b) $L_{\rm loc}=10$. For comparison with the holonomy maps, the displayed region is restricted to the same valid field of view; regions above and to the left of the dashed lines are omitted.}
\end{figure}

\begin{figure}[b]
\begin{center}
\includegraphics[width=15cm]{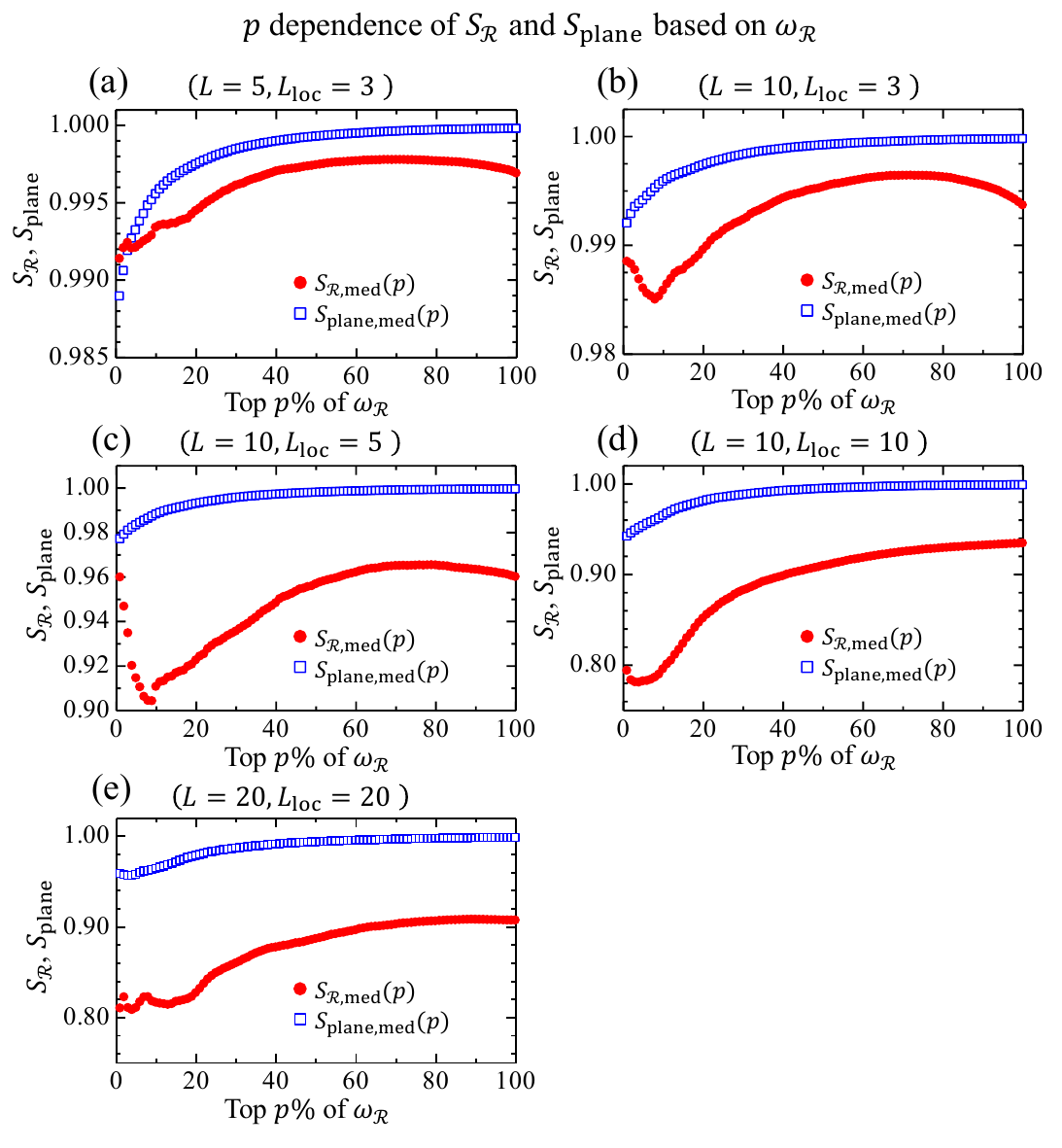}
\end{center}
\caption{Dependence of median local order parameters $S_{\mathcal R}$ and $S_{\rm plane}$ on the percentile threshold $p$. For each $p$, the high-holonomy pixel set $\Omega_{\omega,S}^{(p)}$ consists of pixels in the top $p\%$ of the right-side holonomy angle $\omega_{\mathcal R}$. For $S_{\mathcal R}$, this set determines both the residual rotation axes retained for the local second-moment tensor and the base pixels over which the median is evaluated. For $S_{\rm plane}$, the set defines the base pixels used to evaluate the median. Results are shown for various combinations of loop size $L$ and local window size $L_{\rm loc}$: (a) $L=5$, $L_{\rm loc}=3$; (b) $L=10$, $L_{\rm loc}=3$; (c) $L=10$, $L_{\rm loc}=5$; (d) $L=10$, $L_{\rm loc}=10$; and (e) $L=20$, $L_{\rm loc}=20$.}
\end{figure}

Figure~2 of the main text shows representative spatial distributions of $\omega_{\mathcal L}(x,y)$ and $\omega_{\mathcal R}(x,y)$ for $L=10$. We discuss the rationale for selecting $L=10$ as a representative loop size below. To illustrate loop size dependence, Fig.~S1 shows the results for $L=5$ and $L=20$, which exhibit similar overall spatial distributions to the $L=10$ results in Fig.~2. Notably, the positive and negative structures of $\omega_{\mathcal L}(x,y)$ are largely preserved across different loop sizes. However, as $L$ increases, the residual rotation accumulated along the closed loop also increases, leading to larger typical values for $|\omega_{\mathcal L}|$ and $\omega_{\mathcal R}$. Concurrently, because the loop product incorporates information over a broader area, local variations average out, resulting in a coarse-grained spatial structure. Between $L=5$ and $L=20$, large holonomy localization and the signed structure of $\omega_{\mathcal L}$ remain intact, whereas the holonomy angle magnitude increases and local structures become slightly blurred.

Figure~4(a) of the main text displays the spatial distribution of $S_{\mathcal R}(x,y)$ for the representative condition $L=10$ and $L_{\rm loc}=5$, evaluated using the top $20\%$ of $\omega_{\mathcal R}$ pixels. Figure~S2 shows a two-dimensional histogram of $\omega_{\mathcal R}$ and $S_{\mathcal R}$ under these identical conditions, revealing that $S_{\mathcal R}$ lacks a simple linear relationship with $\omega_{\mathcal R}$. The $\omega_{\mathcal R}$ distribution is heavily biased toward small values, with most pixels corresponding to background regions where $\omega_{\mathcal R}\simeq0$. The top $20\%$ threshold for $\omega_{\mathcal R}$ corresponds to $1.58^\circ$. This filtering excludes low-holonomy background areas, allowing us to evaluate the spatial structure of $S_{\mathcal R}$ in regions with finite rotation residuals. Figure~S3 presents $S_{\mathcal R}(x,y)$ under various conditions, evaluated using residual rotation axes from the top $20\%$ high-$\omega_{\mathcal R}$ pixel set. As $L_{\rm loc}$ increases, the orientational order is evaluated over a broader region, blurring the distribution. Notably, the number of axis samples in the local second-moment tensor depends not only on $L_{\rm loc}^2$ but also on $|\Omega^{\rm valid}_{\mathcal R,L_{\rm loc}}|$, i.e., by the number of high-$\omega_{\mathcal R}$ pixels within the local $L_{\rm loc}\times L_{\rm loc}$ window. Consequently, statistical accuracy depends on both $L_{\rm loc}$ and the local density of retained high-holonomy pixels, varying substantially between cases like $L_{\rm loc}=3$ and $L_{\rm loc}=10$. Nevertheless, regions showing a local decrease in $S_{\mathcal R}(x,y)$ remain consistent across conditions and overlap spatially with areas of locally enhanced $T_{\rm F}$, as shown in Fig.~1(b) of the main text.

Figure~4(b) of the main text illustrates $S_{\rm plane}(x,y)$ for the representative condition $L_{\rm loc}=5$. Figure~S4 similarly presents results for $L_{\rm loc}=3$ and $L_{\rm loc}=10$. These figures demonstrate that $S_{\rm plane}\simeq1$ in the background. Regions where $S_{\rm plane}$ drops below 1 generally overlap with high-holonomy areas. While the stripe-like structure blurs as $L_{\rm loc}$ increases, the spatial distribution remains broadly consistent with that of $T_{\rm F}$ across all examined conditions (see Fig.~1(b) of the main text).

We define $S_{\mathcal R,\mathrm{med}}(p)$ and $S_{{\rm plane},\mathrm{med}}(p)$ as the median values of $S_{\mathcal R}(x,y)$ and $S_{\rm plane}(x,y)$, evaluated over their valid pixel sets ($\Omega^{\rm valid}_{\mathcal R,L_{\rm loc}}$ and $\Omega^{\rm valid}_{{\rm plane},L_{\rm loc}}$) and restricted to the top $p\%$ of $\omega_{\mathcal R}$ pixels. Figure~S5 plots these median values as functions of $p=1,\ldots,100\%$ for various $L$ and $L_{\rm loc}$. For $S_{\mathcal R,\mathrm{med}}(p)$, selecting the top $p\%$ of $\omega_{\mathcal R}$ dictates both the residual-axis samples used in the local second-moment tensor and the base pixels evaluated for the median. Conversely, for $S_{{\rm plane},\mathrm{med}}(p)$, this selection only determines the base pixels used to evaluate the median of the pre-defined plane-order field. The minimum of $S_{{\rm plane},\mathrm{med}}(p)$ occurs at $p=1\%$, increasing toward 1 as $p$ increases. This indicates that in regions with large $\omega_{\mathcal R}$, the local orientational coherence of the temperature-trajectory plane $\mathcal W_{\mathsf P}(x,y)$ is lower than in the background. As $p$ increases, introducing more background pixels, $S_{\rm plane}$ approaches background levels. In contrast, $S_{\mathcal R,\mathrm{med}}(p)$ behaves differently. It typically reaches a minimum at $p\sim 5$--$10\%$ rather than $1\%$, and then gradually increases. This implies that $\omega_{\mathcal R}$ and $S_{\mathcal R}$ capture distinct geometric information: pixels with the largest $\omega_{\mathcal R}$ do not perfectly align with those exhibiting the lowest local orientational order of $\widehat{\boldsymbol a}_{\mathcal R}$. Furthermore, as indicated by Eqs.~(\ref{omega_mathcal_R}) and (\ref{a_rotation}), when $\omega_{\mathcal R}$ is small, $\boldsymbol v_L$ also shrinks, potentially making the direction of $\widehat{\boldsymbol a}_{\mathcal R}$ numerically unstable. Thus, the behavior of $S_{\mathcal R,\mathrm{med}}(p)$ at large $p$ values (which include many low-holonomy background pixels) should be treated purely as a reference trend.

\section*{S5.~Verification of the Left-side Frame Convention and Holonomy Stability}

\vspace{0.3cm}

The left-side holonomy angle $\omega_{\mathcal L}$ is constructed from a representative local two-dimensional frame $\widetilde{U}(x,y)$ attached to the temperature-evolution-mode space at each pixel $(x,y)$. This two-dimensional space has an $\mathrm{O}(2)$ basis freedom, and $\widetilde{U}(x,y)$ is one frame selected under the SVD convention. Since $\omega_{\mathcal L}$ is defined after fixing an orientation convention, the relevant gauge freedom is the local $\mathrm{SO}(2)$ rotation within the orientation-preserving sector. As shown above, the closed-loop holonomy angle is invariant under such rotations. To compare $\omega_{\mathcal L}$ across GPR posterior samples, however, the local frames must be aligned under a common orientation convention. Denoting the frame from the reference series ($b=1$) by $\widetilde{U}^{\rm ref}(x,y)$, for each posterior sample frame $\widetilde{U}^{(b)}(x,y)$ ($b=1,\ldots,1000$) we solve the $\mathrm{SO}(2)$ Procrustes problem:

\begin{equation}
H_{\mathrm{align}}^{(b)}(x,y)
=
\arg\min_{H\in \mathrm{SO}(2)}
\left\|
\widetilde{U}^{\rm ref}(x,y)
-
\widetilde{U}^{(b)}(x,y)H
\right\|_F,
\label{eq:procrustes}
\end{equation}

\noindent where $\|\cdot\|_F$ is the Frobenius norm. This is the orthogonal Procrustes problem: finding the in-plane orthogonal transformation bringing $\widetilde{U}^{(b)}(x,y)$ closest to $\widetilde{U}^{\rm ref}(x,y)$. For $b=1$, $H_{\mathrm{align}}^{(1)}(x,y)=I_2$ up to numerical round-off. The same procedure is applied to all posterior samples including $b=1$ for uniformity.

Specifically, we first compute the SVD

\begin{equation}
\left(\widetilde{U}^{(b)}(x,y)\right)^{\mathsf T}
\widetilde{U}^{\rm ref}(x,y)
=
U_{\rm gauge}^{(b)}(x,y)
D_{\rm gauge}^{(b)}(x,y)
\left(V_{\rm gauge}^{(b)}(x,y)\right)^{\mathsf T},
\label{eq:gauge_svd}
\end{equation}

\noindent where $U_{\rm gauge}^{(b)}(x,y)$ and $V_{\rm gauge}^{(b)}(x,y)$ are $2\times2$ orthogonal matrices, and $D_{\rm gauge}^{(b)}(x,y)$ is a diagonal matrix whose diagonal elements are non-negative singular values. When

\begin{equation}
\det\left[
U_{\rm gauge}^{(b)}(x,y)
\left(V_{\rm gauge}^{(b)}(x,y)\right)^{\mathsf T}
\right]
=+1,
\end{equation}

\noindent the Procrustes solution is

\begin{equation}
H_{\mathrm{align}}^{(b)}(x,y)
=
U_{\rm gauge}^{(b)}(x,y)
\left(V_{\rm gauge}^{(b)}(x,y)\right)^{\mathsf T}.
\label{eq:procrustes_solution}
\end{equation}

\noindent This aligns $\widetilde{U}^{(b)}(x,y)$ at each pixel to the orientation closest to $\widetilde{U}^{\rm ref}(x,y)$. For all valid pixels and all posterior samples, we confirmed that $\det H_{\mathrm{align}}^{(b)}(x,y)=+1$; the alignment is therefore orientation-preserving and free of reflection components. This verification allows $\omega_{\mathcal L}$ to be compared across posterior samples under a common orientation convention. The resulting $\omega_{\mathcal L}$ is a signed in-plane residual rotation defined under fixed real-space coordinates, loop orientation, and an orientation-preserving $\mathrm{SO}(2)$ frame convention. It is invariant under local $\mathrm{SO}(2)$ basis rotations within each posterior sample. The positive and negative signs distinguish the opposite orientations of the loop-induced residual rotation under this fixed convention. However, they do not represent the sign of electric polarization, bound-charge density, or tensile or compressive strain.

Next, we numerically verify that the left-side holonomy is invariant under local orientation-preserving $\mathrm{SO}(2)$ gauge transformations. These random transformations do not alter the two-dimensional subspace itself within the temperature-series space; rather, they randomly rotate the orthonormal basis representing that subspace at each pixel. Specifically, we generate an independent random rotation $H_{\rm rand}(x,y)\in \mathrm{SO}(2)$ at each pixel and apply the local gauge transformation

\begin{equation}
\widetilde{U}^{\rm ref}(x,y)
\mapsto
\widetilde{U}^{\rm ref}(x,y)\,H_{\rm rand}(x,y).
\end{equation}

\noindent Under this transformation, the Grassmann point at each pixel, namely the two-dimensional subspace spanned by $\widetilde{U}^{\rm ref}(x,y)$, is unchanged, and only its representative frame is modified. We then recompute the holonomy angle $\omega_{\mathcal L}^{\rm rand}(x,y)$ from the transformed frame field and compare it with the holonomy angle $\omega_{\mathcal L}^{\rm ref}(x,y)$ obtained from the reference frame. Because $\mathrm{SO}(2)$ is Abelian, the residual conjugation at the base point leaves $R_L(x,y)$ unchanged, as shown in Eqs.~(\ref{eq:left_loop_gauge}) and (\ref{eq:left_loop_gauge2}). Therefore, ideally, the loop holonomy angle should also be unchanged. Below, we verify numerically that this property is satisfied in the implementation within numerical precision.

Let $\Omega_L$ be the set of valid pixels for which the $L\times L$ holonomy loop can be constructed. To evaluate the stability of the rotation-residual magnitude, we compute the Pearson correlation coefficient between the absolute values,

\begin{equation}
\mathcal{C}^{\rm ref}
=
\mathrm{corr}_{(x,y)\in\Omega_L}
\left(
\left|\omega_{\mathcal L}^{\rm ref}(x,y)\right|,
\left|\omega_{\mathcal L}^{\rm rand}(x,y)\right|
\right).
\end{equation}

\noindent $\mathcal{C}^{\rm ref}\approx1$ indicates that $|\omega_{\mathcal L}^{\rm rand}|$ is essentially unchanged from $|\omega_{\mathcal L}^{\rm ref}(x,y)|$ after the random gauge transformation. As shown in Table~S1, $\mathcal{C}^{\rm ref}$ is virtually 1 for all loop sizes $L$, confirming that the left-side holonomy magnitude is invariant under local $\mathrm{SO}(2)$ gauge transformations to within numerical precision.

\begin{table}[tb] 
\centering 
\caption{Numerical validation of the left-side holonomy under random local $\mathrm{SO}(2)$ gauge transformations. Random orientation-preserving gauge transformations were applied to the left-side frame field, and the resulting holonomy was compared against the reference.}
\begin{tabular}{c|cccccc} 
\hline
$L$&$\mathcal{C}^{\rm ref}$&$\mathrm{MAE}$ (deg.)&$\mathrm{RMSE}$ (deg.)&$\mathrm{MaxErr}$ (deg.)&$\mathrm{Agreement}$&$\Delta \mathcal{C}_{\rm sign}$\\ \hline
2&$>0.999$&$3.98\times 10^{-6}$&$7.82\times 10^{-6}$&$3.07\times 10^{-5}$&$>0.999$&$1.14\times 10^{-4}$\\
3&$>0.999$&$5.62\times 10^{-6}$&$9.44\times 10^{-6}$&$3.60\times 10^{-5}$&$>0.999$&$5.19\times 10^{-5}$\\
4&$>0.999$&$6.83\times 10^{-6}$&$1.07\times 10^{-5}$&$4.58\times 10^{-5}$&$>0.999$&$6.07\times 10^{-5}$\\
5&$>0.999$&$7.83\times 10^{-6}$&$1.17\times 10^{-5}$&$4.71\times 10^{-5}$&$>0.999$&$1.23\times 10^{-4}$\\
6&$>0.999$&$8.71\times 10^{-6}$&$1.27\times 10^{-5}$&$5.16\times 10^{-5}$&$>0.999$&$7.30\times 10^{-5}$\\
7&$>0.999$&$9.46\times 10^{-6}$&$1.35\times 10^{-5}$&$5.84\times 10^{-5}$&$>0.999$&$1.88\times 10^{-5}$\\
8&$>0.999$&$1.02\times 10^{-5}$&$1.44\times 10^{-5}$&$6.15\times 10^{-5}$&$>0.999$&$1.43\times 10^{-5}$\\
9&$>0.999$&$1.08\times 10^{-5}$&$1.51\times 10^{-5}$&$6.33\times 10^{-5}$&$>0.999$&$4.59\times 10^{-5}$\\
10&$>0.999$&$1.15\times 10^{-5}$&$1.59\times 10^{-5}$&$6.67\times 10^{-5}$&$>0.999$&$2.99\times 10^{-5}$\\
11&$>0.999$&$1.21\times 10^{-5}$&$1.66\times 10^{-5}$&$7.29\times 10^{-5}$&$>0.999$&$1.91\times 10^{-5}$\\
12&$>0.999$&$1.26\times 10^{-5}$&$1.72\times 10^{-5}$&$7.63\times 10^{-5}$&$>0.999$&$3.28\times 10^{-5}$\\
13&$>0.999$&$1.32\times 10^{-5}$&$1.79\times 10^{-5}$&$7.83\times 10^{-5}$&$>0.999$&$5.34\times 10^{-5}$\\
14&$>0.999$&$1.36\times 10^{-5}$&$1.85\times 10^{-5}$&$7.97\times 10^{-5}$&$>0.999$&$1.48\times 10^{-5}$\\
15&$>0.999$&$1.40\times 10^{-5}$&$1.90\times 10^{-5}$&$8.36\times 10^{-5}$&$>0.999$&$1.74\times 10^{-5}$\\
16&$>0.999$&$1.45\times 10^{-5}$&$1.95\times 10^{-5}$&$8.58\times 10^{-5}$&$>0.999$&$5.59\times 10^{-5}$\\
17&$>0.999$&$1.49\times 10^{-5}$&$2.01\times 10^{-5}$&$8.82\times 10^{-5}$&$>0.999$&$1.61\times 10^{-5}$\\
18&$>0.999$&$1.53\times 10^{-5}$&$2.06\times 10^{-5}$&$9.25\times 10^{-5}$&$>0.999$&$1.52\times 10^{-5}$\\
19&$>0.999$&$1.57\times 10^{-5}$&$2.10\times 10^{-5}$&$9.31\times 10^{-5}$&$>0.999$&$1.69\times 10^{-5}$\\
20&$>0.999$&$1.60\times 10^{-5}$&$2.15\times 10^{-5}$&$9.55\times 10^{-5}$&$>0.999$&$2.16\times 10^{-5}$\\
\hline
\end{tabular} 
\end{table}

We next evaluate the difference between the signed angles $\omega_{\mathcal L}^{\rm ref}$ and $\omega_{\mathcal L}^{\rm rand}$ directly. Because angles are $360^\circ$-periodic, the difference is wrapped into the principal interval:

\begin{equation}
\Delta\omega_{\mathcal L}(x,y)
:=
\mathrm{wrap}_{[-180^\circ,180^\circ)}
\left[
\omega_{\mathcal L}^{\rm ref}(x,y)
-
\omega_{\mathcal L}^{\rm rand}(x,y)
\right].
\end{equation}

\noindent The mean absolute error, root-mean-square error, and maximum error are then

\begin{eqnarray}
\mathrm{MAE}
&=&
\frac{1}{|\Omega_L|}
\sum_{(x,y)\in\Omega_L}
\left|
\Delta\omega_{\mathcal L}(x,y)
\right|,\\
\mathrm{RMSE}
&=&
\sqrt{
\frac{1}{|\Omega_L|}
\sum_{(x,y)\in\Omega_L}
\left[
\Delta\omega_{\mathcal L}(x,y)
\right]^2
},\\
\mathrm{MaxErr}
&=&
\max_{(x,y)\in\Omega_L}
\left|
\Delta\omega_{\mathcal L}(x,y)
\right|.
\end{eqnarray}

\noindent These quantities assess the global and worst-case deviations. Table~S1 shows that all errors are extremely small. This shows that the holonomy angle is unchanged within numerical precision, both globally and locally, before and after the gauge transformation.

To evaluate the stability of the signed spatial pattern, we introduce the agreement rate between the sign fields $\omega_{\mathcal L}^{\rm ref}(x,y)$ and $\omega_{\mathcal L}^{\rm rand}(x,y)$,

\begin{equation}
\mathrm{Agreement}
=
\frac{1}{|\Omega_L|}
\sum_{(x,y)\in\Omega_L}
\mathbf{1}\left[
\mathrm{sign}\left(\omega_{\mathcal L}^{\rm ref}(x,y)\right)
=
\mathrm{sign}\left(\omega_{\mathcal L}^{\rm rand}(x,y)\right)
\right].
\end{equation}

\noindent The evaluated sign represents the orientation of the residual in-plane rotation under fixed real-space coordinates, loop orientation, and orientation-preserving $\mathrm{SO}(2)$ frame convention. It should not be identified with the sign of electric polarization, bound-charge density, or tensile or compressive strain. Reversing the loop orientation reverses the sign of $\omega_{\mathcal L}$. An $\mathrm{SO}(2)$ gauge transformation is a continuous, orientation-preserving rotation that theoretically leaves the signed holonomy angle unchanged. Conversely, a basis transformation with a reflection component in $\mathrm{O}(2)$ could reverse both the local frame orientation and the angle's sign. Thus, the agreement rate serves as a metric to verify if the sign structure of $\omega_{\mathcal L}$ remains stable under random orientation-preserving gauge transformations. An agreement rate close to 1 means that the sign structure is preserved. For pixels where $\omega_{\mathcal L}(x,y)=0$ numerically, we set $\mathrm{sign}(\omega_{\mathcal L}(x,y))=0$. To further assess whether random gauge transformations preserve the nearest-neighbor spatial correlation of the sign field, we define:

\begin{eqnarray}
\mathcal{C}_{\rm sign}^{\rm ref}
&=&
\frac{1}{2}
\biggl[
\frac{1}{|\Omega_x|}
\sum_{(x,y)\in\Omega_x}
\mathrm{sign}\left(\omega_{\mathcal L}^{\rm ref}(x,y)\right)
\mathrm{sign}\left(\omega_{\mathcal L}^{\rm ref}(x+1,y)\right)
\nonumber \\
&&
+
\frac{1}{|\Omega_y|}
\sum_{(x,y)\in\Omega_y}
\mathrm{sign}\left(\omega_{\mathcal L}^{\rm ref}(x,y)\right)
\mathrm{sign}\left(\omega_{\mathcal L}^{\rm ref}(x,y+1)\right)
\biggr],
\label{mathcalC_sign_ref_rand}
\\
\mathcal{C}_{\rm sign}^{\rm rand}
&=&
\frac{1}{2}
\biggl[
\frac{1}{|\Omega_x|}
\sum_{(x,y)\in\Omega_x}
\mathrm{sign}\left(\omega_{\mathcal L}^{\rm rand}(x,y)\right)
\mathrm{sign}\left(\omega_{\mathcal L}^{\rm rand}(x+1,y)\right)
\nonumber \\
&&
+
\frac{1}{|\Omega_y|}
\sum_{(x,y)\in\Omega_y}
\mathrm{sign}\left(\omega_{\mathcal L}^{\rm rand}(x,y)\right)
\mathrm{sign}\left(\omega_{\mathcal L}^{\rm rand}(x,y+1)\right)
\biggr],
\end{eqnarray}

\noindent and their difference 

\begin{equation}
\Delta \mathcal{C}_{\rm sign}
=
\left|
\mathcal{C}_{\rm sign}^{\rm rand}
-
\mathcal{C}_{\rm sign}^{\rm ref}
\right|.
\end{equation}

\noindent The closer $\Delta \mathcal{C}_{\rm sign}$ is to 0, the more the nearest-neighbor spatial correlation of the sign field is preserved under random gauge transformations. As shown in Table~S1, $\mathrm{Agreement}$ approaches 1, and $\Delta \mathcal{C}_{\rm sign}$ is near 0. Consequently, the left-side holonomy sign pattern is robust under local orientation-preserving $\mathrm{SO}(2)$ gauge transformations. This confirms its consistent definition within the selected $\mathrm{SO}(2)$ sector. Ultimately, these tests demonstrate that the left-side holonomy's magnitude, signed angle, and sign structure are invariant under local $\mathrm{SO}(2)$ gauge transformations, within numerical precision.

Next, we fix $\widetilde{U}^{\rm ref}(x,y)$ and evaluate the reproducibility of the left-side holonomy for the gauge-fixed frame fields generated from the GPR posterior samples $(b=1,\ldots,1000)$. For each sample $b$, we compute the absolute-value correlation between $\omega_{\mathcal L}^{\rm ref}(x,y)$ and $\omega_{\mathcal L}^{(b)}(x,y)$ over the valid pixel set $\Omega_L$, where the $L\times L$ loop is computable:

\begin{equation}
\mathcal{C}^{(b)}
=
\mathrm{corr}_{(x,y)\in\Omega_L}\left(
\left|\omega_{\mathcal L}^{\rm ref}(x,y)\right|,
\left|\omega_{\mathcal L}^{(b)}(x,y)\right|
\right).
\end{equation}

\noindent We compute this quantity for $b=1,\ldots,1000$ and define its mean as

\begin{equation}
\overline{\mathcal{C}}^{(1000)}
=
\frac{1}{1000}
\sum_{b=1}^{1000}
\mathcal{C}^{(b)},
\label{C_abs1000}
\end{equation}

\noindent where $b=1$ gives $\mathcal{C}^{(1)}=1$ up to numerical round-off but is included for uniformity. Table~S2 shows that $\overline{\mathcal{C}}^{(1000)}\approx1$ for all loop sizes $L$.

Accounting for the $360^\circ$ periodicity of angles, the wrapped difference in the principal interval is

\begin{equation}
\Delta\omega_{\mathcal L}^{(b)}(x,y)
=
\mathrm{wrap}_{[-180^\circ,180^\circ)}
\left[
\omega_{\mathcal L}^{\rm ref}(x,y)
-
\omega_{\mathcal L}^{(b)}(x,y)
\right].
\end{equation}

\noindent The mean absolute error and root-mean-square error for each posterior sample are

\begin{eqnarray}
\mathrm{MAE}^{(b)}
&=&
\frac{1}{|\Omega_L|}
\sum_{(x,y)\in\Omega_L}
\left|
\Delta\omega_{\mathcal L}^{(b)}(x,y)
\right|,\\
\mathrm{RMSE}^{(b)}
&=&
\sqrt{
\frac{1}{|\Omega_L|}
\sum_{(x,y)\in\Omega_L}
\left[
\Delta\omega_{\mathcal L}^{(b)}(x,y)
\right]^2
}.
\end{eqnarray}

\noindent Averaging over 1,000 posterior samples gives

\begin{eqnarray}
\overline{\mathrm{MAE}}^{(1000)}
&=&
\frac{1}{1000}
\sum_{b=1}^{1000}
\mathrm{MAE}^{(b)},\\
\overline{\mathrm{RMSE}}^{(1000)}
&=&
\frac{1}{1000}
\sum_{b=1}^{1000}
\mathrm{RMSE}^{(b)}.
\end{eqnarray}

\begin{table}[tb] 
\centering 
\caption{Posterior-sample validation of the left-side holonomy and its correlation with the right-side holonomy. Statistics are averaged across posterior samples $b=1,\ldots,1000$. The metrics $\overline{\mathcal{C}}^{(1000)}$, $\overline{\mathrm{MAE}}^{(1000)}$, and $\overline{\mathrm{RMSE}}^{(1000)}$ measure the reproducibility of $|\omega_{\mathcal L}|$, whereas $\overline{\mathcal{C}}_{\mathcal{LR}}^{(1000)}$ represents the posterior-averaged correlation between $|\omega_{\mathcal L}|$ and $|\omega_{\mathcal R}|$.}
\begin{tabular}{c|cccc} 
\hline
$L$&$\overline{\mathcal{C}}^{(1000)}$&$\overline{\mathrm{MAE}}^{(1000)}$~(deg.)&$\overline{\mathrm{RMSE}}^{(1000)}$~(deg.)&$\overline{\mathcal{C}}_{\mathcal{LR}}^{(1000)}$\\ \hline
2&$>0.999$&$5.31\times 10^{-4}$&$1.49\times 10^{-3}$&$0.614$\\
3&$>0.999$&$1.04\times 10^{-3}$&$2.93\times 10^{-3}$&$0.614$\\
4&$>0.999$&$1.61\times 10^{-3}$&$4.46\times 10^{-3}$&$0.617$\\
5&$>0.999$&$2.17\times 10^{-3}$&$5.92\times 10^{-3}$&$0.618$\\
6&$>0.999$&$2.69\times 10^{-3}$&$7.20\times 10^{-3}$&$0.617$\\
7&$>0.999$&$3.15\times 10^{-3}$&$8.23\times 10^{-3}$&$0.618$\\
8&$>0.999$&$3.55\times 10^{-3}$&$9.02\times 10^{-3}$&$0.622$\\
9&$>0.999$&$3.91\times 10^{-3}$&$9.71\times 10^{-3}$&$0.627$\\
10&$>0.999$&$4.26\times 10^{-3}$&$1.04\times 10^{-2}$&$0.629$\\
11&$>0.999$&$4.64\times 10^{-3}$&$1.13\times 10^{-2}$&$0.628$\\
12&$>0.999$&$5.05\times 10^{-3}$&$1.22\times 10^{-2}$&$0.635$\\
13&$>0.999$&$5.48\times 10^{-3}$&$1.32\times 10^{-2}$&$0.648$\\
14&$>0.999$&$5.92\times 10^{-3}$&$1.42\times 10^{-2}$&$0.662$\\
15&$>0.999$&$6.37\times 10^{-3}$&$1.53\times 10^{-2}$&$0.672$\\
16&$>0.999$&$6.82\times 10^{-3}$&$1.62\times 10^{-2}$&$0.678$\\
17&$>0.999$&$7.28\times 10^{-3}$&$1.70\times 10^{-2}$&$0.686$\\
18&$>0.999$&$7.74\times 10^{-3}$&$1.78\times 10^{-2}$&$0.697$\\
19&$>0.999$&$8.21\times 10^{-3}$&$1.87\times 10^{-2}$&$0.708$\\
20&$>0.999$&$8.68\times 10^{-3}$&$1.97\times 10^{-2}$&$0.714$\\
\hline
\end{tabular} 
\end{table}

\noindent As Table~S2 demonstrates, these quantities exhibit a marginal increase in error relative to the random-gauge-transformation test. This slight deviation likely stems from statistical fluctuations within the frame field itself, introduced by posterior samples in the temperature series. Nevertheless, $\overline{\mathcal{C}}^{(1000)}$ remains close to 1, while $\overline{\mathrm{MAE}}^{(1000)}$ and $\overline{\mathrm{RMSE}}^{(1000)}$ remain small. This confirms the high reproducibility of the left-side holonomy across the posterior samples. In other words, these results support the resampling stability of the left-side holonomy against GPR posterior fluctuations, establishing it as a robust geometric descriptor for this analysis.

\begin{table}[tb]
\centering
\caption{Posterior summary of the left-side holonomy statistics. The mean and median are computed over $b=1,\ldots,1000$, with the $b=1$ column representing the reference sample.}
\begin{tabular}{c|ccc|ccc|ccc}
\hline
& \multicolumn{3}{c|}{$\langle |\omega_{\mathcal L}| \rangle$ (deg.)}
& \multicolumn{3}{c|}{$P_{95}(|\omega_{\mathcal L}|)$ (deg.)}
& \multicolumn{3}{c}{$\mathcal{C}_{\rm sign}$} \\
$L$
& Mean & Median & $b=1$
& Mean & Median & $b=1$
& Mean & Median & $b=1$ \\
\hline
2  & 0.034 & 0.034 & 0.035 & 0.156 & 0.156 & 0.158 & 0.665 & 0.665 & 0.665 \\
3  & 0.070 & 0.070 & 0.071 & 0.322 & 0.322 & 0.326 & 0.701 & 0.701 & 0.701 \\
4  & 0.111 & 0.111 & 0.112 & 0.514 & 0.514 & 0.521 & 0.730 & 0.730 & 0.730 \\
5  & 0.152 & 0.152 & 0.153 & 0.716 & 0.715 & 0.722 & 0.755 & 0.755 & 0.755 \\
6  & 0.188 & 0.188 & 0.190 & 0.910 & 0.910 & 0.918 & 0.773 & 0.773 & 0.773 \\
7  & 0.218 & 0.218 & 0.219 & 1.085 & 1.084 & 1.101 & 0.791 & 0.791 & 0.792 \\
8  & 0.240 & 0.240 & 0.242 & 1.251 & 1.251 & 1.264 & 0.804 & 0.804 & 0.804 \\
9  & 0.261 & 0.261 & 0.263 & 1.381 & 1.379 & 1.396 & 0.819 & 0.819 & 0.820 \\
10 & 0.285 & 0.285 & 0.287 & 1.517 & 1.518 & 1.533 & 0.831 & 0.831 & 0.832 \\
11 & 0.317 & 0.317 & 0.320 & 1.652 & 1.653 & 1.669 & 0.839 & 0.839 & 0.840 \\
12 & 0.361 & 0.361 & 0.365 & 1.875 & 1.875 & 1.891 & 0.848 & 0.848 & 0.849 \\
13 & 0.415 & 0.415 & 0.420 & 2.140 & 2.140 & 2.160 & 0.853 & 0.853 & 0.853 \\
14 & 0.473 & 0.473 & 0.478 & 2.431 & 2.430 & 2.456 & 0.858 & 0.858 & 0.858 \\
15 & 0.529 & 0.529 & 0.535 & 2.778 & 2.775 & 2.810 & 0.862 & 0.862 & 0.863 \\
16 & 0.580 & 0.580 & 0.587 & 3.130 & 3.127 & 3.175 & 0.865 & 0.865 & 0.866 \\
17 & 0.627 & 0.627 & 0.634 & 3.405 & 3.403 & 3.447 & 0.870 & 0.870 & 0.870 \\
18 & 0.667 & 0.666 & 0.674 & 3.654 & 3.649 & 3.700 & 0.873 & 0.873 & 0.874 \\
19 & 0.702 & 0.702 & 0.710 & 3.881 & 3.877 & 3.933 & 0.876 & 0.876 & 0.877 \\
20 & 0.739 & 0.739 & 0.747 & 3.993 & 3.994 & 4.048 & 0.879 & 0.879 & 0.880 \\
\hline
\end{tabular}
\end{table}

To assess whether the representative statistics depend specifically on the reference series ($b=1$), we aggregate over all 1,000 posterior samples. For each $b$, we compute the mean absolute holonomy angle $\langle |\omega_{\mathcal L}^{(b)}| \rangle$ and the 95th percentile $P_{95}(|\omega_{\mathcal L}^{(b)}|)$ over valid pixels. The spatial sign-correlation is defined analogously to Eq.~(\ref{mathcalC_sign_ref_rand}):

\begin{eqnarray}
\mathcal{C}_{\rm sign}^{(b)}
&=&
\frac{1}{2}
\biggl[
\frac{1}{|\Omega_x|}
\sum_{(x,y)\in\Omega_x}
\mathrm{sign}\left(\omega_{\mathcal L}^{(b)}(x,y)\right)
\mathrm{sign}\left(\omega_{\mathcal L}^{(b)}(x+1,y)\right)
\nonumber \\
&&+
\frac{1}{|\Omega_y|}
\sum_{(x,y)\in\Omega_y}
\mathrm{sign}\left(\omega_{\mathcal L}^{(b)}(x,y)\right)
\mathrm{sign}\left(\omega_{\mathcal L}^{(b)}(x,y+1)\right)
\biggr],
\label{mathcalC_sign_b}
\end{eqnarray}

\noindent computed for $b=1,\ldots,1000$, with $\mathrm{sign}(\omega_{\mathcal L}^{(b)})=0$ for numerically zero entries. Table~S3 reports the posterior mean, median, and $b=1$ value for $\langle |\omega_{\mathcal L}^{(b)}| \rangle$, $P_{95}(|\omega_{\mathcal L}^{(b)}|)$, and $\mathcal{C}_{\rm sign}^{(b)}$. For all $L$, the posterior mean and median are nearly identical and agree well with the $b=1$ value. The statistics presented in this study are thus not specific to the reference series but are representative of the full set of GPR posterior samples.

\begin{figure}[b]
\begin{center}
\includegraphics[width=15cm]{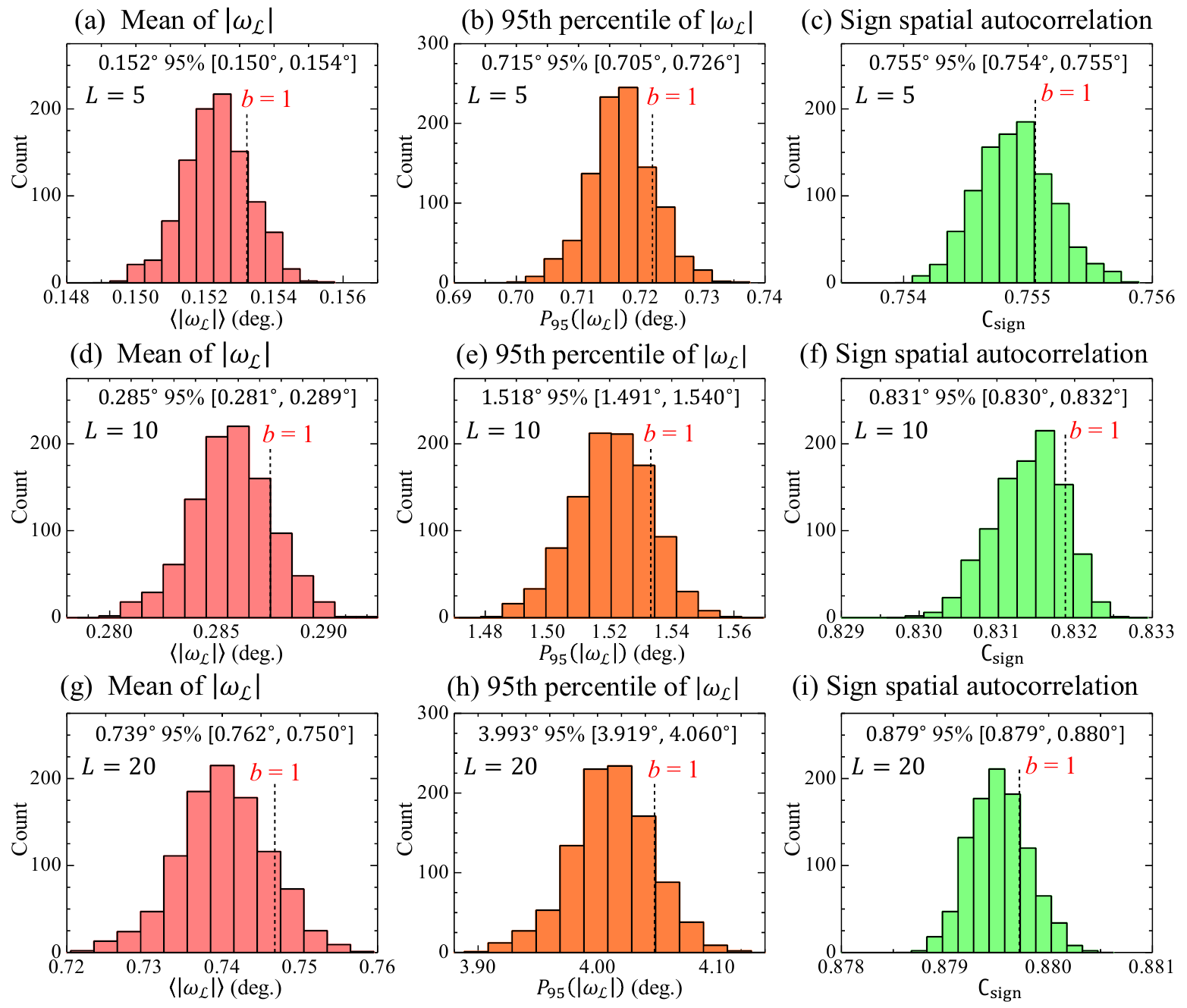}
\end{center}
\caption{Distributions of representative statistics for the signed left-side holonomy $\omega_{\mathcal L}$ across GPR posterior samples. Histograms display the posterior distributions for (a, d, g) the spatial mean of $|\omega_{\mathcal L}|$, (b, e, h) the 95th percentile value $P_{95}(|\omega_{\mathcal L}|)$, and (c, f, i) the nearest-neighbor sign spatial autocorrelation $\mathcal{C}_{\rm sign}$. Loop sizes are (a--c) $L=5$, (d--f) $L=10$, and (g--i) $L=20$. Each panel specifies the posterior mean and central 95\% resampling interval. Vertical dashed lines indicate values from the reference series ($b=1$).}
\end{figure}

Figure~S6 presents histograms of $\langle |\omega_{\mathcal L}^{(b)}| \rangle$, $P_{95}(|\omega_{\mathcal L}^{(b)}|)$, and $\mathcal{C}_{\rm sign}^{(b)}$ for $L=5,10,20$. The distributions are unimodal, and the reference-series ($b=1$) values fall within their representative ranges. The spatial sign correlation satisfies $\mathcal{C}_{\rm sign}>0.75$, indicating strong nearest-neighbor spatial correlation of the sign field. Under the fixed convention, the positive and negative values of $\omega_{\mathcal L}$ therefore form spatially continuous structures inconsistent with a pixel-wise random sign distribution. These results confirm that the magnitude, tail statistics, and sign-field autocorrelation of the left-side holonomy are reproducibly estimated across the posterior samples.

Next, we utilized a shuffle-based null test to evaluate the significance of the observed spatial structure. We generated null fields by retaining the empirical distribution of $|\omega_{\mathcal L}^{\rm ref}(x,y)|$ from the reference sample but randomly shuffling their spatial coordinates. The null hypothesis, $H_0$, posits that the high correlation between the reference and posterior-resampled fields results merely from a random spatial arrangement sharing the same value distribution, rather than an intrinsic spatial structure. Under $H_0$, a shuffled reference field would exhibit a correlation comparable to the posterior-resampled fields. If the observed correlation, $\overline{\mathcal{C}}^{(1000)}$, significantly exceeds the null distribution, it confirms spatial correspondence between the reference and posterior samples beyond random spatial distribution.

Specifically, for each shuffle trial $n=1,\ldots,N_{\rm shuffle}$ (with $N_{\rm shuffle}=500$), we constructed a null field $|\omega_{\mathcal L}^{({\rm Null},n)}(x,y)|$ by randomly reassigning the spatial positions of $|\omega_{\mathcal L}^{\rm ref}(x,y)|$. For each trial $n$, we defined the null correlation as:

\begin{equation}
\mathcal{C}^{\rm Null}(n) =
\mathrm{corr}_{(x,y)\in\Omega_L}
\left(
\left|\omega_{\mathcal L}^{\rm ref}(x,y)\right|,
\left|\omega_{\mathcal L}^{({\rm Null},n)}(x,y)\right|
\right).
\label{C_Null_1000}
\end{equation}

\noindent The empirical $p$ value relative to the null distribution is

\begin{equation}
p_{\rm shuffle}
=
\frac{1}{N_{\rm shuffle}}
\sum_{n=1}^{N_{\rm shuffle}}
\mathbf{1}\left[
\mathcal{C}^{\rm Null}(n)
\ge
\overline{\mathcal{C}}^{(1000)}
\right].
\end{equation}

\noindent To adjust for the 19 loop-size conditions, we applied a Bonferroni-corrected significance threshold of $0.05/19\simeq2.6\times10^{-3}$. A $p_{\rm shuffle}$ below this threshold rejects the null hypothesis, after multiple comparisons correction, that spatial shuffling alone accounts for the observed reproducibility. Our calculations yielded a $p_{\rm shuffle}$ below this threshold for all loop sizes $L=2,\ldots,20$. Consequently, the shuffle-based null model confirms that the preserved spatial structure of the left-side holonomy between the reference and posterior samples is decidedly non-random.

Having established the gauge and resampling stability of $\omega_{\mathcal L}(x,y)$, we now compare it with $\omega_{\mathcal R}(x,y)$ to assess the spatial correspondence between the two holonomies. The two holonomies are constructed from different geometric objects and need not coincide; however, if both are sensitive to the same spatially localized connection mismatches, their spatial distributions should show some consistency. For each posterior sample ($b=1,\ldots,1000$), the absolute-value correlation between $\omega_{\mathcal L}^{(b)}(x,y)$ and $\omega_{\mathcal R}^{(b)}(x,y)$ is

\begin{equation}
\mathcal{C}_{\mathcal{LR}}^{(b)}
=
\mathrm{corr}_{(x,y)\in\Omega_L}
\left(
\left|\omega_{\mathcal L}^{(b)}(x,y)\right|,
\left|\omega_{\mathcal R}^{(b)}(x,y)\right|
\right)
\qquad
(b=1,\ldots,1000).
\end{equation}

\noindent averaged over all 1,000 posterior samples as

\begin{equation}
\overline{\mathcal{C}}_{\mathcal{LR}}^{(1000)}
=
\frac{1}{1000}
\sum_{b=1}^{1000}
\mathcal{C}_{\mathcal{LR}}^{(b)}
\label{C_abs_LR_b},
\end{equation}

\noindent as a measure of left--right correspondence. Table~S2 shows $\overline{\mathcal{C}}_{\mathcal{LR}}^{(1000)}\approx0.61$--$0.71$, indicating a moderate correlation. This suggests that both holonomies are sensitive to spatially localized connection mismatches in the same dataset, even though they are defined for different geometric objects. The correlation tends to increase with $L$ because larger loops enclose a wider area, averaging out local fluctuations and emphasizing broader spatial structures.

Next, to determine whether the correlation between the left-side and right-side holonomies could be explained by a random spatial arrangement, we performed a null test by shuffling the right-side holonomy fields. The observed metric was the posterior-averaged correlation $\overline{\mathcal{C}}_{\mathcal{LR}}^{(1000)}$ between $|\omega_{\mathcal L}^{(b)}(x,y)|$ and $|\omega_{\mathcal R}^{(b)}(x,y)|$, averaged over $b=1,\ldots,1000$. Under the null hypothesis $H_0$, the value distribution of $\omega_{\mathcal R}^{(b)}(x,y)$ is preserved, but its spatial positions are assumed to be independent of $\omega_{\mathcal L}^{(b)}(x,y)$. Specifically, for each sample $b$, the spatial coordinates of $\omega_{\mathcal R}^{(b)}(x,y)$ were randomly shuffled $N_{\rm shuffle}=500$ times. For each shuffle trial $(n=1,\ldots,N_{\rm
shuffle})$, we calculated

\begin{equation}
\mathcal{C}^{\rm Null}(n;b)
=
\mathrm{corr}_{(x,y)\in\Omega_L}\left(
\left|\omega_{\mathcal L}^{(b)}(x,y)\right|,
\left|\left(\omega_{\mathcal R}^{(b)}(x,y)\right)^{\mathrm{rand}}(n)\right|
\right).
\label{eq:C_LR_null}
\end{equation}

\noindent The pooled null distribution of $\mathcal{C}^{\rm Null}(n;b)$ over all $b$ and $n$ is used to define the empirical $p$ value:

\begin{equation}
p_{\mathcal{LR}}
=
\frac{1}{1000\,N_{\rm shuffle}}
\sum_{b=1}^{1000}
\sum_{n=1}^{N_{\rm shuffle}}
\mathbf{1}\left[
\mathcal{C}^{\rm Null}(n;b)
\ge
\overline{\mathcal{C}}_{\mathcal{LR}}^{(1000)}
\right].
\label{eq:p_LR_shuffle}
\end{equation}

\noindent This pooled null distribution served as an empirical reference for the observed posterior-averaged correlation. To account for the 19 loop-size conditions, we applied a Bonferroni-corrected significance threshold of $0.05/19\simeq2.6\times10^{-3}$. A $p_{\mathcal{LR}}$ value below this threshold rejects the null hypothesis, after multiple-comparison correction, that a random spatial arrangement of the right-side holonomy accounts for the observed left--right correlation. Our results showed that $p_{\mathcal{LR}}$ fell below this corrected threshold for all loop sizes $L=2,\ldots,20$. Therefore, this shuffle-based null test indicates that the spatial correspondence between the left- and right-side holonomies cannot be explained by the random spatial arrangement considered here.

\vspace{0.5cm}

\section*{S6.~Comparison of Local Axis Variation and Holonomy}

\vspace{0.3cm}

The residual-rotation-axis field $\widehat{\boldsymbol a}_{\mathcal R}(x,y)$ is obtained from the right-side loop product, whereas the trajectory-plane-normal field $\widehat{\boldsymbol{\mathsf n}}_{\rm plane}(x,y)$ is obtained directly from the right-singular-vector subspace at each pixel. These are not Stokes vectors at individual $T$ points but axis fields extracted from the $T$ evolution of the optical-polarization state. To evaluate how these axis fields vary in real space, we introduce a nearest-neighbor local angular-variation measure and compare it with the holonomy fields. Since the same procedure applies to both, we work with a general axis field $\widehat{\boldsymbol d}(x,y)$, $|\widehat{\boldsymbol d}|=1$, under the identification $\widehat{\boldsymbol d}\sim-\widehat{\boldsymbol d}$, following the gradient-measure approach introduced in our previous study for single-$T$ director fields. We define the angular distance in $\mathbb{R}P^2$ as

\begin{equation}
\theta\bigl((x,y),(x',y')\bigr)
=
\arccos
\left(
\left|
\widehat{\boldsymbol{d}}(x,y)
\cdot
\widehat{\boldsymbol{d}}(x',y')
\right|
\right).
\label{eq:rp2_axis_distance}
\end{equation}

\noindent Using the absolute inner product gives the angular distance for axis data under the identification $\widehat{\boldsymbol{d}}\sim -\widehat{\boldsymbol{d}}$. Equation~\eqref{eq:rp2_axis_distance} is evaluated for nearest-neighbor pairs in the horizontal and vertical directions. For each valid pixel $(x,y)$, let $\mathcal G_{\rm valid}(x,y)$ be the set of nearest-neighbor pixels for which a finite angular distance can be computed. We define

\begin{equation}
g(x,y) =
\frac{1}{|\mathcal G_{\rm valid}(x,y)|}
\sum_{(x',y')\in \mathcal G_{\rm valid}(x,y)}
\theta\bigl((x,y),(x',y')\bigr).
\label{eq:local_axis_gradient}
\end{equation}

\noindent
Here, $g_{\mathcal R}(x,y)$ is obtained by setting $\widehat{\boldsymbol d}(x,y)=\widehat{\boldsymbol a}_{\mathcal R}(x,y)$, while $g_{\rm plane}(x,y)$ is obtained by setting $\widehat{\boldsymbol d}(x,y)=\widehat{\boldsymbol{\mathsf n}}_{\rm plane}(x,y)$. Thus, $g_{\mathcal R}$ and $g_{\rm plane}$ measure the local angular variation of the residual rotation-axis field and the temperature-trajectory-plane normal, respectively. The valid set $\mathcal G_{\rm valid}(x,y)$ need not contain all four cardinal nearest neighbors; after excluding image boundaries and uncomputable regions, it comprises only neighboring pixels with evaluable angular distances. If fewer than four neighbors are available, $g(x,y)$ is still computed using the remaining edges, provided at least one valid nearest-neighbor edge exists. The quantity $g(x,y)$ thus represents the average angular difference between adjacent pixels in the corresponding axis field, where larger values indicate more pronounced local spatial variations. It is important to note that $g(x,y)$ is neither a holonomy nor a gradient of holonomy. It is a non-negative local measure of nearest-neighbor angular variation in the corresponding axis field. In contrast, $\omega_{\mathcal R}$ and $\omega_{\mathcal L}$ are loop-level connection-geometric quantities defined by the residual rotation accumulated after ordered transport around a closed loop. Consequently, $g(x,y)$ does not contain the signed connection information carried by $\omega_{\mathcal L}$. In the subsequent analysis, we compare $g_{\mathcal R}(x,y)$ and $g_{\rm plane}(x,y)$ with the holonomy maps to examine the relationship between local axis variation and loop-level geometric mismatch.

To compare $g(x,y)$ with the $L\times L$-loop holonomy maps across different spatial scales, an independent coarse-graining length $L_g$ is introduced. For a base point $(x,y)$, the set of pixels within the $L_g\times L_g$ local region is

\begin{equation}
\mathcal G_{L_g}(x,y)
=
\left\{
(x',y')
\,\middle|\,
x\leq x' < x+L_g,\;
y\leq y' < y+L_g
\right\}.
\end{equation}

\noindent and the subset for which $g(x',y')$ takes a finite value is

\begin{equation}
\mathcal G_{L_g}^{\rm valid}(x,y)
=
\left\{
(x',y')\in \mathcal G_{L_g}(x,y)
\,\middle|\,
g(x',y')\ {\rm is\ finite}
\right\}.
\end{equation}

\noindent The coarse-grained local variation over valid pixels is then

\begin{equation}
g(x,y;L_g)
=
\frac{1}{
\left|
\mathcal G_{L_g}^{\rm valid}(x,y)
\right|
}
\sum_{(x',y')\in \mathcal G_{L_g}^{\rm valid}(x,y)}
g(x',y').
\label{eq:gL_local_average}
\end{equation}

\noindent Near boundaries or uncomputable regions, the average is taken over whatever valid pixels are available; the full $L_g\times L_g$ window is not required. If $\mathcal G_{L_g}^{\rm valid}(x,y)$ is empty, $g(x,y;L_g)$ is undefined. This normalized box average coarse-grains the local axis-field variation over a spatial scale $L_g$, enabling comparison with holonomy maps evaluated using different loop sizes $L$.

\begin{figure}[b]
\begin{center}
\includegraphics[width=15cm]{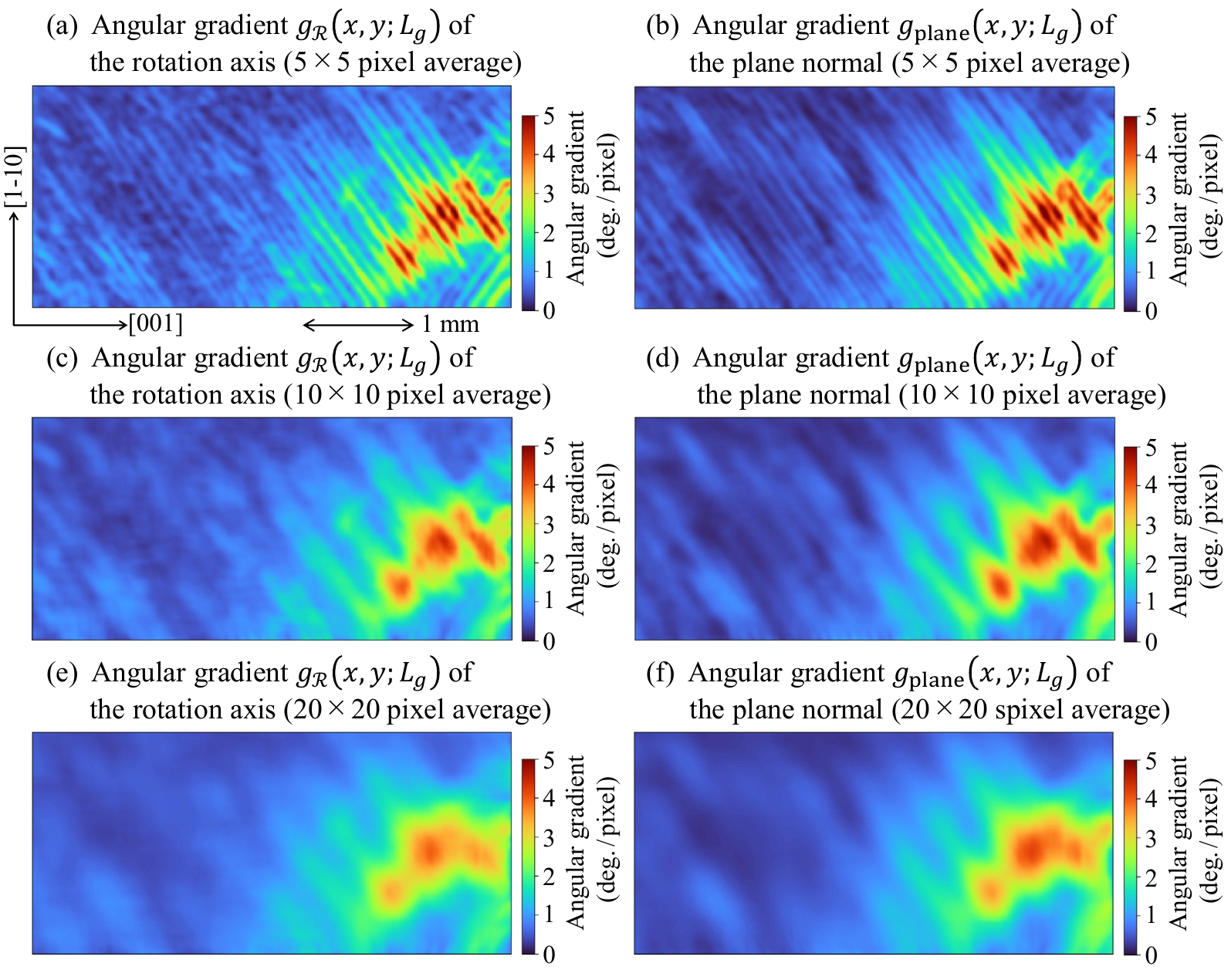}
\end{center}
\caption{Coarse-grained local angular-variation maps of the
local axis fields.
(a, c, e) Local angular variation $g_{\mathcal R}(x,y;L_g)$ calculated from the residual rotation axis $\widehat{\boldsymbol a}_{\mathcal R}(x,y)$. (b, d, f) Local angular variation $g_{\rm plane}(x,y;L_g)$ calculated from the plane-normal director $\widehat{\boldsymbol {\mathsf n}}_{\rm plane}(x,y)$. Maps were obtained by box-averaging the local angular variation over $L_g\times L_g$ pixel windows for (a, b) $L_g=5$, (c, d) $L_g=10$, and (e, f) $L_g=20$. Near image boundaries or uncomputable regions, the box average was evaluated over the available valid pixels within the $L_g\times L_g$ window. Pixels lacking valid data are omitted.}
\end{figure}

Figure~5 of the main text depicts the spatial distributions of the local axis-variation measure $g_{\mathcal R}(x,y)$ (from $\widehat{\boldsymbol a}_{\mathcal R}(x,y)$) and $g_{\rm plane}(x,y)$ (from $\widehat{\boldsymbol {\mathsf n}}_{\rm plane}(x,y)$). Figure~S7 extends this by showing coarse-grained maps $g_{\mathcal R}(x,y;L_g)$ and $g_{\rm plane}(x,y;L_g)$, derived by applying Eq.~(\ref{eq:gL_local_average}). These parameters measure local angular differences between adjacent pixels rather than holonomies along closed loops. The similar spatial enhancement of the two quantities is consistent with the fact that both ultimately originate from geometric objects derived from the same optical-polarization temperature trajectories. The fact that both $g(x,y)$ maps are enhanced in analogous regions indicates that variations in axis direction and trajectory-plane orientation within the temperature trajectory are spatially correlated. Moreover, these local axis-variation distributions align broadly with single-$T$ gradient maps from our previous research, which demonstrated consistent spatial distributions across a wide $T$ range. Thus, the local variation of the axis fields over the full $T$ sweep reflects a stable real-space structure intrinsic to the $T$ evolution, not merely an isolated artifact at a specific $T$.

Next, we assess the correlation between $\omega_{\mathcal R}(x,y)$ and $g_{\mathcal R}(x,y)$. Because $\omega_{\mathcal R}$ utilizes holonomy loops with $L=10$, its computable area is smaller than the full image. We therefore restrict our analysis to $\Omega_{\rm com}$, the common valid pixel set where both quantities are defined. To observe $g_{\mathcal R}(x,y)$ in areas of large $\omega_{\mathcal R}(x,y)$, we isolate the pixel set corresponding to the top $p\%$ of $\omega_{\mathcal R}(x,y)$ on $\Omega_{\rm com}$:

\begin{equation}
\Omega_{\omega}^{(p)}
:=
\left\{
(x,y)\in\Omega_{\rm com}
\,\middle|\,
\omega_{\mathcal R}(x,y)
\ {\rm belongs\ to\ the\ top}\ p\%
\ {\rm within}\ \Omega_{\rm com}
\right\}.
\label{eq:Omega_omega_p}
\end{equation}

\noindent For this correlation analysis, we extract pixel-wise pairs of $\omega_{\mathcal R}(x,y)$ and $g_{\mathcal R}(x,y)$ on $\Omega_{\omega}^{(p)}$ and compute both the Pearson product-moment and Spearman rank correlation coefficients across $p=1,\ldots,100$. The Pearson coefficient measures linear correspondence, whereas the Spearman coefficient evaluates rank-based monotonic correlation, making it particularly useful for nonlinear relationships.

To evaluate the spatial overlap of their high-value regions, we independently identify the top $p\%$ pixel sets for both $\omega_{\mathcal R}(x,y)$ and $g_{\mathcal R}(x,y)$ on $\Omega_{\rm com}$. Analogous to $\Omega_{\omega}^{(p)}$ in Eq.~(\ref{eq:Omega_omega_p}), we define the set for the top $p\%$ of $g_{\mathcal R}(x,y)$ as:

\begin{equation}
\Omega_{g}^{(p)}
:=
\left\{
(x,y)\in\Omega_{\rm com}
\,\middle|\,
g_{\mathcal R}(x,y)
\ {\rm belongs\ to\ the\ top}\ p\%
\ {\rm within}\ \Omega_{\rm com}
\right\}.
\label{eq:Omega_g_p}
\end{equation}

\noindent Unlike the prior correlation analysis (confined to the $\omega_{\mathcal R}$-selected region $\Omega_{\omega}^{(p)}$), this overlap analysis directly compares the independently selected high-value regions, $\Omega_{\omega}^{(p)}$ and $\Omega_g^{(p)}$. Because these sets typically differ, we quantify their spatial alignment using the intersection over union (IoU) and Dice coefficient:

\begin{eqnarray}
{\rm IoU}_{\omega, g}(p)
&=&
\frac{
\left|
\Omega_{\omega}^{(p)}
\cap
\Omega_{g}^{(p)}
\right|
}{
\left|
\Omega_{\omega}^{(p)}
\cup
\Omega_{g}^{(p)}
\right|
},\\
{\rm Dice}_{\omega, g}(p)
&=&
\frac{
2\left|
\Omega_{\omega}^{(p)}
\cap
\Omega_{g}^{(p)}
\right|
}{
\left|
\Omega_{\omega}^{(p)}
\right|
+
\left|
\Omega_{g}^{(p)}
\right|
}.
\end{eqnarray}

\noindent Both metrics gauge the spatial congruence between regions of elevated $\omega_{\mathcal R}(x,y)$ and elevated $g_{\mathcal R}(x,y)$.

\begin{figure}[b]
\begin{center}
\includegraphics[width=11cm]{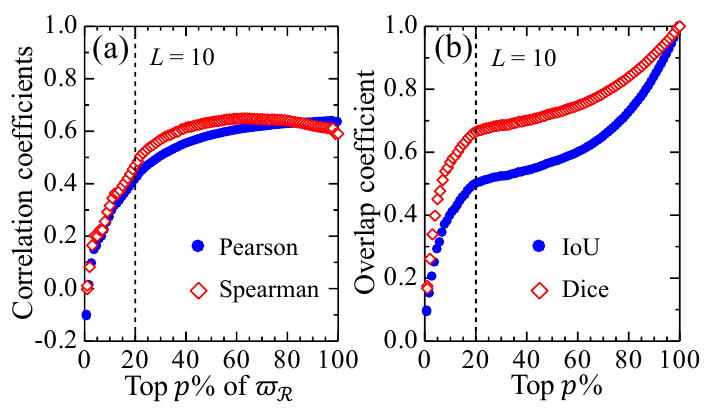}
\end{center}
\caption{Comparison between right-side holonomy strength $\omega_{\mathcal R}(x,y)$ ($L=10$) and local axis variation $g_{\mathcal R}(x,y)$. (a) Pearson and Spearman correlation coefficients between $\omega_{\mathcal R}(x,y)$ and $g_{\mathcal R}(x,y)$ evaluated over the pixel set $\Omega_{\omega}^{(p)}$ corresponding to the top $p\%$ of $\omega_{\mathcal R}(x,y)$ within the common valid set $\Omega_{\rm com}$. (b) Spatial overlap between high-value regions of $\omega_{\mathcal R}(x,y)$ and $g_{\mathcal R}(x,y)$ quantified by intersection over union (IoU) and Dice coefficients; here, $\Omega_{\omega}^{(p)}$ and $\Omega_g^{(p)}$ are independently defined as the top $p\%$ pixel sets of $\omega_{\mathcal R}(x,y)$ and $g_{\mathcal R}(x,y)$ within $\Omega_{\rm com}$. The vertical dashed line marks the representative condition $p=20\%$.}
\end{figure}

Figure~S8 illustrates the $p$-dependence of the correlation, IoU, and Dice coefficients. Figure~S8(a) demonstrates that correlation coefficients are near zero for $p<5\%$ but climb rapidly as $p$ grows. Around $p=20\%$, correlation reaches a moderate $0.4$, eventually saturating near $0.6$ at higher $p$ values. The negligible difference between the Pearson and Spearman coefficients indicates that the observed association between $\omega_{\mathcal R}(x,y)$ and $g_{\mathcal R}(x,y)$ is similarly captured by linear and rank-based correlation measures over a wide $p$ range. In Fig.~S8(b), the IoU and Dice coefficients naturally converge to 1 at $p=100\%$, where both sets encompass all valid pixels. While the Dice coefficient consistently exceeds the IoU, both surge rapidly for $p<20\%$, plateau between $20\% \leq p < 60\%$, and incrementally approach 1 for $p \geq 60\%$. Nevertheless, these correlation and overlap metrics never reflect a perfect one-to-one correspondence, supporting the interpretation that $\omega_{\mathcal R}(x,y)$ is not simply a local angular-variation measure.

In our previous study, $K$-shape clustering identified a stress-concentrated region with enhanced $T_{\rm F}$, covering approximately 25\% of the analyzed pixels, suggesting a potential overlap with the top 20\% $\omega_{\mathcal R}(x,y)$ region. Evaluating this overlap yielded an IoU of 0.396 and a Dice coefficient of 0.567, with 64.5\% of the pixels in the top-20\% $\omega_{\mathcal R}(x,y)$ mask lying within the stress-concentrated mask. This result shows a moderate spatial overlap between the large-holonomy region and the characteristic region extracted by the $K$-shape analysis. Consequently, using the top 20\% threshold for high $\omega_{\mathcal R}$ is consistent with the independent clustering characterization. Complete overlap is not anticipated, however, as $\omega_{\mathcal R}$ is a loop-scale geometric metric ($L=10$), whereas $K$-shape clustering classifies individual pixels based strictly on temperature response. Despite both metrics being sensitive to overlapping stress-concentrated areas, their differing spatial scales and fundamental definitions preclude perfect alignment.

\vspace{0.5cm}

\section*{S7.~Empirical Loop-Size Dependence of Right- and Left-side Holonomies}

\vspace{0.3cm}

We now analyze how $\omega_{\mathcal R}(x,y)$ and $\omega_{\mathcal L}(x,y)$ depend on the holonomy loop size $L$. For brevity, we collectively refer to both as $\omega_{\chi}(x,y;L)$, where $\chi\in\{\mathcal R,\mathcal L\}$. Note that $\omega_{\mathcal R}(x,y;L)$ represents a non-negative angle, while $\omega_{\mathcal L}(x,y;L)$ is signed; however, we utilize the absolute value $|\omega_{\chi}(x,y;L)|$ for both in this analysis. As $L$ expands, nearest-neighbor transformations accumulate along the path roughly proportional to $4L$. This typically increases the holonomy angle's magnitude, though internal loop cancellations can occur. We investigate whether an empirical power-law relationship between mean holonomy strength and $L$ holds within the top $p\%$ pixel set:

\begin{equation}
\overline{|\omega_{\chi}|}(L;p)
\propto
L^{\alpha_{\chi}(p)}.
\label{eq:scaling_alpha}
\end{equation}

\noindent Here, $\overline{|\omega_{\chi}|}(L;p)$ denotes the mean absolute holonomy strength for the top $p\%$ pixel set of $|\omega_{\chi}(x,y)|$ at loop size $L$, and $\alpha_{\chi}(p)$ is an empirical exponent that characterizes the approximate increase of this mean value with $L$. This exponent is used only as a descriptive parameter for loop-size dependence, not as a critical or universal scaling exponent. Importantly, $\alpha_{\chi}(p)$ is estimated from the aggregated mean holonomy strength of the top $p\%$ pixel set, rather than being assigned to individual pixels.

For each loop size $L$, the pixel set belonging to the top $p\%$ of the nonzero holonomy-strength distribution is defined. Because pixels with $\omega_{\chi}=0$ contribute nothing to the log-scale analysis, the valid nonzero pixel set is first defined as

\begin{equation}
\Omega_{L,\chi}^{+}
=
\left\{
(x,y)\in\Omega_L
\,\middle|\,
|\omega_{\chi}(x,y;L)|>0
\right\}.
\end{equation}

\noindent The threshold at each $L$ is the $(100-p)$th percentile within $\Omega_{L,\chi}^{+}$:

\begin{equation}
\omega_{\mathrm{thr},\chi}(L;p)
:=
\operatorname{Percentile}_{100-p}
\left(
\{
|\omega_{\chi}(x,y;L)|
\}_{(x,y)\in\Omega_{L,\chi}^{+}}
\right),
\end{equation}

\noindent and the top $p\%$ pixel set is

\begin{equation}
\Omega_{\mathrm{top},\chi}(L;p)
=
\left\{
(x,y)\in\Omega_{L,\chi}^{+}
\,\middle|\,
|\omega_{\chi}(x,y;L)|
\ge
\omega_{\mathrm{thr},\chi}(L;p)
\right\},
\label{Omega_top_L}
\end{equation}

\noindent where $\Omega_{\mathrm{top},\chi}(L;p)$ is the pixel set where $|\omega_{\chi}|$ falls within the top $p\%$ of the nonzero holonomy-strength distribution at loop size $L$. Thus, $p=100\%$ encompasses all pixels in $\Omega_{L,\chi}^{+}$, representing all valid pixels with nonzero holonomy strength. For a given $p$, $\Omega_{\mathrm{top},\chi}(L;p)$ typically varies with $L$. Using this filtered set, the mean holonomy strength over the selected nonzero pixels is defined as

\begin{equation}
\overline{|\omega_{\chi}|}(L;p)
=
\frac{1}{
\left|
\Omega_{\mathrm{top},\chi}(L;p)
\right|
}
\sum_{(x,y)\in\Omega_{\mathrm{top},\chi}(L;p)}
|\omega_{\chi}(x,y;L)|.
\label{eq:omega_mean_top}
\end{equation}

\noindent and its logarithm is

\begin{equation}
m_{\chi}(L;p)
=
\log
\overline{|\omega_{\chi}|}(L;p)
=
\log
\left[
\frac{1}{
\left|
\Omega_{\mathrm{top},\chi}(L;p)
\right|
}
\sum_{(x,y)\in\Omega_{\mathrm{top},\chi}(L;p)}
|\omega_{\chi}(x,y;L)|
\right].
\label{m_L_P}
\end{equation}

\noindent Therefore, the data points used in the regression analysis are not the pixel-wise values $\log|\omega_{\chi}(x,y;L)|$, but the representative values $m_{\chi}(L;p)$ obtained for each $L$. For fixed $\chi$ and $p$, a sequence of 19 points $\{(\log L,\,m_{\chi}(L;p))\}_{L=2}^{20}$ is constructed. Taking the logarithm of Eq.~(\ref{eq:scaling_alpha}) yields

\begin{equation}
m_{\chi}(L;p)
\simeq
\alpha_{\chi}(p)\log L
+
\beta_{\chi}(p),
\end{equation}

\noindent so that $\alpha_{\chi}(p)$ is estimated as the slope of a linear regression against $\log L$. Since converting $\omega_{\chi}$ from degrees to radians adds only the constant $\log(\pi/180)$ to $m_{\chi}(L;p)$, the slope estimate $\alpha_{\chi}(p)$ is unaffected by the angular unit when a free intercept $\beta_{\chi}(p)$ is included; the analysis was carried out consistently using the same unit system throughout.

For fixed $\chi\in\{\mathcal R,\mathcal L\}$ and $p$, we assume the linear model

\begin{equation}
m_{\chi}(L;p)
=
\alpha_{\chi}(p)\log L
+
\beta_{\chi}(p)
+
\varepsilon_L,
\label{eq:linear_scaling_model}
\end{equation}

\noindent where $\varepsilon_L$ is the deviation of the logarithmic mean holonomy strength from a strictly linear relationship. We apply this procedure independently to both $\omega_{\mathcal R}(x,y)$ and $\omega_{\mathcal L}(x,y)$. Figure~6 of the main text provides log-log plots of $L$ versus $\overline{|\omega_\chi|}$ for both holonomies at $p=10,20,40,60,$ and $100\%$. The regression line slope for each $p$ represents $\alpha_{\chi}(p)$. Because the actual data exhibit weak curvature and gradual fluctuations rather than forming a perfect straight line, our initial step avoids explicitly modeling these deviations. Instead, we apply a least-squares linear fit to the 19 representative values ($L=2,\ldots,20$) for each $\chi$ and $p$. We denote these estimates as $\alpha_{\mathcal R}^{\mathrm{ls}}(p)$ and $\alpha_{\mathcal L}^{\mathrm{ls}}(p)$. The insets in Fig.~6 of the main text depict the $p$-dependence of these parameters.

We also assess this single-line model via Bayesian estimation, where deviations from linearity appear through the residual scale. We introduce weakly informative priors to guarantee numerical stability and to capture linear deviations within the posterior distribution. For the exponent $\alpha_{\chi}(p)$, we apply a broad prior centered on a positive value, and in the implementation, we restrict $\alpha_{\chi}(p)>0$. The intercept $\beta_{\chi}(p)$, representing the overall vertical scale of $m_{\chi}(L;p)$, receives a broad normal prior spanning the expected numerical range of the logarithmic mean holonomy strength. These hyperparameters were pre-determined; we deliberately avoided an empirical Bayes procedure where priors are derived from observations. Consequently, our priors primarily ensure numerical stability and gently regularize parameter scales. Specifically, for a fixed $\chi\in\{\mathcal R,\mathcal L\}$ and $p$, we treat the $m_{\chi}(L;p)$ values ($L=2,\ldots,20$) as observed data, assuming the following likelihood and prior distributions:

\begin{align}
m_{\chi}(L;p)
&\sim
\mathcal{N}
\left(
\alpha_{\chi}(p)\log L+\beta_{\chi}(p),
\,\sigma_{\chi}(p)^2
\right),\\
\alpha_{\chi}(p)
&\sim
\mathcal{N}^{+}(1,\,1^2),\\
\beta_{\chi}(p)
&\sim
\mathcal{N}(0,\,2^2),\\
\sigma_{\chi}(p)
&\sim
\mathrm{Exponential}(\ln 10),
\end{align}

\noindent where $\mathcal{N}^{+}(1,1^2)$ is a normal distribution truncated to $\alpha_{\chi}(p)>0$ and $\mathrm{Exponential}(\ln 10)$ is an exponential distribution with density

\begin{equation}
p(\sigma_{\chi})
=
(\ln 10)\exp[-(\ln 10)\sigma_{\chi}],
\qquad
\sigma_{\chi}>0.
\end{equation}

\noindent The exponential distribution is restricted to positive values and therefore a natural prior for $\sigma_{\chi}(p)$; the rate $\ln 10$ gives a weakly informative but not overly broad distribution. The posterior was obtained by Markov chain Monte Carlo (MCMC) with 20 chains, 3,000 iterations each, and 1,000 warm-up iterations discarded. Convergence was verified by $\widehat{R}<1.05$ for all parameters.

\begin{figure}[b]
\begin{center}
\includegraphics[width=15cm]{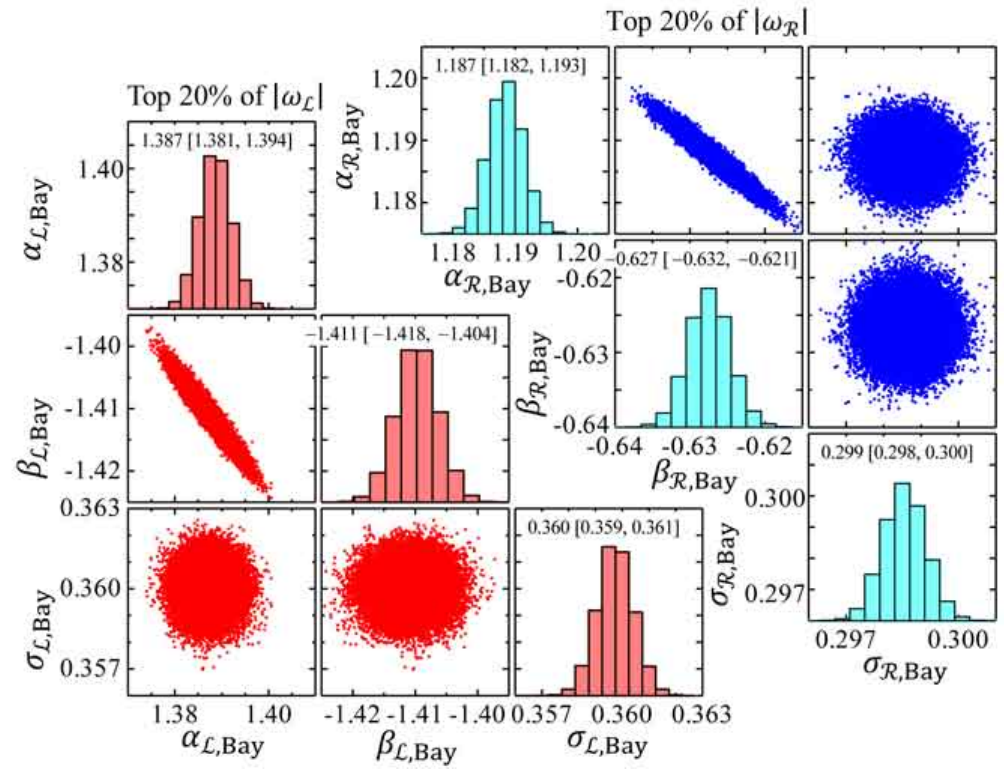}
\end{center}
\caption{Pairwise posterior distributions of Bayesian linear-regression parameters for the empirical loop-size-dependence analysis at $p=20\%$. The lower-left block displays posterior samples for the left-side holonomy ($\alpha_{\mathcal L, {\rm Bay}}$, $\beta_{\mathcal L, {\rm Bay}}$, and $\sigma_{\mathcal L, {\rm Bay}}$), while the upper-right block displays those for the right-side holonomy ($\alpha_{\mathcal R, {\rm Bay}}$, $\beta_{\mathcal R, {\rm Bay}}$, and $\sigma_{\mathcal R, {\rm Bay}}$). Diagonal panels present marginal posterior distributions, and off-diagonal panels illustrate pairwise posterior correlations. Values in the diagonal panels denote the posterior mean and the 95\% Bayesian credible interval.}
\end{figure}

\begin{figure}[b]
\begin{center}
\includegraphics[width=15cm]{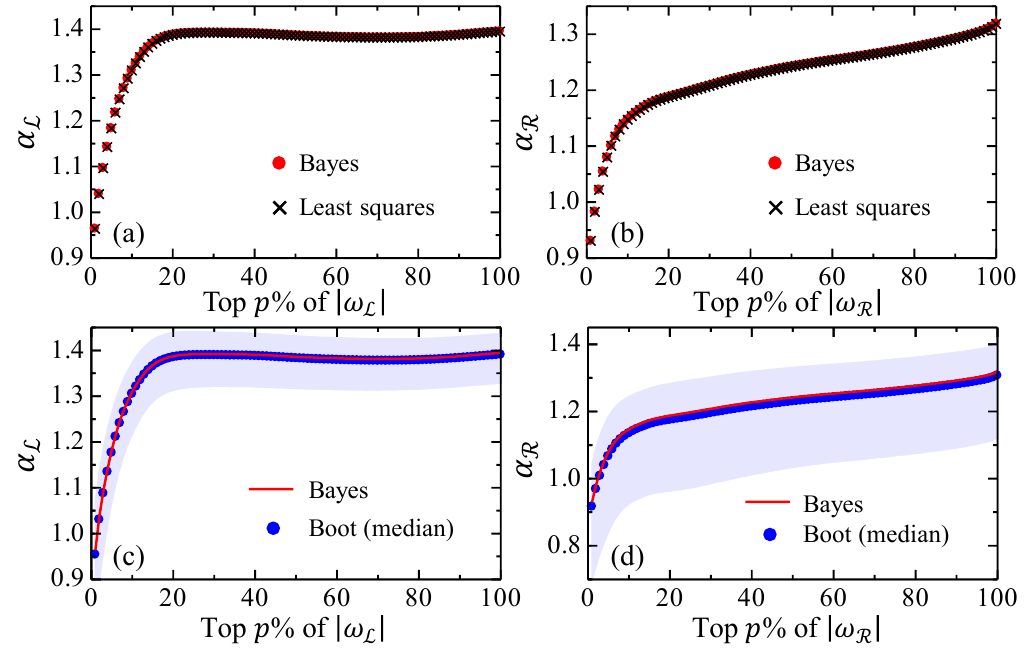}
\end{center}
\caption{Estimation of the empirical exponent $\alpha_{\chi}(p)$, where $\chi\in\{\mathcal R,\mathcal L\}$, for both holonomies. (a, b) Comparison of least-squares estimates with posterior means from Bayesian linear regression for (a) $|\omega_{\mathcal L}|$ and (b) $|\omega_{\mathcal R}|$. (c, d) Comparison of Bayesian estimates against bootstrap medians for (c) $|\omega_{\mathcal L}|$ and (d) $|\omega_{\mathcal R}|$. Shaded regions denote 95\% bootstrap intervals. The horizontal axis represents the top $p\%$ pixel set, filtered by $|\omega_{\chi}|$. Bayesian credible intervals match the marker size and are therefore imperceptible at this scale.}
\end{figure}

From the posterior, $M_s=40{,}000$ samples of $\alpha_{\chi,{\rm Bay}}$ were drawn for each $\chi$ and $p$. The posterior means are

\begin{eqnarray}
\overline{\alpha}_{\mathcal R}^{\rm Bay}(p)
&:=&
\frac{1}{M_s}
\sum_{s=1}^{M_s}
\alpha_{\mathcal R,{\rm Bay}}^{(s)}(p), \\
\overline{\alpha}_{\mathcal L}^{\rm Bay}(p)
&:=&
\frac{1}{M_s}
\sum_{s=1}^{M_s}
\alpha_{\mathcal L,{\rm Bay}}^{(s)}(p).
\end{eqnarray}

\noindent Figure~S9 shows pairwise plots of MCMC posterior distributions at $p=20\%$. All posterior distributions are narrower than the priors, indicating that the parameters are constrained by the representative loop-size data. Figure~S10 compares the Bayesian posterior means with the least-squares estimates for both $\omega_{\mathcal L}$ (panel a) and $\omega_{\mathcal R}$ (panel b). The 95\% Bayesian credible intervals are narrow and comparable to the marker size in Fig.~S10. These intervals reflect posterior uncertainty in fitting the linear model to $\{m_{\chi}(L;p)\}_{L=2}^{20}$ and do not model spatial correlations or dependence among pixels in the top-$p\%$ set. The Bayesian estimates are in close agreement with the least-squares estimates, confirming consistency between the two methods. In both cases, a bend appears near $p=10$--$20\%$, interpreted as a crossover from behavior dominated by a small number of high-holonomy pixels to behavior that includes the broader background.

A bootstrap analysis was also performed as an additional robustness check for $\alpha_{\chi}(p)$. This analysis examines how the fitted slope changes when the scale points in the sequence $\{(\log L,m_{\chi}(L;p))\}_{L=2}^{20}$ are resampled, and thereby evaluates the sensitivity of the regression to weak nonlinear deviations from a single straight line. The bootstrap analysis is not intended to replace the Bayesian estimation; rather, it provides a supplementary check of the stability of $\alpha_{\chi}(p)$ with respect to the choice of representative scale points.

For a fixed $\chi$ and $p$, we analyze 19 data points: $\{(\log L,m_{\chi}(L;p))\}_{L=2}^{20}$. For each bootstrap sample (${\rm boot}=1,\ldots,N_{\mathrm{boot}}$), we draw $L_1^{({\rm boot})},\ldots,L_{19}^{({\rm boot})}$ with replacement from $\{2,\ldots,20\}$ and generate the corresponding sequence $\{(\log L_s^{({\rm boot})},m_{\chi}(L_s^{({\rm boot})};p))\}_{s=1}^{19}$. This resampling iterates $N_{\mathrm{boot}}=1{,}000$ times. We then apply a least-squares linear regression to the resampled sequence:

\begin{equation}
m_{\chi}(L_s^{({\rm boot})};p)
=
\alpha_{\chi}^{({\rm boot})}(p)\log L_s^{({\rm boot})}
+
\beta_{\chi}^{({\rm boot})}(p)
+
\varepsilon_s^{({\rm boot})}
\qquad
(s=1,\ldots,19),
\label{m_L_P_boot}
\end{equation}

\noindent yielding $\alpha_{\chi}^{({\rm boot})}(p)$. The procedure is applied independently to both holonomies, producing bootstrap distributions $\{\alpha_{\mathcal R}^{({\rm boot})}(p)\}$ and $\{\alpha_{\mathcal L}^{({\rm boot})}(p)\}$. The median at each $p$ is

\begin{equation}
\alpha_{\chi,\mathrm{med}}^{({\rm boot})}(p)
=
\operatorname{median}_{{\rm boot}}
\left\{
\alpha_{\chi}^{({\rm boot})}(p)
\right\}
\end{equation}

\noindent and the 2.5\% and 97.5\% percentiles quantify the sensitivity of $\alpha_{\chi}(p)$ to the choice of scale points.

\begin{figure}[b]
\begin{center}
\includegraphics[width=9cm]{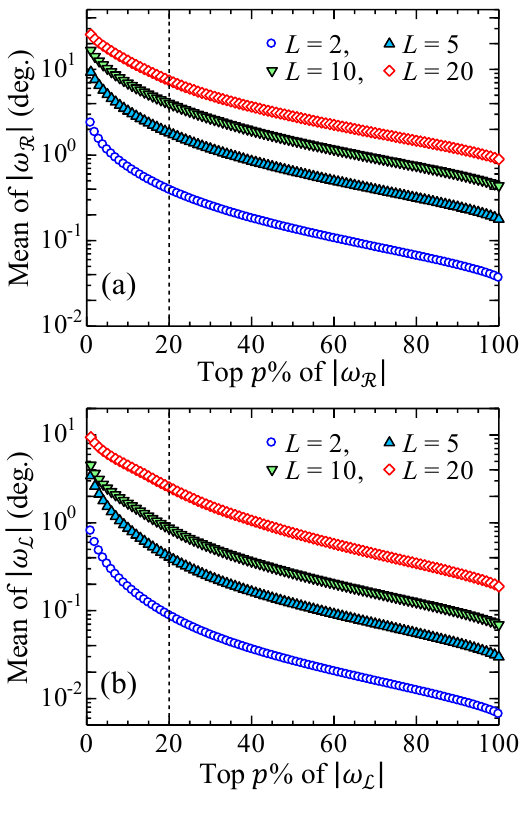}
\end{center}
\caption{Dependence of representative mean holonomy strength $\overline{|\omega_{\chi}|}(L;p)$ on $p$, identifying the top $p\%$ pixel set by $|\omega_\chi|$ within the nonzero holonomy-strength set. (a) Right-side holonomy strength $\overline{|\omega_{\mathcal R}|}(L;p)$ and (b) left-side holonomy strength $\overline{|\omega_{\mathcal L}|}(L;p)$. Results reflect $L=2,5,10,$ and $20$. The vertical dashed line signifies the representative condition, $p=20\%$.}
\end{figure}

Figures~S10(c) and S10(d) depict the $p$-dependence of the bootstrap medians $\alpha_{\mathcal L,\mathrm{med}}^{({\rm boot})}(p)$ and $\alpha_{\mathcal R,\mathrm{med}}^{({\rm boot})}(p)$, calculated from the reference series ($b=1$). These bootstrap medians closely mirror the posterior means $\overline{\alpha}_{\mathcal R}^{\rm Bay}(p)$ and $\overline{\alpha}_{\mathcal L}^{\rm Bay}(p)$ derived from Bayesian estimation. Notably, the 95\% bootstrap interval represents more than the standard error of a regression coefficient; it captures the weak nonlinearity in the $\log L$ dependence of $m_{\chi}(L;p)$, the finite number of scale points, and the sensitivity to the specific regression sequence. Consequently, this interval is typically wider than a standard error from a single linear regression. We interpret the bootstrap distribution as a stability metric for the slope when approximating the scale-dependence sequence $\{(\log L,m_{\chi}(L;p))\}_{L=2}^{20}$ with a single line. Ultimately, the $\alpha_{\chi,\mathrm{med}}^{({\rm boot})}(p)$ curves demonstrate consistent behavior across all three estimation methods. Both the right and left sides exhibit a distinct bend around $p=10$--$20\%$. This bend is interpreted here as a crossover from the low-$p$ behavior dominated by a small number of high-holonomy regions to the high-$p$ behavior that includes broader background regions.

We next examine the representative mean holonomy strength from a complementary viewpoint by fixing $L$ and varying $p$. This is not intended to yield a new estimate of $\alpha_{\chi}(p)$ but to show how the top-$p\%$ selection samples the underlying $|\omega_{\chi}|$ distribution. Figure~S11 shows $\overline{|\omega_{\chi}|}(L;p)$ as a function of $p$. In the small-$p$ region, only a few pixels with large $|\omega_{\chi}|$ contribute, so the mean is high; as $p$ increases, smaller-holonomy pixels are included and the mean decreases. The $p$ dependence evolves gradually across the crossover. The bend in $\alpha_{\chi}(p)$ shown in Fig.~S10 therefore reflects a genuine change in the composition of the selected pixel set, not an artifact of the regression, Bayesian estimation, or bootstrap. From this viewpoint, $p=20\%$ is a practical representative threshold: it captures the high-holonomy region while limiting dominance by extreme outliers and reducing influence from the background.

In summary, the mean absolute holonomy strength increases gradually with $L$, with no clear characteristic loop size or sharp transition in the range $L=2$--20.  Very small loops are more sensitive to pixel-scale variations, whereas larger loops produce stronger coarse graining of the spatial pattern. The choice $L=10$ provides an intermediate scale between pixel-level sensitivity and excessive coarse-graining. It is not intended to identify a critical loop size but to provide a representative scale for visualizing and comparing loop-level connection mismatches.

\end{document}